\documentclass[11pt,a4paper]{article}
\usepackage[ansinew]{luainputenc} 
\usepackage[hmargin=2cm, vmargin=2cm]{geometry}
\usepackage [T1]{fontenc} 
\usepackage{textcomp}
\usepackage{dsfont} 
\usepackage{mathrsfs} 
\usepackage{csquotes} 
\usepackage{amsmath,amsfonts,latexsym, amssymb,amsthm}
\usepackage{fullpage,url,booktabs}  
\usepackage[usenames,dvipsnames]{color}
\usepackage{xcolor}
\definecolor{vertcitation}{RGB}{5,150,5} 
\usepackage[colorlinks,citecolor=vertcitation]{hyperref} 
\usepackage{array} 
\usepackage[sort&compress]{natbib}
\usepackage{stmaryrd} 
\usepackage{xspace}
\usepackage{tikz} 
\usetikzlibrary{matrix}
\usepackage{stackrel} 
\usepackage{cancel} 
\usepackage{comment} 
\newtheorem* {proposition*}{Proposition}
\newtheorem* {theorem*}{Theorem}
\newtheorem* {corollary*}{Corollary}
\newtheorem* {definition*}{Definition}
\newtheorem* {lemma*}{Lemma}
\newtheorem* {assumption*}{Assumption}
\newtheorem {proposition}{Proposition}
\newtheorem {theorem}{Theorem}

\newtheorem {definition}{Definition}

\bibpunct{[}{]}{,}{n}{}{;} 

\newcommand{\prob}{{\rm p}}
\newcommand{\Prob}{{\mathbb{P}}}
\newcommand{\tr}{{\rm Tr}}

\newcommand{\ket}[1]{\ensuremath{|#1\rangle}\xspace}
\newcommand{\bra}[1]{\ensuremath{\langle #1|}\xspace}
\newcommand{\braket}[2]{\ensuremath{\langle #1| #2 \rangle}\xspace}
\newcommand{\ident}{\stackrel{\mbox{{\rm \tiny (def)}}}{=}}
\newcommand{\diag}[1]{\ensuremath\stackrel[#1]{}{\mathrm{Diag} }}
\newcolumntype{C}{>{$}c<{$}}
\newcolumntype{L}{>{$}l<{$}}
\newcolumntype{R}{>{$}r<{$}}
\title{\bf
Information-theoretic interpretation\\ of quantum formalism
}
\author {Michel Feldmann%
\thanks{Electronic address: michel.feldmann@polytechnique.org}
}
\date {}

\begin{document}

\maketitle
\abstract{
We present an information-theoretic interpretation of quantum formalism based on a Bayesian framework and devoid of any extra axiom or principle. Quantum information is construed as a technique for analyzing a logical system subject to classical constraints, based on a question-and-answer procedure.
The problem is posed from a particular batch of queries while the constraints are represented by the truth table of a set of Boolean functions. 
The Bayesian inference technique consists in assigning a probability distribution within a real-valued probability space to the joint set of  queries in order to satisfy the constraints. 
The initial query batch is not unique and alternative batches can be considered at will. They are enabled mechanically from the initial batch, quite simply by transcribing the probability space into a Hilbert space.
It turns out that this sole procedure leads to exactly recover the standard quantum information theory and thus provides an information-theoretic rationale to its technical rules.
In this framework, the great challenges of quantum mechanics become simple platitudes: Why is the theory probabilistic? Why is the theory linear? Where does the Hilbert space come from?
In addition, most of the paradoxes, such as uncertainty principle, entanglement, contextuality, nonsignaling correlation, measurement problem, etc.,  become straightforwards features.
In the end, our major conclusion is that quantum information is nothing but classical information processed by a mature form of Bayesian inference technique and, as such, consubstantial with Aristotelian logic.
} 

\tableofcontents

\section{Introduction}
\label{introduction}

Basically, data are stored in a definite register, but in 1948 C.~E.~Shannon~\cite{shannon} construed a sequence of symbols as a stochastic process, giving rise to information theory.  He thus joined the core concepts of thermodynamics, revealed by the pioneering work of Lèo Szilard on Maxwell's demon dating back to 1929~\cite{szilard, brillouin}, opening a new horizon sometimes viewed as the ultimate explanatory principle in physics~\cite{wheeler1,landauer}. Nowadays, classical information theory focuses essentially on uncertain discrete variables.
 In 1957, E.~T.~Jaynes incorporated the Shannon's concept of entropy in the Bayesian inference theory~\cite{jaynes}. Later, contemplating quantum mechanical formalism, Jaynes noted in 1989~\cite{clearing} that this formalism is strongly reminiscent of the Bayesian model. More explicitly, C.~M.~Caves, C.~A.~Fuchs, and R.~Schack~\cite{caves3} proposed in 2002 in a seminal paper especially endorsed by N. D. Mermin~\cite{mermin5}, to understand quantum probability within a Bayesian framework.  Fuchs coined the term \enquote{QBism}~\cite{fuchs2} for \enquote{Quantum Bayesianism} to describe this view.

Independently, in a pair of papers~\cite{mf3,mf4}, we demonstrated that a surprising way to deeply boost \emph{conventional computation} is to regard calculation as a Bayesian estimation~\cite{boyd} of the Boolean variables involved.  
This means applying probability theory as an alternative tool to solve a mathematical problem, although the uncertainty about the solution sought has nothing to do with that of a conventional random problem.
This nevertheless works because standard probability laws are just the extension of Aristotelian logic rules  to cases where the variables are uncertain, as pointed out by R. T. Cox~\cite{cox} and E. T. Jaynes~\cite{jaynes}.
Technically, this implies \emph{taking probabilities for the very unknowns} of the problem instead of the variables themselves and next equating the calculation to a problem of inference.

\subsection{Motivation}
In this paper, we aim to confront quantum information with \enquote{Bayesian computation}, i.e., calculation employing Bayesian inference, with the primary objective to understand the potential effectiveness of quantum computation.
In quantum information,  data are natively probabilistic and encoded as density operators in a Hilbert space $\mathcal{H}$ whose basis vectors are labelled by the discrete states of a classical register.  

Unlike calculation that consider a unique batch of binary digits and is thus purely static, quantum information describes a multiplicity of viewpoints and, e. g., can directly address the evolution of the system.
To take account of this context, we propose to characterize quantum information by a pair of ingredients: (1) \emph{a register}, to store and compute the input data, and (2) \emph{a set of communication channels} to expose different viewpoints on the system. 
This sole procedure leads to both a profound revision of the very essence of quantum information and to an advance in Bayesian inference techniques. Let us start with an informal draft of the model.

 \subsection{Quantum information in a nutshell}
\label{nutshell}


Consider a memory containing a maximum of (say) $ N $ bits of information accessible by a procedure of questions-and-answers. It can be specified by a  particular batch of $N$ queries, that is to say,  $ N $ Boolean variables, which display $2^N$ distinct classical states. It is clear that this batch  is by no means unique, so it only defines a so-called \emph {observation window}  termed \enquote{source observation window}.
Ideally, we would like all queries to be mutually independent, but generally this cannot be determined in advance.

The problem arises when the specifications are not complete but only based on the observation of a limited number of  Boolean functions. At best, the memory can only be evaluated by Bayesian estimation.
Technically, any Bayesian probability is then a linear combination of the $2^N$ classical state probabilities. 
As a result, the input data  is  a specific set of linear functions of classical state probabilities, referred to as \emph{observables}.  The full input is called \emph{Bayesian prior}. 
For convenience, the memory is called \emph{Bayesian theater}.

The first task of quantum information is the analysis of the source window, that is, the likelihood of the $2^N$ states compatible with the prior, i.e., technically, the estimation of their probability. In general, depending on this prior, there is more than one solution and even a continuous set of feasible probability distributions. It can be shown that the locus of these solutions is a specific simplex, say $\mathcal{W}$, in the real-valued vector space spanned by the  $2^N$ states.
A particular solution, called \emph{working distribution}, say $w$, can be singled out on the simplex by its barycentric coefficients, which we term \emph{contextual distribution}. 
Remarkably, it turns out that  the conventional \emph{quantum state} of standard quantum information is the equivalent of the pair $(w, \mathcal{W})$, composed of the working distribution $w$  and the simplex $\mathcal{W}$. This pair $(w, \mathcal{W})$ is thus called \emph{simplicial quantum state}. This equivalence can be extended: When there is only one solution, the simplicial quantum state is called a \emph{pure state} and otherwise it is termed a \emph{mixed state}.

By construction, the only expectation values that can be assessed in the source window are the linear combinations of the $2^N$ state probabilities but this is far from exhausting the set of all possible observables on the full memory. 
Therefore, to complement this ensemble, it is necessary to reformulate the issue with \emph{other Boolean batches of queries}, constituting some kind of factor analysis. 
The way to construct each batch of relevant Boolean variables is the main novelty of quantum information.

The second task is indeed to review every compatible batch of  Boolean variables.
Amazingly enough, it turns out that this is possible in a purely mechanical way simply by transcribing the probability space into a new complex-valued mathematical object, namely, a \emph{Hilbert space}.  
In standard physics, a Hilbert space is introduced from scratch thanks to a pivotal theorem, namely Gleason's theorem. However, this mathematical theorem provides a rather obscure concept of contextuality, which is at the origin of standard quantum \enquote{paradoxes}.
In the present model, the Gleason's theorem is not used. Instead,  the Hilbert space is naturally introduced from the probability space by a simple algebraic procedure. Therefore,  contextuality is no longer abstract but corresponds simply to a change of binary query batch.
Then, it is remarkable that the simplicial quantum state $(w, \mathcal{W})$ is now effectively represented by a perfectly standard quantum state in the Hilbert space, that is a specific matrix $\rho$, called \emph{density operator}, while the observables themselves are represented by \emph{Hermitian operators}. 
The major consequence is that any observation window using a particular batch of Boolean variables corresponds to a particular basis of the Hilbert space. Therefore, changing the Boolean variable batch, that is changing the window in our terminology, is straightforward. As a result, every observable expressed on the memory with any variable batch can so be assessed.

This construction offers new insights on quantum information theory. Most of the usual paradoxes find perfectly rational grounds and furthermore, some banal consequences falsify the common belief.
To mention only one, the most significant observation window corresponds to a basis where the density operator is diagonal in the Hilbert space. It turns out that this window corresponds to a set of \emph{mutually independent} binary queries. We call  this window a \emph{principal window} as opposed to the other windows which are thus \emph{twisted}. 
In the principal window,  the full probability problem proves to be entirely \enquote{classical} with its usual acceptation.

 \subsection{Main new results}
Listed below are the main new insights provided by the model in both quantum information and Bayesian inference theories.  Some of these are very surprising because they are at odds with current beliefs.

\paragraph{Nature of quantum information.}
The major point  already mentioned is that quantum information  is nothing but classical information processed by an elaborate Bayesian inference technique.
This means that quantum information is the relevant tool for managing the responses to an ensemble of binary queries. 
Technically, each binary query is expressed by a Boolean variable and the responses are stored in a memory whose storage capacity (in bits) is the number of non-redundant dichotomic queries. 

\paragraph{Major feature of Bayesian analysis.}
The Bayesian representation of a specific Boolean variable is very different from its deterministic representation. The main reason is that in this latest case,  any  Boolean variable involved is determined in advance. By contrast, in the Bayesian representation, this Boolean variable  has no reason to coincide with a specific query of the current window.  As a result, it is represented by a set of $N$ probabilities corresponding to the $N$  queries of the question-and-answer procedure.  Furthermore, in general, several weighted Boolean variables are simultaneously involved which is of course impossible in the deterministic case. 
Each particular query batch, i.e., each observation window, so introduces a partial point of view on the system. This multiplicity of points of view can be regarded as the signature of a Bayesian representation.

\paragraph{Entanglement.}

Entanglement is in no way a characteristic of the system itself, but only expresses that the current binary queries are not mutually independent. In other words, entanglement is the aftermath of a \emph{twisted} information window. 
This seems surprising since it is generally believed that entanglement is intrinsic and therefore cannot be changed by changing the observation window. But this is only true for local operations and classical communication (LOCC) and not in general. Indeed, technically it is always possible to diagonalize the density operator!
As a result, among all observation windows, there is at least one optimal batch in which the queries are mutually independent. In this particular window, called \enquote{principal window} as opposed to \enquote{twisted window}, the problem is strictly classical.  
Therefore, the concept of \emph{entanglement} is a Bayesian artifact that expresses the non-independence of the current batch of variables. Entanglement is not an intrinsic resource.  A striking consequence is that a pure state is in fact strictly deterministic in a principal window. 

\paragraph{Measurement.}
A measurement is defined naturally as the Bayesian estimate of an observable, which solves the so-called \enquote{measurement problem} as previously stated by Caves et al~\cite{caves3}. 
Retrieving all the information stored in the memory usually requires several observation windows, but in return, this often generates some redundancy expressed by the uncertainty principle.

\paragraph{Uncertainty principle.}
An astonishing consequence is that the iconic \emph{uncertainty principle} expresses simply the obvious fact that it is impossible, by using two observation windows, to retrieve more information than is stored in the memory.
Quantitatively, the uncertainty principle is expressed by standard entropic bounds, namely the Maassen-Uffink~\cite{maassen} and the more precise Frank-Lieb~\cite{frank} inequalities. Now, the present model provides a concrete and intuitive basis for these relationships.  This is not a physical property of the quantum world.

\paragraph{Window contextuality.}
The \emph{window contextuality} is the free choice of a particular batch of binary variables and gives rise to the famous \enquote{paradoxes}, like violation of Bell's inequalities,  perfectly rational in the present model.
More generally, the model provides a concrete and intuitive basis for the contextually  dependent aspects of quantum objects. The changes of binary queries, \emph{a priori} complicated in the probability space, are simply expressed by unitary operators acting on the Hilbert space.

%
\paragraph{Gauge principle.}
Changing the observation window from the source requires constructing an auxiliary Hilbert space and transcribing the Bayesian probability state into a complex-valued operator. This transcription is not unique and different transcriptions lead to equivalent implementations which preserve the intrinsic symmetries of the source. In combination with transcription artifacts, this implies the existence of a \enquote{gauge group}. 
Therefore, the so-called \enquote{gauge principle} of particle physics finds a natural root in this framework.
A noteworthy new result is that the gauge group is just another expression of the Bayesian prior, in agreement with the  deep insight by Steven Weinberg that \enquote{specifying the symmetry group of Nature may be all we need to say about the physical world}~\cite{weinberg}.
 The method provides an \emph{explicit derivation} of the general gauge group as a combination of unitary and antiunitary operators.  While antiunitary operators play an important but somewhat mysterious role in standard quantum information, they are now naturally introduced into the current model. Details are left outside the scope of this article.

\paragraph{Miscellaneous.}
The other new results are rather technical details. 
Among other instances, we generalize the entropic inequalities between a pair of bases to entropic inequalities between a pair of POVMs.
Moreover, as an illustrative example, we clarify certain paradoxes of the \enquote{non-local} PR-box while the Tsirelson bound proves to be strictly limited to bipartite systems.

\paragraph{}

Finally, this interpretation indicates that beyond physics, the scope of quantum information is actually universal. In physics, it suggests finding the origin of most concepts in the corpus of information theory, thus paving the way to a huge field of investigation. In data science, Bayesian inference should form the foundation of artificial neural networks.
More generally, all disciplines dealing with deep cross-correlations, such as physics, biology, evolution, cognition or linguistics, should benefit from the use of quantum formalism, which turns out to be  the more elaborate technique of Bayesian inference.

\subsection{Overview}
In Sec. (\ref{background}) we describe the basics of the model and define the concept of \enquote{Bayesian algebra} in a source system.This is the key point to introduce a probability distribution over the classical the states, what we call the \enquote{Born method}. It happens that the natural formulation is a linear programming (LP) problem, introduced in Sec. (\ref{sourcewindow}). This leads to identify the essence of a quantum state with a specific feasible LP problem. In the source system, the initial framework is a real-valued probability space, convenient to describe the current viewpoint on the register and to compute various observable expectations. But an alternative structure is possible, namely, a Hilbert space. The transcription is detailed in Sec. (\ref{transcription}): this opens a new landscape where different viewpoints over the register become accessible \emph{via} quantum channels, to begin with a survey of source problems. General systems, describing all the possible viewpoints are considered in Sec. (\ref{dynamic}). 
Observables defined from distinct viewpoints generate overlapping information and technically do not commute. This is particularly the case of complementary windows, which lead to the uncertainty relations. Diagonalization of the density operator allows to fully characterize the gauge group and beyond Noether gauge invariants and antiunitary operators. The model is next illustrated by some examples in Sec. (\ref{examples}). Several speculative points are finally discussed in Sec. (\ref{discussion}). Ultimately, after referring to the earlier approaches, we conclude in Sec. (\ref{conclusion}) on the universal nature of quantum information.

\section{Background}
\label{background}
\subsection{Classical register}
A classical register is a finite set $\mathsf{X}$ capable of storing classical information.
We will only deal with binary degrees of freedom.

\begin{definition}[Discrete degree of freedom]
\label{degreefreedom}
 A discrete degree of freedom is one dichotomic choice. 
 \end{definition}

Now, a register will be made of a finite number of subregisters, $\mathsf{X}_i$, each capable of storing one classical bit.
We take into account the input variables, but also the auxiliary variables that may be necessary to formulate the problem. Let finally $N$ be  the actual number of involved binary variables. The number of \emph{classical states} is thus $2^N$. 
It is also possible to regard the register  $\mathsf{X}$ as a discrete variable taking values in the set $\llbracket 0, 2^N-1\rrbracket$.   

\subsection{Boolean algebra}
\label{booleanalgebra}
First,  we must assign a query to each degree of freedom. 
We identify the classical register with a binary Boolean algebra, still denoted by  $\mathsf{X}$, with a batch of $N$ Boolean variables $\mathsf{X}_i$, for $i\in \llbracket 1,N\rrbracket$.
We adopt the symbol \enquote{$1$} for \enquote{valid} and \enquote{$0$} for \enquote{invalid}.
We name \emph{complete assignment}, $x$, a full assignment to the $N$ variables and \emph{partial assignment} an assignment to less than $N$ variables. We note   $\overline{\mathsf{X}}_i$ the negation of $\mathsf{X}_i$.
Finally, we call \emph{literal} a variable or its negation. 
Obviously, this choice is a matter of gauge since we could rename $\overline{\mathsf{X}}_i= \mathsf{Y}_i$ and $\mathsf{X}_i= \overline{\mathsf{Y}}_i$. 
Let us term \enquote{discrete Boolean gauge} this choice. 
This initial allocation is done \emph{once and for all} and its simultaneous inversion for all variables is simply a change of terminology. 
\begin{definition}[Discrete Boolean gauge]
\label{sourcegauge}
The discrete Boolean gauge is the initial allocation of a Boolean variable or its negation to all $N$ degrees of freedom.
\end{definition}

Given two Boolean formulas $\mathsf{f}_1$ and $\mathsf{f}_2$, it is convenient to note  $(\mathsf{f}_1 ; \mathsf{f}_2)$ (with a semicolon) the conjunction $\mathsf{f}_1\wedge\mathsf{f}_2$ and  $(\mathsf{f}_1 , \mathsf{f}_2)$ (with a comma) the disjunction $\mathsf{f}_1\vee\mathsf{f}_2$. We name \emph{partial requirement} a partial register of \emph{literals}, that is a conjunction of literals, e.g., $({\mathsf{X}}_i;\overline{\mathsf{X}}_j;{\mathsf{X}}_k)$ and \emph{complete requirement} (or \emph{classical state}), $\omega$, a conjunction of $N$ literals, e.g., $\omega=({\mathsf{X}}_1;\overline{\mathsf{X}}_2;\dots;{\mathsf{X}}_N)$, which is satisfiable by a complete assignment $x_\omega$, e.g., $x_\omega=(1; 0;\dots; 1)$. Clearly, there are $2^N$ different complete assignments and therefore $2^N$  complete requirements.
In multivariate information analysis~\cite{yeung} these complete requirements are called \emph{atoms} and the particular atom labelled $0=(0,0,\dots,0)$ is referred to as the \emph{empty atom}, $\varpi_0$.  Clearly, the fact that a particular atom is the empty atom depends on the discrete Boolean gauge, Definition (\ref{sourcegauge}).
Throughout this paper,  we will use indifferently the terms \enquote{complete requirement}, \enquote{classical state} or \enquote{atom}. 
Let $\Omega \ident \{ \omega\}$ denote the set of classical states.

On the other hand, with up to $N$ variables, it is possible to construct $2^{2^N}$ different Boolean formulas, $\mathsf{f}:\Omega\to\{0,1\}$, described, e.g., as  full disjunctive normal forms, i.e.,  reunion of  complete requirements.  Thus, any Boolean function   can be described as a disjunctions $(\omega_1,\omega_2,\dots,\omega_\ell)$ of $\ell\le 2^N$ classical states $\omega_i$. In particular, the \emph{tautology} $I:\Omega\to\{1\}$ corresponds to the reunion of all $2^N$ classical states. We will also consider the set of $2^N-1$ non-empty atoms  $X\ident\Omega-\{\varpi_0\}$.


\subsection{Bayesian algebra}

We propose to treat any \emph{Boolean function} as a \emph{random event} and account for the constraints by a set of equations between the \emph{probabilities} of the relevant  \emph{requirements} (partial or complete), as explained below in Sec. (\ref{LP}). 
For this, we use the Bayesian theory of inferences~\cite{jaynes} and regard henceforth the Boolean variables $\mathsf{X}_i$ as \emph{random variables} taking values on the alphabet $\{0,1\}$.
We will name \emph{Bayesian algebra} such a mathematical object composed of a classical Boolean algebra endowed with a Bayesian probability structure.

In general, the hypotheses are specified by a set of constraints.  We regard these constraints as a Bayesian \emph{prior}, that is an \emph{ensemble of definite conditions}, say ($\Lambda$), e.g, a set of Boolean formulas compelled to be valid or invalid. 
Now, \emph{the probability of any event will be conditional on $({\Lambda})$}. For instance, in the conventional binary addition of two integers $U$ and $V$~\cite{mf4}, the prior $({\Lambda})$ is the statement that the two integers $U$ and $V$ sum to a third integer $S$.

\paragraph{Kolmogorov probability space.}
The basic sample set is the ensemble $\Omega = \{ \omega\}$ of all  mutually exclusive $2^N$ complete requirements, labelled by the $2^N$ complete assignments $x_\omega$. 
Since the cardinality of $\Omega$ is finite, the power set $\mathfrak{P}(\Omega)$, of cardinality $2^{2^N}$, is a sigma-algebra $\mathcal{T}$, identical to  the ensemble of all Boolean functions. 
This means that an \emph{event} is just a Boolean formula, that is a finite set of atoms.
Next, we have to introduce an unknown probability measure $\Prob$ on $\mathcal{T}$ conditional on $({\Lambda})$. Finally, the Kolmogorov probability space associated with the prior $(\Lambda)$ is $(\Omega, \mathcal{T}, \Prob)$. 
When convenient, it is also possible to regard the register  $\mathsf{X}$ as a single random variable taking values in the alphabet $\llbracket 0, 2^N-1\rrbracket$. 

In general there is a number of probability distributions $\Prob$ compatible with a prior $(\Lambda)$. We will define later these different possibilities as the \enquote{source contextuality}. 

\paragraph{Notation.}
Throughout this paper, we will specifically name $\emph{unknowns}$ the conditional \emph{probability} of complete or partial requirements,  not to be confused with the variables or Boolean functions subject to randomness.
Except when mentioned otherwise, we will use a shorthand to describe the unknowns, 
namely  
 $\Prob(i)$  for $\Prob({\mathsf{X}_i}=1|\Lambda)$,  
 $\Prob(-i)$ for $\Prob({\mathsf{X}}_i=0|\Lambda)$, 
 $\Prob(i;-j)$ for $\Prob(\mathsf{X}_i=1;{\mathsf{X}}_j=0|\Lambda)$, 
etc. (for $i,j\dots\in\llbracket 1, N\rrbracket$).
Similarly, we will use $\Prob(\omega)$ for $\Prob(\omega=1|\Lambda)$.
We will often call \emph{partial probability} an unknown like $\Prob(i;-j)$ with less than $N$ literals and \emph{complete probability} an unknown $\Prob(\omega)$ with $N$ literals.  An unknown  labeled $k$ without  further detail will be denoted by $p_k$, e.g., we may have $p_k= \Prob(i;-j)$. An array of unknowns will be denoted by $p=(p_k)$.

For clarity, we use most of the time the term \enquote{classical} in its usual acceptation, as opposed to \enquote{quantum}, although this term remains vague at this stage. By exception, we will propose in Sec.~(\ref{qversusc}) a precise definition widely different.

\paragraph{Source observation window.} Up to Sec. (\ref{transcription}), we ignore communication channels and only consider a single viewpoint.
This means that we are given a classical register and  investigate what we can infer from the known assumptions. All parameters, either input data in the prior $(\Lambda)$ or observable entries $(\mathrm{q}_\omega)$,  rely to a single batch of binary variables, what we call a single \emph{observation window}. We will discuss later the possibility of reformulating the same problem by using other batches of queries, that is,  in our terminology,  other \enquote{observation windows}. This defines the concept of general system and requires the construction of transition mappings between successive windows: Eventually, the reunion of all windows within a global atlas, that we call a \enquote{Bayesian theater} will make use of a complex Hilbert space endowed with a density operator. We will refer to the initial static issue as the \emph{source window}. \enquote{Observation windows} and \enquote{Bayesian theaters} will be defined more precisely in Sec. (\ref{defwindow}).

\paragraph{Universal equations.}
Since the probability laws are just an extension of Aristotelian logic the following relations are universal:
\begin{align}
\label{Gconstraints}  &\Prob(\pm i;\pm j;\pm k;\dots) \ge 0\\
\label{simplet} 1 &= \Prob(i) +\Prob( -i)\\
\label{doublet} \Prob(i) &= \Prob(i; j) +\Prob(i; -j)\\
\label{triplet} \Prob(i;j) &= \Prob(i; j;k) +\Prob(i; j;-k)
\end{align}
 etc., where $i,j,k,\dots $ are signed integers and $|i|,| j|, |k|,\dots\in\llbracket 1,N \rrbracket$ are distinct. It is easy to establish that we have $\binom{N}{1}=N$ distinct equations like Eq. (\ref{simplet}),   $4\binom{N}{2}$ distinct equations like Eq. (\ref{doublet}),  $12\binom{N}{3}$ distinct equations like Eq. (\ref{triplet}), etc. 
Note that accounting for Eqs. (\ref{simplet}, \ref{doublet}, \ref{triplet}, etc.), Eq. (\ref{Gconstraints}) implies that
\begin{equation}
\label{bounds}
\Prob(\pm i;\pm j;\pm k;\dots) \le 1
\end{equation}
and
\begin{align}
\label{zero} \Prob(i)=0 &\Rightarrow \Prob(i;j)=0 \Rightarrow \Prob(i;j;k)=0 \mathrm{~etc}\dots\\
\label{one} \dots\Prob(i;j;k)=1 &\Rightarrow \Prob(i;j)=1 \Rightarrow \Prob(i)=1.
\end{align}

Due to these universal equations, the LP problem considered in the next section is specific. It can be called \enquote{Bayesian LP system}. Its solutions are in the range $[0, 1]$ and their specific polytopes will be proved to be a simplex.

\section[Source system]{Source observation window}
\label{sourcewindow}

We start with  a particular batch of queries, referred to as the \enquote{source window}.
The logical problem at hand is defined by a set of hypotheses to be satisfied.
In the present Bayesian model, they are viewed as a \emph{prior},  say $({\Lambda})$.
In general, when the problem is well posed, the conditions are unambiguous and the prior is composed of deterministic Boolean formulas, that is \emph{events} of the sigma-algebra $\mathcal{T}$.
In the probability space, beyond Boolean formulas which can take only two values, a more general concept lies in \enquote{observables}.
\begin{definition}[Observable] 
\label{defobservable}
An observable $Q$ is a real-valued functions  of the classical states on the register, defined  as
\begin{align}
\label{expecobs}
Q:\quad \Omega \to \mathbb{R} : \quad \omega\mapsto Q(\omega)=\mathrm{q}_\omega.
\end{align}
\end{definition}
We will denote the array $(\mathrm{q}_\omega)$ by $\mathrm{q}$.
Specifically, we will consider the \emph{indicator function} $F(\omega)$ of a Boolean formula $\mathsf{f}=(\omega_1,\omega_2,\dots,\omega_\ell)$, defined as $F(\omega)=1$ if $\omega \in \{\omega_1,\omega_2,\dots,\omega_\ell\}$ and $0$ otherwise. 
We will often write $F(\omega)=\mathrm{f}_\omega$ and denote the array $(\mathrm{f}_\omega)$ by $\mathrm{f}$.
In particular, the indicator function of the tautology is $\mathrm{f}_{\scriptscriptstyle{I}}=(1,1\dots,1)$.

\subsection{Linear programming problem}
\label{LP}

The Bayesian inference of the variables at issue is to decide how the prior knowledge affects the probabilities $p_i$ of the relevant requirements. 

In Ref. \cite{mf3}, we have proposed that the prior be simply incorporated by assigning a probability of $1$  to observables compelled to be valid and a probability $0$ to observables compelled to be invalid. 
To ensure consistency, we need also to incorporate a number of universal equations, Eqs. (\ref{simplet}, \ref{doublet}, \ref{triplet}, etc.).
To this end, encode any logical constraint as a linear \emph{specific equation}.  In this way, the prior $(\Lambda)$ happens to be naturally expressed as a linear system. For instance, a partial requirement $({\mathsf{X}}_i;\overline{\mathsf{X}}_j;{\mathsf{X}}_k)$, compelled to be valid or invalid in the Boolean algebra, is trivially encoded as $\Prob(i;-j;k)=1$ or $0$ respectively.  A Boolean function defined as a disjunction of classical states $\mathsf{f}=(\omega_1,\omega_2,\dots,\omega_\ell)$  and compelled to be valid or invalid in the Boolean algebra, is encoded as $\sum_i \Prob(\omega_i)=1$ or $0$, because the classical states, $\omega_i$, are disjoint, etc. When convenient, we can also consider linear combinations of event probabilities, that is to say, \emph{observables} instead of only Boolean functions. 

Subsequently, the full prior, comprising both the specific equations and the relevant universal constraints is formulated as a linear programming (LP) problem in stack variables~\cite{murty} within a convenient real-valued vector space  in the form,

\begin{align}
\label{lpeq}
\begin{aligned}
Ap&=b\\
\mathrm{subject~to~~~} p&\ge 0
\end{aligned}
\end{align}
where $p=(p_i)$ is a real-valued positive unknown vector, $A=(\mathrm{a}_{j,i})$ a real matrix and $b=(b_j)$ a real vector, while $p\ge 0$ stands for $\forall i, p_i\ge 0$. The LP system is specific because the unknowns are all in the range $[
0,1]$, thanks to the universal equations.
The number of unknowns $p_i$, say $n$, is based on the particular formulation, that is the partial and complete probabilities explicitly involved. In Bayesian computation, it is crucial to have a minimum set of unknowns and indeed, $n$ can always be polynomial in $N$ for problems of  \textsf{NP}-complexity class. On the contrary, for a theoretical discussion, and also to take into account evolving systems, it is necessary to take the full set of complete probabilities as unknowns, even if the number $n=2^N$ is exponential in $N$. We will adopt this choice from Sec. (\ref{realspace}).
Let $m>0$ denote the number of rows of the matrix, so that $A$ is a $m\times n$ matrix. We will assume that the non-independent rows have been eliminated and that $m$ is also the rank of the system.

It remains to complete the computation by solving this LP problem,  Eq. (\ref{lpeq}). 
A feasible solution is a numerical vector of unknowns, $p$,  that satisfies the prior $(\Lambda)$, that is Eq. (\ref{lpeq}), and therefore defines a probability distribution $\Prob$ on the sample set $\Omega$ and thus a probability measure on the sigma-algebra $\mathcal{T}$. 

 If the problem is inconsistent, the system is unfeasible.
\emph{A priori},  if the problem is well posed and admits a solution, one might expect the system to provide a deterministic solution. However, there are LP problems that do not accept deterministic solutions but are nonetheless feasible and even this circumstance is by no means exceptional: This is the case not only of quantum information but also arithmetic in Bayesian computation! 
In fact,  this only means that the initial batch of Boolean variables is not the best suited to the problem because the constraints implies that they are not mutually independent.
\begin{proposition}
\label{mutualindependence}
When the  LP problem accept a deterministic solution, the binary variables $\mathsf{X}_i$ of the source window are mutually independent.
\end{proposition}
\emph{Proof.} A deterministic solution is a separable joint probability~\cite{mf3} which implies that the variables $\mathsf{X}_i$ regarded as random variables are mutually independent (see also Sec. \ref{separability} below).
$\Box$
\paragraph{}

When the LP system is feasible but does not accept a deterministic solution, such a deterministic solution exists nevertheless  but  in another window, namely a \enquote{principal window} defined in Sec. (\ref{canoniwindow}).

In general, the rank $m$ of the matrix $A$ is less than $n$ and thus, there is a continuous set of solutions.
This arises when for some reason the Bayesian prior $(\Lambda)$ is not specific enough. For example, in Bayesian computation, the problem may have multiple solutions, or in quantum mechanics, a set of data may be fundamentally out of control of the experimenter.
Thus, the particular probability distribution to be used depends on the context. In other words, the \enquote{Born method} basically leads to context-dependent systems.
 Let us recall precisely what we term \enquote{contextuality}.
\begin{definition}[Contextuality]
\label{defcontextuality}
A system is context-dependent when the probability distribution involved depends on an exogenous choice.
\end{definition}
Given that contextuality has also other causes in general systems (Sec. \ref{transcription}, below), we will refer to this property as the \emph{source contextuality}.
\begin{definition}[Source contextuality]
\label{defsourcecontextuality}
Source contextuality  expresses the possibility of choosing a particular feasible probability distribution among the solutions of the source LP problem.
\end{definition}

A particular solution is chosen by a selection rule. 
In linear programming, this solution is usually selected by maximizing an objective function. Specifically,  in Bayesian computation~\cite{mf3,mf4}, we use optimization to select the deterministic distributions when possible. 

Therefore, in quantum information, a specific selection rule is needed. This rule will be said to  fix a particular \enquote{context}. 
Thus, source contextuality is a piece of intrinsic information specified at the outset in addition to the Bayesian prior.

\subsection[Real probability space]{Real probability space $\mathcal{P}$}
\label{realspace}

We now assume that the unknowns $p=(p_\omega)$ are specifically the $2^N$ complete probabilities of the classical states, i.e., $p_\omega= \Prob(\omega=1|\Lambda)$ with $\omega\in\Omega$. This can easily be achieved by eliminating the partial probabilities using Eqs. (\ref{doublet}, \ref{triplet}, \dots). 
Then $p\in Span(\omega|\omega\in\Omega)= \mathbb{R}^{\Omega}$. We will denote by  $\mathcal{P}$ this real-valued vector space $\mathbb{R}^{\Omega}$ and $\mathcal{P}^*$ its dual space, both of dimension $n=d= 2^N$. As long as static issues are concerned, no metric is required. We will indifferently refer to  $\mathcal{P}$ as the \enquote{real probability space} or the \enquote{LP space}.

\paragraph{Notation.}
When there is no risk of confusion, we will use the same symbols $\omega,\omega',\omega_i,\dots$ to designate either the classical states in $\Omega$ or the different labels in $\mathcal{P}$ and $\mathcal{P}^*$.
 
- We note $\tilde{\omega}\in\mathcal{P}$, with $\omega\in\Omega$, the basis vectors in $\mathcal{P}$, i.e., $\tilde{\omega} =(p_{\omega'})$ with $p_{\omega'}=\delta_{\omega'\omega}$. A basis vector describes a deterministic probability distribution.
The full basis  is denoted by $\tilde{\Omega}\ident\{\tilde{\omega}\}$ or simply $\Omega$ when no confusion can occur.

- A covector in the dual space $\mathcal{P}^*$ is denoted  $\mathrm{q} =(\mathrm{q}_{ \omega})$ with $\omega\in\Omega$. 
A covector defines an observable on the register, $Q(\omega)=\mathrm{q}_{ \omega} $.

- A dual form $(\mathcal{P}^*,\mathcal{P})\to \mathbb{R}$ is denoted  $\langle \mathrm{q} p \rangle$, where $\mathrm{q}\in\mathcal{P}^*$ and $p\in\mathcal{P}$.
 
- We will note $\tilde{\omega}^*$ the canonical basis covectors in $\mathcal{P}^*$ defined by $\langle\tilde{\omega}^* \tilde{\omega}'  \rangle=\delta_{\omega\omega'}$. 

- An observable defined by a covector $\mathrm{q} =(\mathrm{q}_{ \omega})$ with $\mathrm{q}_{ \omega} \ge 0\ (\forall \omega \in \Omega)$ is called \emph{non-negative}.

- A Boolean function $\mathsf{f}$ defines  an observable $F(\omega)$, that is a non-negative dual form whose associated  covector $\mathrm{f}=(\mathrm{f}_\omega)$ is the indicator function of $\mathsf{f}$ in $\Omega$. In particular, a basis covector $\tilde{\omega}^*$ defines a Boolean function and thus an observable $F(\omega')= \langle\tilde{\omega}^* \tilde{\omega}'  \rangle$ that we will also denote  $\tilde{\omega}^*$ for simplicity when no confusion can occur.

\paragraph{Expectation.}
The value $\langle Q \rangle$ of a dual form $\langle \mathrm{q} p \rangle$ with respect to the probability distribution $\Prob(\omega) = p_\omega$, is trivially the \emph{expectation value} of the observable $Q(\omega)=\mathrm{q}_\omega$.
\begin{align*}
 \langle Q \rangle = \sum_{\omega\in\Omega}  Q(\omega)\ \Prob(\omega) = \sum_{\omega\in\Omega} \mathrm{q}_\omega p_\omega = \langle \mathrm{q} p \rangle
\end{align*} 

\begin{theorem}[Bayesian formulation]
Any LP system,  Eq. (\ref{lpeq}), can be expressed as the following Bayesian problem, 
\begin{align}
\label{expectation}
(\Lambda):\ \mathrm{Given~} m-1 \mathrm{~observables~} A_\ell \mathrm{~assign~} \Prob \mathrm{~on~}\Omega \mathrm{~subject~to~}\langle A_\ell\rangle=b_\ell, 
\end{align}
where $\ell\in\llbracket1,m-1\rrbracket$.
In addition, it is possible to assume that the expectation of the observables $A_\ell$ is zero, that is $b_\ell=0$.
\end{theorem}
\emph{Proof.}
In Eq. (\ref{lpeq}), without loss in generality, assume that one row is the normalization constraint  that is the tautology. We reserve the index $\ell=0$  to this normalization equation, namely, $A_0=I$, $\mathrm{a}_{0,\omega}=1, \forall \omega\in\Omega$ and $b_0=1$.
Clearly, each row, now labeled $\ell$, defines a covector, $\mathrm{a}_\ell =\sum_{\omega} \mathrm{a}_{\ell,\omega} \tilde{\omega}^*, (\ell\in\llbracket 0,m-1\rrbracket)$. It  can be regarded as a constraint on the expectation of an observable $A_\ell(\omega)=\mathrm{a}_{\ell,\omega}$. 
Therefore,  $\sum_\omega \mathrm{a}_{\ell,\omega} p_\omega =b_{\ell}$ means $\langle A_\ell\rangle=b_\ell$. 
 
Now, Eq. (\ref{lpeq}) can be reformulated as follows: Assign a probability distribution $\Prob$ on $\Omega$, given that the expectation of $m$ independent  observables $A_\ell$ are subject to $\langle A_\ell \rangle = b_\ell$. 
Since normalization is implicit in probability theory, Eq. (\ref{lpeq}) can be expressed as Eq. (\ref{expectation}).
We can assume that $b_\ell=0$ for $\ell>0$ because otherwise, we can replace $A_\ell$ by $A_\ell-b_\ell I$.
The converse is obvious. Now, the system, Eq. (\ref{expectation}) depicts a standard Bayesian problem~\cite{jaynes}.
Also, the LP problem is specifically called a \enquote{Bayesian LP problem}.
$\Box$
\paragraph{}
Let us first address the simplest problem, in which the prior is reduced to the normalization equation.

%

\subsubsection{Tautology}

Irrespective of the particular prior $(\Lambda)$, consider the following Bayesian LP system in the probability space $\mathcal{P}$,
\begin{align}
\label{tautoLP}
\begin{aligned}
\sum_{\omega\in\Omega} p_\omega &=1\\
\mathrm{subject~to~~~} p_\omega&\ge 0
\end{aligned}
\end{align}
Any solution $p =(p_\omega)$ of this system
describes a potential probability distribution $\Prob$ on $\Omega$.  
The $d$ classical deterministic states $\omega\in\Omega$ label both the basis vector  $\tilde{\omega}\in \mathcal{P}$ and the extreme points of a convex polytope, $\mathcal{W}_{\scriptscriptstyle I}$, of dimension $d-1$ with $d$ vertices, that is a $(d-1)$-simplex, known as \enquote{probability simplex} or \enquote{Choquet simplex} in convex geometry. In the present context, we will call this polytope, $\mathcal{W}_{\scriptscriptstyle I}$, the $d$-dimensional \emph{tautological simplex}. 
\begin{definition}[Tautological simplex $\mathcal{W}_{\scriptscriptstyle I}$]
The  \enquote{tautological simplex} in the $d$-dimensional vector space $\mathcal{P}$ is the $(d-1)$-simplex 
\label{tautosimplex}
\begin{equation}
\mathcal{W}_{\scriptscriptstyle I}= \mathrm{conv}(\tilde{\omega}\ |\ \omega\in\Omega)\subset Span(\omega\ |\ \omega\in\Omega)=\mathcal{P}= \mathbb{R}^{\Omega}
\end{equation}
\end{definition}
\begin{proposition}
\label{barycentre}
The entries $p_{\omega}$ in Eq. (\ref{tautoLP}) represent both the $d$ components of $p$ in $\mathcal{P}$ and the $d$ barycentric coordinates of the point $p$ on the tautological simplex $\mathcal{W}_{\scriptscriptstyle I}$. In other words, the distinction between barycentric and contravariant components vanishes on $\mathcal{W}_{\scriptscriptstyle I}$.
\end{proposition}
\emph{Proof.} Since $\sum_{\omega\in\Omega} p_\omega=1$, the two formulations mean $p=\sum_\omega p_\omega\  \tilde{\omega}$. 
Note that beyond the points $p\in\mathcal{W}_{\scriptscriptstyle I}$ on the simplex, this identity is also valid for direction vectors $v=\sum_\omega v_\omega\  \tilde{\omega}\in\mathcal{W}_{\scriptscriptstyle I}$ with $\sum_{\omega\in\Omega} v_\omega=0$.
$\Box$
\paragraph{}

Since $\mathcal{W}_{\scriptscriptstyle I}$ is a simplex, the barycentric coordinates are uniquely defined.
 The set of its extreme points $\tilde{\Omega}=\{\tilde{\omega}\}$ forms its \emph{Choquet boundary} and describes the deterministic distributions.
 \begin{proposition}
 \label{largest}
 The tautological simplex is the largest set of solutions satisfying Eq. (\ref{tautoLP}).
 \end{proposition}
 \emph{Proof.}
 Obvious because $p\in\mathcal{P}$ implies $p=\sum_{i=1}^d p_i \tilde{\omega}_i$ and  Eq. (\ref{tautoLP}) means that $p\in\mathcal{W}_{\scriptscriptstyle I} $. 
$\Box$
\begin{proposition}
\label{basicexpectation}
Any basic subspace of $\mathcal{P}$ is specified by a Boolean function compelled to be valid. 
\end{proposition}
\emph{Proof.}
Let $p$ be located on the simplex $\mathcal{W}_{\scriptscriptstyle I}$ and thus $p_\omega \ge 0,\ \forall \omega\in\Omega$.
Let ${\mathsf{f}}$ be a Boolean function that can be expressed as a disjunctions of $\ell$ classical states $\omega_i$, say ${\mathsf{f}}=(\omega_1,\omega_2,\dots,\omega_\ell)$. Let $\overline{\mathsf{f}}$ be its negation, expressed as a disjunctions of the $d-\ell$ other classical states $\omega'_j$, say $\overline{\mathsf{f}}=(\omega'_1,\omega'_2,\dots,\omega'_{d-\ell})$. Let $\overline{F}$ be its indicator function and $\overline{\mathrm{f}}= (\overline{\mathrm{f}}_\omega)$ the corresponding covector.
In addition, assume that $\langle \overline{F} \rangle=\langle \overline{\mathrm{f}}\, p \rangle=0$, i.e., $p_{\omega'_j}=0$ for all $d-\ell$ indexes $j$ involved. Since $p_{\omega'_j}=0$ describes a basic subspace of $\mathcal{P}$ of dimension $d-1$, the equation $\langle \overline{F} \rangle=0$ depicts a basic subspace of $\mathcal{P}$  of dimension $d-(d-\ell)=\ell$. This $\ell$-dimensional subspace is thus also characterized by $\langle {F} \rangle=1$, that is the Boolean function ${\mathsf{f}}=(\omega_1,\omega_2,\dots,\omega_\ell)$ compelled to be valid.
Conversely, any basic subspace is the direct sum of one-dimensional subspaces $\mathcal{P}_i$, each spanned by a basis vector $\tilde{\omega}_i$ so that the direct sum $\mathcal{P}_1\oplus\mathcal{P}_2\dots\oplus\mathcal{P}_\ell $ is specified by $\mathsf{f}=(\omega_1,\omega_2,\dots,\omega_\ell)=1$.
$\Box$

\subsubsection{General Bayesian LP system}
Return now to the current LP system, Eq. (\ref{expectation})  associated with the prior $(\Lambda)$.
Suppose that the system is feasible and consider the set of solutions.
It is convenient to single out two subspaces containing the solutions.

\paragraph{Affine subspace $P_{\scriptscriptstyle \Lambda}$ and effective probability space $\mathbb{W}_{d-m+1} $.}
Consider first the \emph{affine} set of all solutions, that is a an affine subspace $P_{\scriptscriptstyle \Lambda}\subset\mathcal{P}$ of dimension $d-m$ such that $\alpha p_1+(1-\alpha)p_2\in P_{\scriptscriptstyle \Lambda}$ for every $p_1\in P_{\scriptscriptstyle \Lambda}$, $p_2\in P_{\scriptscriptstyle \Lambda}$ and $\alpha\in\mathbb{R}$. Second, consider their \emph{linear} span, that is a  particular $(d-m+1)$-dimensional subspace $\mathbb{W}_{d-m+1}\subseteq \mathcal{P}$  such that $\alpha_1 p_1+\alpha_2p_2\in \mathbb{W}_{d-m+1} $ for every $p_1\in \mathbb{W}_{d-m+1}$, $p_2\in \mathbb{W}_{d-m+1}$, $\alpha_1\in\mathbb{R}$ and $\alpha_2\in\mathbb{R}$.
\begin{definition}[Affine subspace $P_{\scriptscriptstyle \Lambda}$]
\label{probplan}
The affine subspace $P_{\scriptscriptstyle \Lambda}$ is the affine set of the solutions.
\end{definition}
\begin{definition}[Effective probability space  $\mathbb{W}_{d-m+1} $]
\label{defeffprobspace}
The effective probability space  $\mathbb{W}_{d-m+1}$ is the linear span of the solutions.
\end{definition}

\paragraph{Specific polytope $\mathcal{W}_{\scriptscriptstyle \Lambda}$.}
In fact, from standard LP theory, the locus of the solutions is a specific polytope $\mathcal{W}_{\scriptscriptstyle \Lambda}$. This polytope is compact and convex and will prove to be a simplex in Proposition (\ref{propspecificsimplex}) just below.
It is characterized by the set  of its extreme points, that is its vertices $w_\mathrm{k}=\sum_{i=1}^d w_{k,\omega_i}\tilde{\omega}_i$, with $w_{k,\omega_i}\ge0$ and $\sum_{i=1}^d w_{k,\omega_i}=1$.
 
We have from a simple inspection
\begin{equation}
\label{imbrications}
\mathcal{W}_{\scriptscriptstyle \Lambda}=\mathrm{conv}(w_\mathrm{k})= P_{\scriptscriptstyle \Lambda}\cap \mathcal{W}_{\scriptscriptstyle I}= \mathbb{W}_{d-m+1}\cap \mathcal{W}_{\scriptscriptstyle I}.
\end{equation}
Still from standard LP theory, the maximum number of vertices is $\binom{d}{m}$ so that \emph{a priori} the actual number, say $r$, might be very large for large $d$.
When  $m=d$, there is a single solution and the specific polytope is reduced to an isolated point, i.e.,   $r=1$, that can be regarded as a particular simplex with a single vertex.
 More generally, when  the number of simplices $r$ is equal to $d-m+1$ the polytope $\mathcal{W}_{\scriptscriptstyle \Lambda}$ is a standard simplex and the vertices $\{w_\mathrm{k}\}$ constitute a basis of the effective probability space  $\mathbb{W}_{d-m+1} $. 
 Remarkably, it turns out that only these cases can be encountered in the present Bayesian LP system. 
 They deserve therefore a special name.
\begin{definition}[Simplicial  system]
\label{defsimplicial}
A simplicial system is a LP problem whose specific polytope is either an isolated point or a simplex.
\end {definition}
\begin{proposition}
\label{propspecificpolytope}
The specific polytope $\mathcal{W}_{\scriptscriptstyle \Lambda}$ of any Bayesian LP system, Eq. (\ref{expectation})  is  pointwise identical to the tautological simplex of the effective probability space  $\mathbb{W}_{d-m+1} $ when using the set of $r$ vertices  $\{w_\mathrm{k}\}$ as basis vectors. 
\end{proposition}
\emph{Proof.}
From Definition (\ref{defeffprobspace}), the effective probability space is the linear span of the extreme points of the polytope $\mathcal{W}_{\scriptscriptstyle \Lambda}$, that is $\mathbb{W}_{d-m+1} = Span (w_k \ |k\in\llbracket 1, r\rrbracket)$.
If $r>d-m+1$, it is possible, from Carathéodory's theorem,  to extract $d-m+1$ vertices, $w_j$ for say $j\in\llbracket 1, d-m+1\rrbracket$ after reordering the simplices if necessary, such that $\mathbb{W}_{d-m+1} $  is actually the linear span of only $d-m+1$ extreme points, that is $\mathbb{W}_{d-m+1} = Span (w_j\  |j\in\llbracket 1, d-m+1\rrbracket)$, while the set $\{w_j \},\,j\in\llbracket 1, d-m+1\rrbracket)$ is a basis in the $(d-m+1)$-dimensional effective probability space $\mathbb{W}_{d-m+1}$.

It is possible to complement this basis $\{w_j \},\,j\in\llbracket 1, d-m+1\rrbracket)$ in $\mathcal{P}$ with $m-1$ vectors, $v_\ell,\,\ell\in\llbracket 1, m-1\rrbracket)$. Choose specifically $v_\ell =\sum_{i=1}^d \mathrm{a}_{\ell,i} \tilde{\omega}_i$
where the coefficients $\mathrm{a}_{\ell,i}$ are the entries of the matrix obtained from the Bayesian formulation with $b_\ell=0$ in Eq. (\ref{expectation}). Since $m$ is the rank of the LP system, these vectors are independent by construction.
Now, any point $p\in\mathcal{P}$ can be expanded as
$$p =\sum_{i=1}^d p_i\, \tilde{\omega}_i=\sum_{j=1}^{d-m+1} x_j w_j+\sum_{\ell=1}^{m-1} y_\ell\, v_\ell.$$
The effective probability space $\mathbb{W}_{d-m+1}$ is characterized by the linear system
\begin{equation}
\label{ellzero} 
  y_\ell=b_\ell=0 \qquad\forall\ell\in\llbracket 1, m-1\rrbracket
\end{equation}
The restriction to $\mathbb{W}_{d-m+1}$ of the tautology, expressed as $I(p) =\sum_{i=1}^d p_{\omega_i}$ in the old basis $\{\tilde{\omega}_i\}$,  is expressed in the basis  $\{w_j, v_\ell \}$ with $y_\ell=0$ as 
$$I(p)=A_0(x) =\sum_{i=1}^d \sum_{j=1}^{d-m+1} w_{j,\omega_i} x_j=\sum_{j=1}^{d-m+1}x_j$$
because $I(w_j) =\sum_{i=1}^d w_{j,\omega_i} =1$. Then $A_0(x)=1$ states that $p$ is located on the affine subspace $P_{\scriptscriptstyle \Lambda}$. In addition, $x\ge0$ specifies that $p\in\mathcal{W}_{\scriptscriptstyle \Lambda}$.

Now,   in  $\mathbb{W}_{d-m+1}$  there is no longer any specific constraint. Therefore, the solutions of the initial LP system just require that $A_0(x) =\sum_{j=1}^{d-m+1}x_j=1$ with $ x_j\ge0$.
As a result,  the specific polytope $\mathcal{W}_{\scriptscriptstyle \Lambda}$ is the tautological simplex in $\mathbb{W}_{d-m+1}$, (Definition \ref{tautosimplex}), with exactly $r=d-m+1$ vertices playing each the same role.
 $\Box$
 
 \begin{proposition}
\label{propspecificsimplex}
Any Bayesian LP system, Eq. (\ref{expectation}), is simplicial. 
\end{proposition}
\emph{Proof.} This is a trivial corollary of Proposition (\ref{propspecificpolytope}). $\Box$
 
 \begin{definition}[Specific simplex $\mathcal{W}_{\scriptscriptstyle \Lambda}$]
 The solutions of the LP system, Eq. (\ref{expectation}) are located on a simplex $\mathcal{W}_{\scriptscriptstyle \Lambda}$, called \enquote{specific simplex}, with $r=d-m+1$ vertices.
 \end{definition}
 In other words, 
 \begin{proposition}
 \label{LPsimplex}
 The LP system, Eq. (\ref{expectation}) may be alternatively specified by the following Bayesian equation,
 $$ \mathrm{Assign}\quad \Prob(\omega)=p\quad\mathrm{subject~to}\quad p\in\mathcal{W}_{\scriptscriptstyle \Lambda}.$$
 \end{proposition}


\subsubsection{Source contextuality}
In general, there are a number of solutions to the current Bayesian system Eq. (\ref{expectation}) located on the specific simplex $\mathcal{W}_{\scriptscriptstyle \Lambda}$. The choice of a particular solution specifies the \enquote{source context}.

\paragraph{Default context.}

Suppose first that there is no extra constraint, which we call the \enquote{default context}.
 The standard Bayesian solution is then the \emph{most likely distribution}, determined by the maximum entropy principle~\cite{caticha}, that is a generalization of the Laplace's principle of indifference. This requires to consider a \emph{uniform probability density} $\varphi_c$ of dimension $d-m$ in the affine subspace $P_{\scriptscriptstyle \Lambda}$, normalized to unity on the convex hull of the specific polytope.
 
\begin{definition}[Hull density]
 \label{hulldensity}
 We will call \enquote{hull density} a continuous density of dimension  $(d-m)$ on the specific simplex.
 \end{definition}

\begin{definition}[Center of mass, $\tilde{c}$]
\label{wlambda}
The center of mass $\tilde{c}$ is the mean point with respect to a uniform hull density. 
\end{definition}
From Choquet theory~\cite{choquet},  in \emph{simplicial systems} the center of mass is also uniquely defined as
$\tilde{c}=\frac{1}{r} \sum_{k=1}^{r}w_k,$ 
where $r= d-m+1$ is the number of vertices.
In other words, the center of mass $\tilde{c}$ can be defined indifferently either by a uniform hull \emph{density} or a  uniform \emph{discrete} probability distribution, say $\mu_k=1/r$ with $k\in\llbracket1,r\rrbracket$, on the $r$ vertices. 

The center of mass, $\tilde{c}=(c_{\omega})$ is the most likely probability distribution  of the current system Eq. (\ref{expectation}) without extra constraints. It will be noted $\Prob(\omega=1|\Lambda_c)= c_{\omega}$.
Beyond this context by default, we need to define any other particular context.

\paragraph{Other contexts.}
 \emph{A priori}, any arbitrary context should be obtainable by assigning a non-uniform probability hull density on the specific polytope. However, if we insist to have a true probability density, that is always positive, this is only feasible in the vicinity of the default context.
Derivation of the general hull density is easy but left out of the scope of this article.
Indeed, it is always possible to  specify an arbitrary context by means of a discrete true probability distribution on the vertices of the specific simplex, which we will call a \enquote{simplicial quantum state}.

 \subsection{Representation of quantum states}
\label{secsimplqu}
It is remarkable that \emph{the pair} composed of a LP system and a selection rule among the feasible solutions, that is in the present framework a  contextual probability distribution on the vertices of the specific simplex,
represents actually a standard \enquote{quantum state} restricted to the source window. 
\subsubsection{Working distribution}
Technically, we need only to specify the mean point $w_{\scriptscriptstyle \Lambda}\in\mathcal{W}_{\scriptscriptstyle \Lambda}$ of the auxiliary distribution because the details will be derived from the framework.
Let us name this mean point the \enquote{working distribution}. 

  \begin{definition}[Working distribution]
 \label{mixedistri}
 The working distribution $w_{\scriptscriptstyle \Lambda}\in\mathcal{W}_{\scriptscriptstyle \Lambda}$  is the mean point with respect to an auxiliary probability distribution on the specific simplex.
 \end{definition}
 The working distribution $w_{\scriptscriptstyle \Lambda}$ will describe the current probability distribution of the quantum state.
 Of course it is possible to choose the default context but in general we will specify $w_{\scriptscriptstyle \Lambda}$ different from the center of mass of the simplex.

Before proceeding further, it is convenient for clarity to give a special name to the entropy of the working distribution in the sample set $\Omega$, as opposed to the entropy of the auxiliary distribution that we will compute later.
 
\begin{definition}[Window entropy]
\label{defwindowentropy}
The window entropy $\mathbb{H}(\Omega)$ or $\mathbb{H}(w_{\scriptscriptstyle \Lambda})$  is the Shannon entropy $S_w$ of the working distribution $w_{\scriptscriptstyle \Lambda}$.
\begin{equation} 
S_w=\mathbb{H}(\Omega)=\mathbb{H}(w_{\scriptscriptstyle \Lambda})\ident\sum_{\omega\in\Omega} -w_{{\scriptscriptstyle \Lambda}, \omega} \log_2 w_{{\scriptscriptstyle \Lambda}, \omega} .
\end{equation} 
\end{definition} 
The window entropy  is rather a Bayesian parameter and has little to do with a real uncertainty. By contrast, the so-called \enquote{simplicial entropy} defined in the next section will directly represent a form of uncertainty. 
 
 Now, the actual state, referred to as \enquote{simplicial quantum state}, cannot be limited to the working distribution $w_{\scriptscriptstyle \Lambda}$ and the full LP system is required, because otherwise this would arbitrarily introduce biased information.

 \subsubsection{Simplicial quantum states}
 
Let $\mathcal{W}_{\scriptscriptstyle \Lambda}$ be the specific simplex and $w_{{\scriptscriptstyle \Lambda}}\in\mathcal{W}_{\scriptscriptstyle \Lambda}$ the working distribution.
 Let $w_i$ be its vertices and $\Sigma_\mu=\{\mu_i\}$ the set of barycentric coordinates of $w_{{\scriptscriptstyle \Lambda}}$. We have with $r=d-m+1$,
  $$w_{{\scriptscriptstyle \Lambda}} =\sum_{i=1}^{r}\mu_i w_i\quad\mathrm{where}\quad\mu_i\ge 0\quad\mathrm{and}\quad \sum_{i=1}^{r}\mu_i =1$$
Therefore, $w_{{\scriptscriptstyle \Lambda}}$ is the center of mass of the vertices $\{w_i\}$ weighted by $\{\mu_i\}$.

 \begin{definition}[Simplicial quantum state]
 \label{qrepg}
 A simplicial quantum state is the pair $(\Sigma_\mu,\mathcal{W}_{\scriptscriptstyle \Lambda})$  of a contextual probability distribution $\Sigma_\mu=\{\mu_i\}$  and the  specific simplex $\mathcal{W}_{\scriptscriptstyle \Lambda}$. The  working distribution is the mean point $w_{{\scriptscriptstyle \Lambda}} =\sum_{i=1}^{r}\mu_i w_i$ where $r=d-m+1$.
We will refer to a simplicial quantum state indifferently by the pairs  $(\Sigma_\mu,\mathcal{W}_{\scriptscriptstyle \Lambda})$ or  $(w_{\scriptscriptstyle \Lambda},\mathcal{W}_{\scriptscriptstyle \Lambda})$.
 \end{definition}
Let us compute the entropy of the contextual distribution with respect to the simplicial distribution.
\begin{definition}[Simplicial entropy $S_\mu$ in $\mathcal{W}_{\scriptscriptstyle \Lambda}$]
\label{simplicialentropy}
The simplicial entropy of a simplicial quantum state $(\Sigma_\mu,\mathcal{W}_{\scriptscriptstyle \Lambda})$ is the Shannon entropy of the simplicial distribution
\begin{equation} 
S_\mu\ident \mathbb{H}(\Sigma_\mu)=\sum_{i=1}^{r} -\mu_i \log_2 \mu_i.
\end{equation}
We will use indifferently the terms simplicial entropy or contextual entropy.
\end{definition}

We have $S_\mu\le\log r$. For instance,  we have $S_\mu\le\log d$ if $m=1$ and $S_\mu=0$ if $r=1$. 
Among the LP problems of rank $m$, the maximum simplicial entropy $S_\mu=\log r$ is attained when $w_{\scriptscriptstyle \Lambda}$ is the center of mass $\tilde{c}$ of $\mathcal{W}_{\scriptscriptstyle \Lambda}$.

To sum up, we encountered two forms of entropy, the window entropy $\mathbb{H}(\Omega)$ on the sample set and the simplicial entropy $\mathbb{H}(\Sigma_\mu)$ on the simplex. 
The two forms of entropy  obviously differ in the source window, for instance the simplicial entropy of a  pure state (defined just below) is zero, which is not the case in general for the window entropy.
However,  they will merge in a \enquote{principal window} (Proposition \ref{commonentropy} below).
At last, they are both bounded above by the storage capacity of the register, i.e., $N$ bits.

\paragraph{}
The simplicial entropy is closely related to the von Neumann entropy of standard quantum information.  It turns out that the von Neumann entropy is actually the lower bound of all simplicial entropies over all windows, defined in general systems, Sec. (\ref{dynamic}).
This will lead to a more substantial interpretation of the von Neumann entropy in terms of information theory in Theorem  (\ref{fondamental}) below.

 \subsubsection{Pure states}
 When the simplex $\mathcal{W}_{\scriptscriptstyle \Lambda}$ is reduced to an isolated point, we have a \emph{pure state}.  This means that the rank $m$ of the LP-system, Eq. (\ref{lpeq}) is equal to the dimension of the space, $m=d$ and thus $r=d-m+1=1$. There is a single feasible solution, $w_{\scriptscriptstyle \Lambda}=(w_{{\scriptscriptstyle \Lambda},\omega})$ and the polytope $\mathcal{W}_{\scriptscriptstyle \Lambda}=\mathcal{W}_{\scriptscriptstyle \Lambda}\subset\mathcal{W}_{\scriptscriptstyle I}$ is trivially identical to the working distribution $w_{\scriptscriptstyle \Lambda}$. At last there is a single probability distribution $\Prob$,
\begin{equation*}
\label{purestate}
\Prob(\omega=1|\Lambda_\mu)\ident w_{{\scriptscriptstyle \Lambda},\omega}
\end{equation*} 
 The simplicial entropy is  zero. Finally, the expectation of any observable $Q(\omega)=\mathrm{q}_\omega$ reads trivially
 \begin{equation}
 \label{Qpure}
 \langle Q \rangle = \langle \mathrm{q} w_{\scriptscriptstyle \Lambda} \rangle =\sum_{\omega\in\Omega}\mathrm{q}_\omega w_{{\scriptscriptstyle \Lambda},\omega}.
 \end{equation}

 The definition of a pure state can be extended to the case where the polytope is not reduced to an isolated point, but the contextual distribution $\Sigma_\mu$ is deterministic, because the working distribution is then a definite vertex of the simplex and the simplicial entropy is also zero. In the two cases, the working distribution is then an extreme point of the polytope.  This can be used as a definition.
 
 \begin{definition}[Pure and mixed simplicial quantum states]
  \label{pureandmixed}
 A simplicial quantum state is pure when the working distribution is an extreme point of the specific simplex. Otherwise, the state is mixed.
 \end{definition}

 \subsubsection{Mixed states}
 When the rank $m>0$ is less than $d$ 
the prior does not uniquely determine the solution of the system and therefore the working probability $w_{\scriptscriptstyle \Lambda}$ is defined by the contextual distribution $\Sigma_\mu$. 
In that case, from Definition (\ref{pureandmixed}) the simplicial state that accounts for both the specific simplex and the particular context is termed \enquote{mixed}.
 
 Let $\mu_i$ be the simplicial coordinates of $w_{\scriptscriptstyle \Lambda}$ in $\mathcal{W}_{\scriptscriptstyle \Lambda}$. We have,
 \begin{equation}
\label{mixedstate}
\Prob(\omega=1|\Lambda_\mu)\ident w_{{\scriptscriptstyle \Lambda},\omega} =\sum_{i=1}^{d-m+1} \mu_i w_{i,\omega}
\quad\mathrm{~with~}\sum_{i=1}^{d-m+1} \mu_i =1
\end{equation} 
 As a result, for any observable $Q(\omega)=\mathrm{q}_\omega$, we have
 \begin{equation}
 \label{mixedexpectation}
 \langle Q \rangle = \langle \mathrm{q} w_{\scriptscriptstyle \Lambda} \rangle =\sum_{i=1}^{d-m+1}\mu_i  \langle \mathrm{q} w_i \rangle =\sum_{\omega\in\Omega}\sum_{i=1}^{d-m+1}\mu_i \mathrm{q}_\omega w_{i,\omega}
 \end{equation}
 This equation is also valid for pure states, with $m=d$, $\mu_1=1$ and $w_1=w_{\scriptscriptstyle \Lambda}$.

\subsection[Measurement]{Measurement with respect to a simplicial quantum state}

 Let us now turn to the measurement of an observable with respect to a simplicial quantum state $(w_{\scriptscriptstyle \Lambda}, \mathcal{W}_{\scriptscriptstyle \Lambda})$, i.e., the expectation value with respect to the joined probability distribution on $(\Sigma_\mu, \Omega)$ composed of both the simplicial distribution $\{\mu_i\}$ and the LP solutions of $\mathcal{W}_{\scriptscriptstyle \Lambda}$. Since the two probabilities are independent, the global expectation is the expectation with respect to the working distribution. For simplicity, we take this result as a definition.
\begin{definition}[Quantum expectation $\langle Q\rangle$]
\label{qexpectation}
 The quantum expectation of an observable $Q(\omega)=\mathrm{q}_\omega$  is the  expectation $\langle Q\rangle=\langle \mathrm{q}  w_{\scriptscriptstyle \Lambda}\rangle$ with respect to the working distribution $w_{\scriptscriptstyle \Lambda}$.
 \end{definition}
Let us compute the probability of an event or the expectation of an observable. 

\subsubsection[Boolean function]{Measurement of a Boolean function} 
 Let $\mathsf{f}=(\omega_1,\omega_2,\dots,\omega_\ell)$ be a Boolean function, that is a disjunction of $\ell$ classical states $\omega_i$.   Since complete requirements are disjoint, the probability of $\mathsf{f}$ with respect to the probability distribution $w_{\scriptscriptstyle \Lambda}$ is the sum of the probabilities of its complete requirements $\omega_i$,
 \begin{equation*}
 \Prob(\mathsf{f}=1|\Lambda_\mu)=\sum_{i=1}^\ell w_{{\scriptscriptstyle \Lambda},\omega_i}.
\end{equation*}
Let $F$ be the indicator of the Boolean function and $\mathrm{f}=(\mathrm{f}_\omega)$ denote its associated covector. 
We have then from Eqs. (\ref{mixedstate}, \ref{mixedexpectation}),
\begin{equation}
 \label{classicalprojection}
 \Prob(\mathsf{f}=1|\Lambda_\mu)=\langle\mathrm{f} w_{\scriptscriptstyle \Lambda} \rangle = \langle F \rangle= \sum_{i=1}^{d-m+1}\sum_{\omega\in\Omega} \mu_i\mathrm{f}_\omega  w_{i,\omega}.
\end{equation}

\paragraph{Expectation of an observable.} 
 Let $\mathrm{q}=(\mathrm{q}_\omega)$ be a covector, corresponding to an observable $Q$. 
We saw, Eq. (\ref{mixedexpectation}), that
\begin{equation}
 \label{observablexpectation}
 \langle Q \rangle=\langle\mathrm{q} w_{\scriptscriptstyle \Lambda} \rangle =  \sum_{i=1}^{d-m+1}\sum_{\omega\in\Omega} \mu_i\mathrm{q}_\omega  w_{i,\omega}.
\end{equation}

\subsubsection{Projective measurement} 
Let $\Gamma=\{ \gamma\}$ denote a finite set. Define an ensemble of mutually disjoint Boolean functions $\{\mathsf{f}_\gamma, \gamma\in\Gamma\}$ such that the reunion of all $\mathsf{f}_\gamma$ is the tautology. Equivalently,  let $\{\mathrm{f}_\gamma=(\mathrm{f}_{\gamma,\omega}),\  \gamma\in\Gamma\}$ be the indicators $F_\gamma$ of $\mathsf{f}_\gamma$ in $\mathcal{P}^*$, such that $\sum_{\gamma} \mathrm{f}_{\gamma, \omega} =1$ for all $\omega\in\Omega$, i.e., $\sum_\gamma {F_\gamma}=I$. 

A standard measurement is defined as

$$\gamma\in\Gamma\mapsto \prob(\gamma)=\Prob(\mathsf{f}_\gamma=1|\Lambda_\mu)=\langle\mathrm{f_\gamma} w_{\scriptscriptstyle \Lambda} \rangle =\langle{F_\gamma}  \rangle\ge 0. $$
From Proposition (\ref{basicexpectation}),  a projective measurement means expanding the working distribution $w_{\scriptscriptstyle \Lambda}$ with respect to the set of subspaces defined by the Boolean functions $\mathsf{f}_\gamma$.
In particular,  when $\Gamma=\Omega$, $\{\mathsf{f}_\omega= \tilde{\omega}, \omega\in\Omega\}$,  $ \prob(\omega)= \Prob(\omega)$.

\subsubsection{General measurement} 
Let $\Gamma=\{ \gamma\}$ denote a finite set. Define an abstract resolution of the tautology, that is a set of non-negative forms in $\mathcal{P}^*$,\{q$_\gamma=(\mathrm{q}_{\gamma, \omega})$\} (with $\gamma\in\Gamma)$, such that $\sum_{\gamma} \mathrm{q}_{\gamma, \omega} =1$ for all $\omega\in\Omega$, i.e., $\sum_\gamma \mathrm{q}_\gamma=I$. 
Since $\mathrm{q}_{\gamma, \omega}$ is not necessarily $0$ or $1$, $\mathrm{q}_\gamma$ is not necessarily associated with a Boolean function, but corresponds to a positive observable $Q_\gamma$ and $\sum_\gamma {Q_\gamma}=I$.  
A general measurement is defined by

$$\gamma\in\Gamma\mapsto \prob(\gamma)=\langle\mathrm{q}_\gamma w_{\scriptscriptstyle \Lambda} \rangle =\langle{Q_\gamma}  \rangle. $$
This is similar to a particular positive-operator valued measure (POVM)  in quantum information, \emph{when the involved observables commute}. 

\subsection{Pair of registers}
\label{pairofregisters}

The combination of two registers brings together most of the peculiarities of quantum information. This will be briefly discussed in Sec. (\ref{qversusc}).
In the following, we review the consequences of the \enquote{Born's method} in the current source window.

Consider a global classical register $\mathsf{X}_c$ composed of two distinct subregisters $\mathsf{X}_a$ and $\mathsf{X}_b$. Let $(\Lambda_c)$ denote a global Bayesian prior.
Let $N_a$, $N_b$ and $N_c=N_a+N_b$ be the numbers of binary variables in $\mathsf{X}_a$, $\mathsf{X}_b$ and $\mathsf{X}_c$ respectively, still referred to as $\mathsf{X}_i, i\in\llbracket, 1,N_c\rrbracket$. Let $\mathcal{P}_a$, $\mathcal{P}_b$  and $\mathcal{P}_c$ denote the probability spaces corresponding to $\mathsf{X}_a$, $\mathsf{X}_b$ and $\mathsf{X}_c$ of dimension $d_a=2^{N_a}$, $d_b=2^{N_b}$ and  $d_c=2^{N_c}$ respectively. We have $\mathcal{P}_a \otimes \mathcal{P}_b=\mathcal{P}_c$, $N_a+N_b=N_c$ and $d_a\times d_b=d_c$. 
Let $\Omega_a$,   $\Omega_b$ and $\Omega_c$ be the sample sets of the probability distributions, so that  $\Omega_c$ is  the Cartesian product $\Omega_a\times\Omega_b=\Omega_c$. The classical states $\omega_a\in\Omega_a$,   $\omega_b\in\Omega_b$ and  $\omega_c\in\Omega_c$ also index the basis vectors in $\mathcal{P}_a$, $\mathcal{P}_b$  and $\mathcal{P}_c$.
Any classical state $\omega_c\in\Omega_c$ is the conjunction of  two partial classical states $\omega_a\in\Omega_a$ and $\omega_b\in\Omega_b$ belonging respectively to the two subregisters, i.e., $\omega_c=(\omega_a;\omega_b)$, where e.g., $\omega_a$ is both a complete requirement in $\mathsf{X}_a$ and a partial requirement in $\mathsf{X}_c$. Therefore, the atoms of the system are the $d_c$ classical states $\omega_c$.
%
On the other hand, the basis vectors $\tilde{\omega}_c\in\mathcal{P}_c$ are the tensorial products  $\tilde{\omega}_a\otimes\tilde{\omega}_b$ of the basis vectors in $\mathcal{P}_a$ and $\mathcal{P}_b$.
At last, the registers $\mathsf{X}_a$, $\mathsf{X}_b$ and $\mathsf{X}_c$ can also be viewed as random variables, taking values in $\llbracket 0,d_a-1\rrbracket$,  $\llbracket 0,d_b-1\rrbracket$ and $\llbracket 0,d_c-1\rrbracket$ respectively.

\emph{Notation.} 
The classical states,  e.g. in $\Omega_c$, are noted $\omega_{c,i},\, i\in\llbracket 1, d_c\rrbracket$.  To lighten the writing  when no confusion can occur, we use simply $\omega_c\in\Omega_c$.
The basis vectors are,  e.g. in $\mathcal{P}_c$, $\tilde{\omega}_{c,i},\, i\in\llbracket 1, d_c\rrbracket$, or simply $\tilde{\omega}_c,\, \forall \omega_c\in\Omega_c$. The entries of a vector, e.g. $w_c\in\mathcal{P}_c$, are noted $w_{c,\omega_c}$, $w_{c,i}$ or  $\mathbb{P}_c(\omega_c)$ where appropriate and the vector itself is noted $w_c=(w_{c,\omega_c})$ so that $w_a\otimes w_b=(w_{a,\omega_a} \times w_{b,\omega_b})\in\mathcal{P}_c$.

\subsubsection[Separability and entanglement]{Separability and entanglement of a single probability distribution}
\label{separability}

Consider a single probability distribution $w_c=\Prob_c(\omega_c)$ of the full LP problem, for instance, but at this stage not necessarily, the working distribution of a simplicial quantum state in $\mathcal{P}_c$.
The distribution, $w_c=\Prob_c(\omega_c)$, is \emph{separable} with respect to the partition ($\mathsf{X}_a$, $\mathsf{X}_b$) if $w_c$ is the Kronecker product $w_c=w_a\otimes w_b$ of two probability distributions, $w_a=(\Prob_a(\omega_a))$ and $w_b=(\Prob_b(\omega_b))$ belonging to $\mathcal{P}_a$ and $\mathcal{P}_b$ respectively, provided that $w_a$, $w_b$ and $w_c$  be normalized.
This is a standard problem in joint multivariate analysis, where separable random variables are termed \enquote{independent}.
\begin{definition}[Separability, entanglement]
A probability distribution, $\Prob_c(\omega_a;\omega_b)$ on a global register, $\mathsf{X}_c =(\mathsf{X}_a,\mathsf{X}_b$),  is separable with respect to a partition into the two distinct subregisters $\mathsf{X}_a$ and $\mathsf{X}_b$, iff
\begin{align}
\label{separable}
\begin{aligned}
&\Prob_c(\omega_a;\omega_b) =\Prob_a(\omega_a)\times \Prob_b(\omega_b),\\
\mathrm{subject~to}\quad
\sum_{\omega_a\in\Omega_a}&\Prob_a(\omega_a)
=\sum_{\omega_b\in\Omega_b}\Prob_b(\omega_b)
=\sum_{\omega_c\in\Omega_c}\Prob_c(\omega_c)=1.
\end{aligned}
\end{align}
The two distributions $\Prob_a(\omega_a)$ and $\Prob_b(\omega_b)$ are then the marginals of $\Prob_c(\omega_a;\omega_b)$ on $\Omega_a$ and $\Omega_b$ respectively.
In the language of random variables,  $\mathsf{X}_a$ and $\mathsf{X}_b$ are independent.
Otherwise, the joint distribution is \emph{entangled} and the random variables  $\mathsf{X}_a$ and $\mathsf{X}_b$ are correlated.
\end{definition}
For instance, consider a pair of distinct classical registers, each subject to particular constraints leading to two distinct LP problems. If we decide to regard the pair of independent registers as a unique register, the system is clearly separable.
Even if the system is not separable as a whole, it may arise that some solutions are separable.   In particular, any deterministic distribution $w_c=\tilde{\omega}_c$ is separable~\cite{mf3}. This mean that if $\Prob(\omega_a;\omega_b)\in\{0,1\}$, then the marginals $\Prob(\omega_a)\in\{0,1\}$ and $\Prob(\omega_b)\in\{0,1\}$ are both deterministic. In short, entanglement is impossible in the deterministic realm and the deterministic states are always separable.

However, in general a current solution of the global LP system, $\Prob_c(\omega_c)=\Prob_c(\omega_a;\omega_b)$ is not separable, i.e.,  is entangled. 
The two standard marginal distributions on $\Omega_a$ and $\Omega_b$ are respectively
\begin{align}
\label{partialprob}
\begin{aligned}
&\Prob_a(\omega_a)\ident\Prob_c(\omega_a)=\sum_{\omega_b\in\Omega_b}\Prob_c(\omega_a;\omega_b)
\quad;\quad 
\Prob_b(\omega_b)\ident\Prob_c(\omega_b)=\sum_{\omega_a\in\Omega_a}\Prob_c(\omega_a;\omega_b) 
\\
&\mathrm{where}\quad
\sum_{\omega_a\in\Omega_a}\Prob_a(\omega_a)
=\sum_{\omega_b\in\Omega_b}\Prob_b(\omega_b)
=\sum_{\omega_c\in\Omega_c}\Prob_c(\omega_c)=1.
\end{aligned}
\end{align}
On the other hand, the concept of marginal distribution is related to the joint distribution $\Prob_c(\omega_a;\omega_b)$ by the \emph{conditional probability}  $\Prob_c(\omega_a|\omega_b)$ thanks to Bayes' law, 
$$\Prob_c(\omega_a;\omega_b)=\Prob_b(\omega_b)\times\Prob_c(\omega_a|\omega_b). $$
When the marginal $\Prob_b(\omega_b)$ is zero,  the joint distribution $\Prob_c(\omega_a;\omega_b)$ is also zero.
When $\Prob_c(\omega_a;\omega_b)$ is separable, $\Prob_c(\omega_a|\omega_b)=\Prob_a(\omega_a)$.

From the probability distribution $\Prob_c(\omega_c)$  on $\Omega_c$, it is easy to derive a particular separable probability distribution $\Prob_c'(\omega_c)$ still on $\Omega_c$ as the product of the two marginal distributions $\Prob_a(\omega_a)$  and $\Prob_b(\omega_b)$, namely,
\begin{equation}
\label{separablejoint}
\Prob_c'(\omega_a;\omega_b)\ident \Prob_a(\omega_a)\times \Prob_b(\omega_b). 
\end{equation}
It turns out that the amount of entanglement of $\Prob_c$ can be characterized by the \emph{relative entropy} $S( \Prob_c \Vert \Prob_c')$ between the actual distribution $\Prob_c$ and the separable distribution $\Prob_c'$  in the sample set $\Omega_c$,  as (in bits)
\begin{equation}
\label{relativeentropy}
S( \Prob_c\ \Vert\ \Prob_c') = \sum_{\omega_c\in\Omega_c}\Prob_c(\omega_c)\log_2\frac{\Prob_c(\omega_c)}{\Prob_c'(\omega_c)}
\ge 0.
\end{equation}
\begin{proposition}
The global probability $\Prob_c$ is separable with respect to the partition ($\mathsf{X}_a$, $\mathsf{X}_b$) if and only its relative entropy with respect to the product $\Prob_c'(\omega_c)=\Prob_a(\omega_a)\times \Prob_b(\omega_b)$ of the marginal distribution in $\mathcal{P}_a$ and $\mathcal{P}_b$ is zero, that is, $S( \Prob_c \Vert \Prob_c')=0$.
\end{proposition}
\emph{Proof.} We have $S( \Prob_c \Vert \Prob_c')\ge 0$ because a relative entropy is always non-negative. In addition, $S( \Prob_c \Vert \Prob_c')$ is the minimum value over all possible  relative entropies $S( \Prob_c\Vert \Prob_c'')$ for all separable distributions $\Prob_c''(\omega_a;\omega_b) =\Prob_a''(\omega_a)\times\Prob_b''(\omega_b) $, since we have from Eqs. (\ref{partialprob}, \ref{relativeentropy})~\cite{mf2},
 $$S( \Prob_c\ \Vert\ \Prob_a\times\Prob_b)-S( \Prob_c\ \Vert\ \Prob_a''\times\Prob_b'')=-S( \Prob_a\ \Vert\  \Prob_a'')-S( \Prob_b\ \Vert\ \Prob_b'')\le 0. $$
Therefore, $0\le S( \Prob_c\ \Vert\ \Prob_a\times\Prob_b)\le S( \Prob_c\ \Vert\ \Prob_a''\times\Prob_b'')$.
The minimum of $S( \Prob_c\Vert \Prob_c'')$ is zero iff $\Prob'_a=\Prob_a$, $\Prob'_b=\Prob_b$ and $\Prob_c=\Prob_a\times \Prob_b$. $\Box$
\paragraph{} To sum up, we have the following result:
\begin{proposition}
\label{propentanglement}
A global probability distribution $w_c$ governing a pair of distinct classical registers subject to a global prior is generally entangled with respect to the pair of registers. The amount of entanglement is characterized by the relative entropy between the global distribution and the product of its marginal distributions, Eq. (\ref{relativeentropy}).
When the relative entropy is zero, the distribution $w_c$ is separable and equal to the product of its marginals.
\end{proposition}

Recall  that  the relative entropy between the joint distribution and the product of its marginals is specifically termed \emph{mutual information} in standard information theory. Therefore, the relative entropy $S( \Prob_c\Vert \Prob_c')$ can be expressed equivalently in terms of  mutual information $\mathbb{H}(\Omega_a;\Omega_b)$ with respect to the global probability $\Prob_c$ in the sample set $\Omega_c$ as, 
\begin{align}
\label{mutualinformation}
\begin{aligned}
{S}( \Prob_c\ \Vert\ \Prob_c')=\mathbb{H}(\Omega_a;\Omega_b) &= \mathbb{H}(\Omega_a) - \mathbb{H}(\Omega_a|\Omega_b) =\mathbb{H}(\Omega_b)-\mathbb{H}(\Omega_b|\Omega_a) \\
 &=\mathbb{H}(\Omega_a)+\mathbb{H}(\Omega_b)-\mathbb{H}(\Omega_a,\Omega_b)
 \end{aligned}
\end{align}
where e.g. $\mathbb{H}(\Omega_c)=\mathbb{H}(w_c)$ is the window entropy. In addition, this expression is a special case for bipartite systems of the so-called \enquote{total correlation} defined by S. Watanabe~\cite{watanabe} in communication theory (see also Ref. \cite{mf2}).

Entanglement is a trivial consequence of the \enquote{Born method} even in the classical realm. This is also a general feature of standard quantum information.

\emph{Notation.}
In the present framework,  we use the concept of \enquote{information} as a quasi-synonym of \enquote{negentropy}%
\footnote{Negentropy, as defined by L. Brillouin~\cite{brillouin}, is just the opposite of the entropy $\mathbb{H}$. However, it is convenient to consider the information of complete registers as positive and thus we define the information $\mathbb{I}$ of a probability distribution on a  $N$-bit sample set as $N-\mathbb{H}$ (in bits) instead of  $-\mathbb{H}$.}
 and adopt the symbol  $\mathbb{I}(.)$. 
However, in standard information theory, this symbol denotes the so-called signed \emph{information measure}~\cite{yeung} (condensed in \enquote{$I$-measure}) in the sigma-algebra (often pictured by a Venn diagram). By convention, any event is then regarded as a particular set of atoms $\omega_c$. With this convention, $\Omega_c=\Omega_a\cup\Omega_b=(\Omega_a,\Omega_b)$.
The $I$-measure is the unique extension to the sigma-algebra of the standard entropy defined on complete sample sets and specifically denoted by $\mathbb{H}(.)$ in that case. 
For clarity and without introducing ambiguity, we note here $\mathbb{H}(.)$  \emph{both} the positive $I$-measure of complete sample sets and the signed $I$-measure of other events.%
\footnote{
Strictly speaking, in the context of $I$-measure,  $\mathbb{H}(\Omega_c)$ should be written  $\mathbb{H}(X_c)$ where $X_c = \Omega_c-\varpi_c$ is a random variable and $\varpi_c$ is the empty atom in $\Omega_c$, i.e., the negation of all binary variables, while $A-B$ stands for $A\cap B^C$ but we retain for simplicity the notation  $\mathbb{H}(\Omega_c)$ and the similar expressions since $\mathbb{H}(\varpi_c)=0$.
} 
In particular, we note $\mathbb{H}(\Omega_a;\Omega_b) = \mathbb{H}(\Omega_a\cap\Omega_b)$ the mutual information usually noted ${I}(\Omega_a:\Omega_b)$ in quantum information theory.
 We reserve the symbol $S(.)$ either to the relative entropy ${S}( \Prob_c\ \Vert\ \Prob_c')$ or (below) to compute entropy in a Hilbert space.$\Box$
\paragraph{}

Proposition (\ref{propentanglement}) holds for the working distribution $w_c$ of a simplicial quantum state, 
but  the simplex $\mathcal{W}_c$ does not intervene as such.
To overcome this drawback, we will now construct a form of  \enquote{marginalization} of the complete simplicial quantum states.

\subsubsection{Partial simplicial quantum state}
\label{partialsystem}

The restriction of a global LP system to a subregister  will be termed \enquote{partial LP system}.
In essence, the problem is to reconstruct the effective probability subspace in the subregister.
Technically, the reduction is implemented with respect to the current working distribution at work in the global system, that is on the simplicial quantum state, but the reduced specific simplex is actually independent of the working distribution.
We will use indifferently the terms \enquote{partial}, \enquote{reduced} and \enquote{marginal} when no confusion can occur.

While the concept of  \emph{separable distributions} is not ambiguous, the situation is more subtle in LP systems.
For convenience, set the following definitions, where every vertex of the specific simplex is viewed as a single probability distribution.

\begin{definition}[Separable simplex]
\label{separablesimplex}
A simplex is separable with respect to a partition between two subregisters if all of its vertices are separable. Otherwise,  the simplex is twisted.
\end{definition}

\begin{definition}[Separable LP system]
\label{separableLP}
A LP system is separable with respect to a partition between two subregisters if its specific simplex is separable.  Otherwise, the LP system is twisted.
\end{definition}

\begin{definition}[Separable simplicial quantum state]
\label{separablesimplicial}
A simplicial quantum state $(w_c,\mathcal{W}_c)$  is separable with respect to a partition between two subregisters if its specific simplex $\mathcal{W}_c$ is separable, irrespective of the working distribution $w_c$. 
Otherwise, the simplicial quantum state is twisted.  For pure simplicial quantum state, $(w_c,\mathcal{W}_c)$ with $\mathcal{W}_c=\{ w_c\})$ twisted state and entangled state are synonymous.
\end{definition}

\begin{definition}[Product state]
\label{productstate}
A simplicial quantum state $(w_c,\mathcal{W}_c)$ is a product state with respect to a partition between two subregisters 
if it results merely from the simple concatenation of the two registers  $\mathsf{X}_a$ and  $\mathsf{X}_b$, meaning that the registers are defined independently, each subjected to its own constraint set.
\end{definition} 

\begin{definition}[Completely divisible state]
\label{divisiblestate}
A simplicial quantum state $(w_c,\mathcal{W}_c)$ is  completely divisible if it results from the concatenation of $N$ independent 1-bit registers  $\mathsf{X}_i$, each subjected to its own constraint set.
\end{definition}

\paragraph{Reduction of a pure state.}
Assume first that the Bayesian system $(\Lambda_c)$ in  $\mathcal{P}_c$ accepts a unique solution, i.e., depicts a pure state $w_c=(w_{c,(\omega_a;\omega_b)})$. The rank of the LP system is $m_c=d_c$. As a simplicial quantum state, its simplex is $\{w_c\}$ and the state is noted $(w_c, \{w_c\})$ or just $w_c$ for simplicity. The rank of the state is $r_c=d_c-m_c+1=1$ and the effective probability space $\mathbb{W}_c = \mathrm{Span}(w_c)$ is of dimension 1.

 \begin{proposition}[Reduction of a pure simplicial quantum state]
 \label{partialLPpure}
 The restriction to  $\mathcal{P}_a$ of a global pure state, $w_c(\omega_c)=\Prob_c(\omega_a;\omega_b)\in\mathcal{P}_c$, is a partial simplicial quantum state $(w_a, \mathcal{W}_a)$ whose specific simplex $\mathcal{W}_a$ is the convex hull of the points $\tilde{v}_{\omega_b}\in\mathcal{P}_a$
 \begin{align}
\label{vlambdaa} 
\mathcal{W}_{a} = \mathrm{conv}\,(\tilde{v}_{\omega_b})  
\quad;\quad
\tilde{v}_{\omega_b} &\ident
\sum_{\omega_a\in\Omega_a}  {\Prob_c(\omega_a|\omega_b)}\ \tilde{\omega}_a 
\end{align}
 Its rank $r_a$ is thus the rank of the set of vectors $\{\tilde{v}_{\omega_b}\}$ and the rank $m_a$ of the associated LP system is $m_a=d_a-r_a+1$. 
 The working distribution $w_a$ is the marginal in $\mathcal{P}_a$ of the probability distribution $w_c$ in $\mathcal{P}_c$.
  
  When the global pure state $w_c$ is separable, $r_a=1$ and the partial simplicial quantum state  is also a pure state $(w_a, \{w_a\})$. 
 \end{proposition} 
 \emph{Proof.}
 The restriction of the pure state $w_c\in\mathcal{P}_c$ to $\mathcal{P}_a$ comprises by definition its marginal,  $w_a =(w_{a,\omega_a})$, Eq. (\ref{partialprob}), as  
\begin{align}
\label{marginalpure}
\begin{aligned}
w_{a}&\ident\sum_{\omega_a\in\Omega_a}\sum_{\omega_b\in\Omega_b} \Prob_c(\omega_a;\omega_b)\,\tilde{\omega}_a 
= \sum_{\omega_b\in\Omega_b} \Prob_c(\omega_b)\sum_{\omega_a\in\Omega_a} \Prob_c(\omega_a|\omega_b) \,\tilde{\omega}_a
\end{aligned}
\end{align}
where
\(
\Prob_c(\omega_b)\ident
\sum_{\omega_a\in\Omega_a} w_{c,(\omega_a;\omega_b)} =w_{b,\omega_b}= \Prob_b(\omega_b).
\)
Let $v_{\omega_b,\omega_a} \ident \Prob_c(\omega_a|\omega_b)$, that is
\begin{align}
\label{vvlambda}
v_{\omega_b,\omega_a} =
\begin{cases}
{w_{c,(\omega_a;\omega_b)} }/{w_{b,\omega_b}}&\mathrm{if~}w_{b,\omega_b}\ne0\\
0&\mathrm{if~}w_{b,\omega_b}=0.
\end{cases}
\end{align}
Construct the vector set $\{\tilde{v}_{\omega_b}\,|\,\omega_b\in\Omega_b\}=\{(v_{\omega_b,\omega_a})\}$ in $\mathcal{P}_a$. Then, each vector $\tilde{v}_{\omega_b}\ne0$ is a probability distribution in $\mathcal{P}_a$. Define $\nu_{\omega_b} = \Prob_c(\omega_b)$ and let $r_a$ denote the rank of  $\{\tilde{v}_{\omega_b}\}$.
As a result,  from Eq. (\ref{marginalpure}), we have
\begin{equation}
\label{purepartial} 
w_a= \sum_{\omega_b\in\Omega_b}\nu_{\omega_b}\,\tilde{v}_{\omega_b}\in\mathcal{P}_a
\end{equation}
In other words, the working distribution in  $\mathcal{P}_a$ is determined by the barycentric coefficients $\nu_{\omega_b} = \Prob_c(\omega_b)$. Since by hypothesis the outcomes $\omega_b$ are no more involved in the partial states,  the coefficients $\nu_{\omega_b}$ are regarded henceforth as exogenous.
As a result, the set of feasible solutions in  $\mathcal{P}_a$ is the full polytope $\mathrm{conv}(\tilde{v}_{\omega_b})$ and its extreme points $\{w_{ai}\}$ are a subset of $\{\tilde{v}_{\omega_b}\}$.
This polytope is actually the tautological simplex $\mathcal{W}_{a}$ in the effective probability space $\mathbb{W}_{a}=\mathrm{Span}(\tilde{v}_{\omega_b})$ with basis  $\{w_{ai}\}$ in $\mathcal{P}_a$.
Thus, the pair of this simplex $\mathcal{W}_{a}$ and the initial marginal distribution $w_a$, Eq. (\ref{marginalpure}), defines a simplicial quantum state $(w_a, \mathcal{W}_{a})$ in the probability space $\mathcal{P}_a$.

 Since the global simplex $\mathcal{W}_{c}$ is reduced to a single point in isolation, there is only one choice for $w_c$ and therefore there is a unique partial LP system. 
 When $w_c$ is separable, $\Prob_c(\omega_a|\omega_b) = \Prob_c(\omega_a)$ irrespective of $\omega_b$ and
\begin{align*}
\tilde{v}_{\omega_b} &=
\sum_{\omega_a\in\Omega_a}  {\Prob_c(\omega_a|\omega_b)}\ \tilde{\omega}_a
=\sum_{\omega_a\in\Omega_a}  {\Prob_c(\omega_a)}   \ \tilde{\omega}_a =w_a
\end{align*}
so that the simplex $\mathcal{W}_{a}$ is reduced to the marginal distribution in isolation $\{w_a\}$.
 $\Box$

 \begin{proposition}
 A pure separable simplicial quantum state is a product state.
 \end{proposition}
 \emph{Proof.}
 The two independent LP systems are trivially e.g., in $\mathcal{P}_a$,  $\langle\tilde{\omega}_a\rangle=\Prob_a(\omega_a)$ and in $\mathcal{P}_b$,  $\langle\tilde{\omega}_b\rangle=\Prob_b(\omega_b)$. The concatenation leads in $\mathcal{P}_c$ to  $\langle\tilde{\omega}_c\rangle=\Prob_c(\omega_c)$ with $\omega_c=(\omega_a;\omega_b)$ so that $\Prob_c(\omega_c)=\Prob_a(\omega_a)\times\Prob_b(\omega_c)$. $\Box$

\paragraph{Reduction of a mixed state.}
\label{marginalcomplet}
Assume now that the Bayesian system $(\Lambda_c)$ in the probability space $\mathcal{P}_c=\mathcal{P}_a\otimes \mathcal{P}_{b}$ accepts a set of solutions located on a simplex $\mathcal{W}_c$  of $r_c$ vertices $w_{ci}, \,i\in\llbracket1,r_c\rrbracket$.
Every vertex  $w_{ci}$ determines a probability distribution $\Prob_{ci}(\omega_c)=w_{ci,\omega_c}$ on the sample set $\Omega_c$.
The simplex is complemented by a working distribution $w_c$, so that $\Prob_{c}(\omega_c)=w_{c,\omega_c}$ and the global simplicial quantum state is $(w_c,\mathcal{W}_c)$. 
\begin{proposition}[Reduction of a simplicial quantum state]
\label{partialLP}
The restriction to $\mathcal{P}_a$ of a global simplicial quantum state $(w_c, \mathcal{W}_c)\subset\mathcal{P}_c$ with $r_c$ vertices $w_{ci},\,i\in\llbracket 1, r_c\rrbracket$ where
\begin{equation}
\label{eqwc}
w_{ci}=\sum_{\omega_c\in\Omega_c}  w_{ci,\omega_c}\tilde{\omega}_c\quad;\quad
w_c=\sum_{i=1}^{r_c} \mu_iw_{ci}
\quad\mathrm{with}\quad
\mu_i>0
\quad\mathrm{and}\quad 
\sum_{i=1}^{r_c} \mu_i=1,
\end{equation} 
is a simplicial quantum states,  $(w_a, \mathcal{W}_a)$. The partial working distributions $w_a\in\mathcal{P}_a$ is the marginal of the global working distribution $w_c\in\mathcal{P}_c$. 
The simplex $\mathcal{W}_a\subset\mathcal{P}_a$ is the convex hull  $\mathcal{W}_a=\mathrm{conv}(v_{i\omega_b})$ of the set of vectors $\tilde{v}_{i\omega_b}=\sum_{\omega_a\in\Omega_a}\Prob_{ci}(\omega_a|\omega_b)\,\tilde{\omega}_a\in\mathcal{P}_a$ for $i\in\llbracket1,r_c\rrbracket$ and  $\omega_b\in\Omega_b$. 
The number of vertices $r_a$ is the rank of the set of vectors $\tilde{v}_{i\omega_b}$ in $\mathcal{P}_a$.
The simplex $\mathcal{W}_a$ is independent of the contextual distribution $\{\mu_i\}$ while the working distribution $w_a$ depends linearly on $\{\mu_i\}$.
Similar results are obtained by permuting the indexes \enquote{$a$} and \enquote{$b$}.
In general, even for separable simplicial quantum states, $w_c\ne w_a\otimes w_b.$
\end{proposition}
\emph{Proof.}
Let $w_a$ denote the marginal of $w_c$ in $\mathcal{P}_a$.
Clearly, Eq. (\ref{marginalpure}) is still valid,
\begin{align*}
\begin{aligned}
w_{a}&\ident\sum_{\omega_a\in\Omega_a}\sum_{\omega_b\in\Omega_b} w_{c,(\omega_a;\omega_b)}\,\tilde{\omega}_a 
= \sum_{\omega_b\in\Omega_b} \Prob_c(\omega_b)\sum_{\omega_a\in\Omega_a} \Prob_c(\omega_a|\omega_b) \,\tilde{\omega}_a,
\end{aligned}
\end{align*}
but now, $w_c=\sum_{i=1}^{r_c} \mu_iw_{ci}$ and thus,
\begin{align}
\label{marginalmixed}
\begin{aligned}
w_a &= \sum_{i=1}^{r_c} \mu_i \sum_{\omega_a\in\Omega_a}\sum_{\omega_b\in\Omega_b} w_{ci,(\omega_a;\omega_b)}\,\tilde{\omega}_a \\
&=\sum_{i=1}^{r_c}\sum_{\omega_b\in\Omega_b}  \mu_i \Prob_{ci}(\omega_b)\sum_{\omega_a\in\Omega_a}\Prob_{ci}(\omega_a|\omega_b) \,\tilde{\omega}_a,
\end{aligned}
\end{align}
so that $w_a$ depends linearly on $\mu_i$.

For every pair $(i,\omega_b)$ with $i\in\llbracket1,r_c\rrbracket$ and  $\omega_b\in\Omega_b$ define a $\mu_i$-dependent positive coefficient $ \nu_{i\omega_b}$ as $\nu_{i\omega_b}= \mu_i \Prob_{ci}(\omega_b)\in\mathbb{R}$ and a vector  $\tilde{v}_{i\omega_b}\in\mathcal{P}_a$ independent of $\mu_i$ as
 \begin{equation}
 \label{vlambdab}
 \tilde{v}_{i\omega_b}=\sum_{\omega_a\in\Omega_a}\Prob_{ci}(\omega_a|\omega_b)\,\tilde{\omega}_a,
 \end{equation}
 where $\Prob_{ci}(\omega_a|\omega_b)=0$ when $\Prob_{ci}(\omega_b)=0 $.
 As a result,
 $$
 w_a=\sum_{i=1}^{r_c}\sum_{\omega_b\in\Omega_b} \nu_{i\omega_b} \tilde{v}_{i\omega_b}\quad \mathrm{where}\quad
 \sum_{i=1}^{r_c}\sum_{\omega_b\in\Omega_b} \nu_{i\omega_b}=\sum_{i=1}^{r_c}\mu_i\sum_{\omega_b\in\Omega_b}  \Prob_{ci}(\omega_b)= 1.$$
Let $r_a$ be the rank of the vector set $\{\tilde{v}_{i\omega_b}\}$ in $\mathcal{P}_a$. Now, construct the subspace
\begin{equation}
 \label{partialeffective}
\mathbb{W}_{r_a}=\mathrm{Span}(\tilde{v}_{i\omega_b}|\,{i\in\llbracket1,r_c\rrbracket,  \omega_b\in\Omega_b}).
\end{equation}
 and in addition, construct the polytope
 \begin{equation}
 \label{partialsimplex}
 \mathcal{W}_a=\mathrm{conv}(\tilde{v}_{i\omega_b}|\,{i\in\llbracket1,r_c\rrbracket,  \omega_b\in\Omega_b}).
 \end{equation}
As in the case of a pure state, $\mathcal{W}_{a}$ is the specific polytope of a partial LP system of rank $m_a = d_a-r_a+1$ in $\mathbb{W}_{r_a}\subseteq\mathcal{P}_a$, and, from Proposition (\ref{propspecificsimplex}),  $\mathcal{W}_{a}$ is a simplex.
Its vertices $\{w_{aj}\,|\,j\in\llbracket1,r_a\rrbracket\}$ are a subset of $\{\tilde{v}_{i\omega_b}\}$. 
As a result, $(w_a, \mathcal{W}_{a})$ is a simplicial quantum state constituting the reduced state in $\mathcal{P}_a$ of the global simplicial quantum state $(w_c, \mathcal{W}_{c})$. Furthermore, the simplex $\mathcal{W}_{a}$ is the union of all partial simplices of the global states $w'_c\in\mathcal{W}_{c}$ regarded are as pure states $(w'c,\{w'_c\})$.
 
Since the vectors $\tilde{v}_{i\omega_b}$, Eq. (\ref{vlambdab}) are independent  of  the global contextual distribution $\{\mu_i\}$, the simplex $ \mathcal{W}_{a}$ is also independent of $\{\mu_i\}$, that is, every vertex, $w_{aj}$ where $j\in\llbracket1,r_a\rrbracket$ is independent of $\{\mu_i\}$. 
By contrast, since the global working distribution is linearly dependent on $\mu_i$,  the partial simplicial coefficients, say $\mu_{aj}$, also depend linearly on $\mu_i$. 

The same procedure can be used in  $\mathcal{P}_b$.
By construction, the three working distributions $w_a$, $w_b$ and $w_c$ depend linearly on $\mu_i$, so that the Kronecker product $w_a\otimes w_b$ is quadradic on $\mu_i$. As a result, in general $w_c\ne w_a\otimes w_b$.
$\Box$
%
\begin{proposition}[Separable state]
\label{centerseparable}
When the global state is separable, the rank ratio $r_c/r_a$ is integer and the partial mass center $c_a$ is the marginal $\tilde{a}$ of the global mass center $\tilde{c}$.
\end{proposition}
\emph{Proof.}
When $\mathcal{W}_c$ is separable, $\Prob_{ci}(\omega_a|\omega_b)=\Prob_{ci}(\omega_a)$, so that, from Eq. (\ref{vlambdab}),  irrespective of $\omega_b$, $\tilde{v}_{i\omega_b}$ is the marginal $v_{ai}$ of  $w_{ci}$ while, from Proposition (\ref{partialLPpure}), the reduction in $\mathcal{P}_a$ of any extreme point $(w_{ci},\{w_{ci}\})$ in isolation is a pure state $(v_{ai},\{v_{ai}\})$. As a result, the marginal $v_{ai}$ for $i\in\llbracket1,r_c\rrbracket$ of every extreme point $w_{ci}$ is an extreme point $w_{aj}$ for $j\in\llbracket1,r_a\rrbracket$ of the partial simplex $\mathcal{W}_a$ so that the local vertices $w_{aj}$ of $\mathcal{W}_a$ are all the marginal of one or several global vertices. Since the contextual distribution is not involved, from Proposition (\ref{propspecificpolytope}) the vertices play the same role and by symmetry $r_c/r_a$ must be integer. The marginal of the center of mass $\tilde{c}=(1/r_c)\sum_{i=1}^{r_c}w_{ci}$ is thus  $\tilde{a}=(1/r_c)\sum_{i=1}^{r_c}v_{ai}= (1/r_a)\sum_{j=1}^{r_a}w_{aj}= c_a$.
$\Box$
\paragraph{Construction of a global simplicial quantum state from a pair of reduced states.}
Given two arbitrary simplicial quantum states in  $\mathcal{P}_a$ an  $\mathcal{P}_b$, it is always possible to construct a compatible  global  state in  $\mathcal{P}_c$. 
\begin{proposition}
\label{compatiblestate}
There is always a non-empty set of global simplicial quantum states compatible with an arbitrary pair of partial simplicial quantum states.
\end{proposition}
\emph{Proof.}
The set of compatible global simplicial quantum state contains the product state and is thus non-empty.
$\Box$
\paragraph{}
In conclusion, the restriction of a global simplicial quantum state to a subregister is always possible.
Even if  the global state $(w_c, \mathcal{W}_c)$ is pure,  the partial states  $(w_a, \mathcal{W}_a)$ and  $(w_b, \mathcal{W}_b)$ are generally mixed, with the exception of separable pure states $(w_c, \{w_c\})$. In other words, the simplicial entropy of the subsystem can be greater than the entropy of the full system and therefore \emph{the simplicial entropy is not extensive}. Again, this property is a simple consequence of the  \enquote{Born method} and corresponds to the partial trace in standard quantum information theory.


\subsubsection[Non-signaling correlations]{Local consistency and non-signaling correlations}

Consider two correlated subregisters $\mathsf{X}_a$, $\mathsf{X}_b$ and the partial sample sets $\Omega_a$, $\Omega_b$. The joint distribution $\Prob_c(\omega_c)$ is defined in the Cartesian product $\Omega_c=(\Omega_a,\Omega_b)$. From the definition of a partial subsystem, a local observer has only access to the variables of one subsystem and can only take into account the corresponding marginal probabilities. In other words, \emph{each subsystem endowed with its marginal probability distribution is self-consistent} and can be considered in isolation. 
\begin{proposition}
The correlations between two partial subsystems subject to a global Bayesian prior are non-signaling.
\end{proposition}
\emph{Proof.} From Proposition (\ref{compatiblestate}), whatever the second subsystem, the two partial subsystems are compatible. Therefore, any measurement in a subsystem is unable to provide information on the other subsystem. 
$\Box$
\paragraph{} 
 Implicitly, the variables involved in the system comprise all input,  output and ancillary data. 
The  non-signaling property is less trivial when some input variables are implicit and considered as parameters. Then, for clarity,  the actual variable set can be complemented so that the implicit variables become genuine variables as opposed to only parameters (see e.g., Example \ref{tripletstate} below).

We proved this result first in the context of the EPR paradox~\cite{mf} (the free choice of a working distribution  was called \enquote{argument} and the complete setup termed \enquote{stochastic gauge system}). 
The expression \enquote{non-signaling correlations} was coined by Barrett \emph{et al}~\cite{barrett} after a proposal by Popescu and Rohrlich to regard \enquote{nonlocality} as an axiom of quantum physics~\cite{popescu}.

Eventually, this is also an important feature of the partial trace in quantum information.

\subsubsection[Purifying a simplicial quantum state]{\enquote{Purification} of $(w_a, \mathcal{W}_a)$ into $\mathcal{P}_c$}
\label{purification}
We saw that computing a partial LP system is similar to calculating the partial trace in quantum formalism. 
This suggests to consider the equivalent of a purification of the simplicial quantum state  $(w_a, \mathcal{W}_a)$ in $\mathcal{P}_a$ with $r_a>1$ vertices into a pure state $w_c$ in $\mathcal{P}_c$. 

Consider the LP system of rank $m_a$ in $\mathcal{P}_a$ with $m_a=d_a-r_a+1$ extreme points, $w_i$. It is possible to  construct a \enquote{purification} of $(w_a, \mathcal{W}_a)$ in $\mathcal{P}_c$.
\begin{proposition}[\enquote{Purification}]
A simplicial quantum state $(w_a,\mathcal{W}_a)$ in a probability space $\mathcal{P}_a$ can be considered as the partial system a pure state $w_c$ in a probability space $\mathcal{P}_c=\mathcal{P}_a\otimes\mathcal{P}_b$.
\end{proposition}
\emph{Proof.} 
Start from
\begin{equation}
\label{wa}
w_a=\sum_{i=1}^{r_a}\mu_i w_i\in\mathcal{W}_a\subset\mathcal{P}_a.
\end{equation}
where $\mu_i$ are the simplicial coordinates of $w_a$. 
Define an auxiliary space $\mathcal{P}_b$ and suppose that $d_b\ge r_a$.
Construct an arbitrary set of $r_a$ independent vectors $v_i$ in the tautological simplex $\mathcal{W}_{I_b}$ in $\mathcal{P}_b$,  i.e., $v_i\in\mathcal{W}_{I_b}\subset\mathcal{P}_b$ for $i\in\llbracket 1,r_a\rrbracket$.
Construct a probability distribution $w_{c}=(w_{c,\omega_c})=(w_{c,(\omega_a; \omega_b)})\in \mathcal{P}_c=\mathcal{P}_a\otimes\mathcal{P}_b$ as
\begin{align*}
{w_c}=\sum_{i=1}^{r_a} \mu_{i}w_{i} \otimes v_i
\quad\mathrm{i.e.}\quad
w_{c,(\omega_a; \omega_b)}= 
\sum_{i=1}^{r_a} \mu_{i}w_{i,\omega_a}v_{i,\omega_b}
\end{align*}
We have clearly,
$$\sum_{\omega_c\in\Omega_c}w_{c,\omega_c}=\sum_{\omega_a\in\Omega_a}\sum_{\omega_b\in\Omega_b}w_{c,(\omega_a; \omega_b)}=\sum_{i=1}^{r_a}\mu_{i}\sum_{\omega_a\in\Omega_a}w_{i,\omega_a}\sum_{\omega_b\in\Omega_b} v_{i,\omega_b}=1$$
so that $w_c$ is indeed a probability distribution in $\mathcal{P}_c$ and from Eq. (\ref{wa})
$$\sum_{\omega_b\in\Omega_b}w_{c,(\omega_a; \omega_b)}=\sum_{i=1}^{r_a}\mu_{i} w_{i,\omega_a}\sum_{\omega_b\in\Omega_b}v_{i,\omega_b}=\sum_{i=1}^{r_a}\mu_{i} w_{i,\omega_a}=w_{a,\omega_a}.$$
Then, $w_a\in\mathcal{P}_a$ is effectively the marginal of $w_c\in\mathcal{P}_c$.
The \enquote{purification}  is completed. $\Box$
\paragraph{} 

 Depending upon the particular set of distributions $\{v_i\}$ in $\mathcal{P}_b$ there is a number of possible solutions. 
For simplicity, it is possible to select $v_i$ specifically among the basis vectors in $\mathcal{P}_b$.
Label $\omega_b\in\llbracket 1, d_b\rrbracket$ the basis vectors $\tilde{\omega}_b$ in $\mathcal{P}_b$.
Consider the set of $r_a$ basis vectors $\tilde{\omega}_b\in\mathcal{P}_b$ for $\omega_b\in\llbracket 1,r_a\rrbracket$.
For ease of exposition, rename $\omega_b$ the dummy subscript $i$ in Eq. (\ref{wa}).  Rewrite $w_a=\sum_{\omega_b=1}^{r_a}\mu_{\omega_b} w_{\omega_b}$ and set $v_{\omega_b}=\tilde{\omega}_b\in\mathcal{P}_b$ for $\omega_b\in\llbracket 1,r_a\rrbracket$.
Construct the specific probability distribution $w_{c}=(w_{c,(\omega_a; \omega_b)})\in \mathcal{P}_c=\mathcal{P}_a\otimes\mathcal{P}_b$ as
\begin{align}
\label{wc0}
\begin{aligned}
{w_c}=\sum_{\omega_b=1}^{r_a} \mu_{\omega_b}w_{\omega_b} \otimes \tilde{\omega}_b
\quad\mathrm{then}\quad
w_{c,(\omega_a; \omega_b)}= 
\begin{cases}
{\mu_{\omega_b}}w_{\omega_b,\omega_a}&\mathrm{~if~}\omega_b\in\llbracket 1, r_a\rrbracket\\
0&\mathrm{~otherwise.}
\end{cases}
\end{aligned}
\end{align}

Partial systems and  \enquote{purifications} in real probability spaces are formally equivalent  to partial traces and purifications in Hilbert spaces.

\section[Construction of a Hilbert space]{Transcription of the probability space into a Hilbert space}

When solving a constrained logic problem, a particular LP system was expressed in a probability space, $\mathcal{P}$. By construction, $\mathcal {P} $ is specific to the current batch of $N$ binary queries.

\subsection{Window contextuality}
\label{transcription}

On the other hand, the choice of a batch of queries is arbitrary, and depends in principle on the free choice of the observer. This choice therefore introduces a form of contextuality which we will call \enquote{window contextuality}.

\begin{definition}[Window contextuality]
\label{defwindowcontextuality}
Window contextuality corresponds to the free choice of a particular batch of dichotomic queries.
\end{definition}

Recall that \enquote{source contextuality}, Definition (\ref{defsourcecontextuality}), corresponds to the exogenous assignment of a specific working distribution among the feasible solutions on the specific simplex.
\paragraph{}

Now, there is a close connection between the  particular batch of dichotomic queries and the sample set $\Omega$ in the source window.

\begin{proposition}
There is a one-to-one correspondence between the sample set $\Omega$ defined in the source window and  the source batch of dichotomic queries.
\end{proposition}

\emph{Proof.} By definition, the basic sample set $\Omega$ is the ensemble $\{ \omega\}$ of the $2^N$  mutually exclusive classical states  describing the joint probability distribution of all source queries.
$\Box$
\paragraph{}

For simplicity, when no confusion can occur, we will name $\Omega$ both the probability sample set and the corresponding query batch. 
Of course, it is possible to change $\Omega$ while leaving invariant the logical system.
How to implement such a change while keeping the probability distribution defined by the Bayesian prior? It turns out that this is possible purely mechanically simply by  introducing an exogenous tool, namely, a Hilbert space.
\subsection{Conservation of probability}

By hypothesis, all batches of queries concern the same logical system. Therefore, each observation window $\Omega$ depicts a particular resolution of the tautology of  total probability 1. Namely
\begin{equation}
\label{tautores}
\forall \Omega\,:\, \sum_{\omega\in\Omega} \Prob(\omega)=1
\end{equation}

 \begin{proposition}
Any resolution of the tautology defines a particular observation window.
\end{proposition}

\emph{Proof.} Any resolution of the tautology defines a sample set $\Omega$ and thus an observation window. 
$\Box$
\paragraph{}

Now,  to change the observation window, just change the sample set $\Omega$.

\subsection{Changing the observation window}

For convenience, let us introduce an equivalent formulation to Eq. (\ref{tautores}).

\begin{equation}
\label{tores}
\forall \Omega\,:\, \sum_{\omega\in\Omega} \Big|\sqrt{\Prob(\omega)}e^{i\theta}\Big|^2=1
\end{equation}
where $\theta(\omega)$ is an arbitrary gauge parameter. 

This suggests to introduce a Hermitian metric in a convenient space, namely, a finite dimensional Hilbert space, as a tool to change the sample set $\Omega$. 
This might seem arbitrary but  \enquote{math is also art to add unexpected elements to solve problems more easily} (quoting a well known mathematician, Claude Dellacherie). For example, in geometry, we think of drawing a segment, and the demonstration takes shape.
Here, the trick is that unitary channels~\cite{holevo} acting on Hilbert spaces allow to assign consistently the probability distributions describing different observation windows while respecting the initial constraints. In addition, we need to conserve the value of the observables, that is technically to maintain the relationship between the space and its dual.

In standard quantum information, a Hilbert space of infinite dimension is arbitrarily introduced from crash. Next, a founding rule called \enquote {Born rule} is deduced from Gleason's theorem.
In the current model, there is an altogether elementary algebraic equivalent of Gleason's theorem. 

We do start  from the source probability space $ \mathcal {P} $ and transcribe the problem into another space, namely a Hilbert space, just requiring that the relationship between each space and its dual be preserved. The change of observation window is obtained by unitary operators acting on the Hilbert space.
Thereby, the contextuality thus introduced is in no way \enquote{abstract} as in standard quantum information but indeed based on the free choice of a batch of dichotomics queries.
Incidentally, this leaves no room for the paradoxical speculations of standard quantum information.
In every observation window, the guideline is simply to conserve in the transcription the value of  dual forms in both the probability space $ \mathcal {P} $ and its counterpart in the Hilbert space.
 
\begin{proposition}[Hilbert space]
\label{reformule}
It is always possible to reformulate each Bayesian LP problem, initially expressed in a probability space $\mathcal{P}$, by using a finite dimensional Hilbert space $\mathcal{H}$ while conserving the value of dual forms.
\end{proposition}

\emph{Proof.}
From definition (\ref{defobservable}), the expectation $\langle Q\rangle$ of an observable $(\mathrm{q}_\omega)$ is just the dual form $\langle Q\rangle=\langle \mathrm{q}p\rangle$ of the probability distribution $\Prob(\omega)=(p_\omega)$ in  $\mathcal{P}$.
Let us construct a complex-valued vector space, say $\mathcal{H}$, derived from the sample set $\Omega$ as the \emph{complex} span of the classical states $\omega$.
Next, from Eq. (\ref{tores}), represent each probability vector $\mathbb{P}(\omega)$ in  $\mathcal{P}$ by a rank 1-projector in $\mathcal{H}$ as
\begin{equation}
\label{vec}
\ket{\Prob(\omega)e^{i\theta}}\bra{\Prob(\omega)e^{i\theta}}=\ket{\Prob(\omega)}\bra{\Prob(\omega)}
\end{equation}

Dual forms are conserved provided that any observable in   $\mathcal{P}$ is represented by a diagonal operator $\mathsf{Q}=Diag(\mathrm{q}_\omega)$ in   $\mathcal{H}$. Hence, by simple inspection, its expectation remains by construction 
$$\langle Q\rangle \ident\langle \mathrm{q}p\rangle= \tr(\ket{\Prob(\omega)}\bra{\Prob(\omega)}\mathsf{Q})=\bra{\Prob(\omega)}\mathsf{Q}\ket{\Prob(\omega)}.$$
$\Box$
\paragraph{}

By construction, the transcription preserves both the simplex and the working distribution. 
\emph{Gleason's theorem is not used}. This excludes any possibility that quantum mechanics harbors an extra-logical part, surreptitiously introduced by Gleason's theorem, as certain authors suspect.

\paragraph{}
The Bayesian theater is now planted.
The main result is posited by the following theorem whose demonstration will be given throughout this paper.
\begin{theorem}
The Bayesian inference resolution of a constrained logical problem can be formulated indifferently using any batch of variables from an ensemble of related batches.
It is possible to switch from one variable batch to another by unitary channels acting on an auxiliary Hilbert space. 
In general, only part of the information contained in the prior can be extracted by specific measurements using a single batch of variables.
The complete ensemble of variable batches enables to extract the totality of the information and thus the totality of the relevant variable batches is thereby obtained by unitary channels.
\end{theorem}
\emph{Hints.} 
We have seen that the current problem can be transcribed into a Hilbert space. The proof that other batches of variables express the same problem will be given constructively by reverse transcription, in Sec. (\ref{reversetrans}).
A particular observable is well-defined only when expressed in terms of a specific variable batch because it is precisely a linear function $\Omega\to\mathbb{R}$ from the corresponding specific sample set $\Omega$, Definition (\ref{defobservable}).
That the complete ensemble of related variables is obtained from all windows of the Hilbert space will be proved by Proposition (\ref{completude}), based  on the comprehensive distribution of the prior information.
$\Box$
\paragraph{}

Although communication channels are well known, this particular treatment of a classical batch of Boolean variables is ignored both in classical information theory and conventional Bayesian analysis.

We will first describe the transcription of the source LP problem defined in a real-valued probability space into a complex-valued Hilbert space while preserving at this stage the initial batch of Boolean variables. 
The transcription is performed with respect to a particular source context, i.e., preserves both the simplex and the working distribution.

\subsection{Transcription of simplicial quantum states}
\label{statetranscription}
In this section, we use the subscript \enquote{$a$} for ease of exposition. We will resume our current notations in Sec. (\ref{Bornrule}) below.

Consider a source window as defined in the previous section, i.e., the simplicial representation of a quantum state $(w_a, \mathcal{W}_a)$ or equivalently $(\Sigma_\mu, \mathcal{W}_a)$, Definition (\ref{qrepg}),
\begin{equation*}
w_a=\sum_{i=1}^{r_a}\mu_i w_i\in\mathcal{W}_a\subset\mathcal{P}_a\quad;\quad \mu_i\in\Sigma_\mu
\end{equation*}
where  $\mathcal{W}_a\subset \mathcal{P}_a$ is a simplex with $r_a=d_a-m+1$ vertices $w_i\in\mathcal{W}_a$ ($i\in\llbracket 1, r_a\rrbracket$) and $w_a$  a working distribution in a real-valued probability space $\mathcal{P}_a$, while  $\Sigma_\mu=\{\mu_i\}$ denotes the set of simplicial coefficients, i.e., $\mu_i> 0$ and $\sum_{i=1}^{r_a} \mu_i=1$.

Now, we propose to construct a Hilbert space $\mathcal{H}_a$ as the \emph{complex} span of the sample set $\Omega_a$ with a standard Hermitian metric as, 
$$\mathcal{H}_a = Span(\omega_a | \ \omega_a\in{\Omega_a}).$$
We note $\ket{\omega_a}$ for $\omega_a\in\Omega_a$ the $d_a$ basic vectors in $\mathcal{H}_a$. 
For simplicity, when no confusion can occur, we note also $\Omega_a$ this particular basis so that $\{\ket{\omega_a}\}=\Omega_a$.
Except when mentioned otherwise, all linear operators $\mathsf{M}\in\mathcal{L}(\mathcal{H}_a)$ map $\mathcal{H}_a$ to $\mathcal{H}_a$. 
We note $\mathsf{M}^\dagger$ the adjoint of a linear operator $\mathsf{M}$ with respect to the Hermitian metric. Let $\mathrm{D}(\mathcal{H}_a)\subset\mathcal{L}(\mathcal{H}_a)$ be the set of density operators acting on $\mathcal{H}_a$, that is the set of positive Hermitian matrices of trace $1$.

In the previous section, we constructed a simplicial quantum state from a LP problem using the scheme
$$ \mathrm{Bayesian~prior~}\Lambda_a \to \mathrm{~simplex~}\mathcal{W}_a \mathrm{~in~} \mathcal{P}_a \to \mathrm{~simplicial~quantum~state~}( w_a, \mathcal{W}_a ),$$
The construction requires to set the working distribution $w_a$ within the simplex $\mathcal{W}_a$.  
This is an intrinsic input and in no way a gauge entity.
Now, we propose the following transcription scheme:
$$ \mathrm{~simplicial~quantum~state~}(w_a, \mathcal{W}_a) \mathrm{~in~} \mathcal{P}_a \to\mathrm{~density~operator~} \rho_a \mathrm{~in~}\mathrm{D}(\mathcal{H}_a)$$

We will find that the transcription is not unique in general and requires a gauge selection among a set of equivalent transcriptions.

\subsubsection[Pure state]{Transcription of a pure state}
\label{sectranspure}
When $r_a=1$, the simplex is reduced to a single distribution $w_a$ in the real space $\mathcal{P}_a$ of dimension $d_a$.  This distribution can be transcribed as a projection operator $\ket{a}\bra{a}$ acting on $\mathcal{H}_a$, where $\ket{a}$ is a unit vector:
\begin{align}
\label{transpure}
w_a\mathrm{~is~transcribed~as~} \rho_a=\ket{a}\bra{a}\mathrm{~with~} |a_{\omega_a}|^2=w_{a,\omega_a}.
\end{align}
\begin{proposition}
\label{puregauge}
It is possible to transcribe a pure simplicial quantum state from a probability space $\mathcal{P}_a$ into a Hilbert space $\mathcal{H}_a$ by constructing a unit vector  $\ket{a}\in\mathcal{H}$ complying with Eq. (\ref{transpure}). The density matrix acting on $\mathcal{H}$ is  the projector $\rho_a=\ket{a}\bra{a}$. 
\end{proposition}
\emph{Proof.} 
A pure state corresponds to a simplex reduced to a single vertex. This vertex defines a probability distribution vector  $w_a$ in the probability space $\mathcal{P}$. Now, just apply Proposition (\ref{reformule}) in $\mathcal{H}_a$. Note that in standard physics, this is also the direct application of  Gleason's theorem. 
$\Box$
\paragraph{}
We find convenient to call \enquote{Gleason's vector} the vector $\ket{a}$.
\begin{definition}[Gleason's vector]
\label{defgleason}
A Gleason's vector is any unit vector $\ket{a}\in\mathcal{H}_a$ obtained by transcription of a pure state.
\end{definition}
From  Eq. (\ref{transpure}), the entries of the working distribution $w_a$ in  $\mathcal{P}_a$ coincide with the diagonal entries of the density operator $\rho_a$ in  $\mathcal{H}_a$.
Therefore,  the reverse transcription of the current pure state from the density operator $\rho_a$ acting on the Hilbert space $\mathcal{H}_a$ to the working distribution $w_a$ in the probability space $\mathcal{P}_a$ is trivial.

\paragraph{Gauge selection.} Obviously, the transcription, Proposition (\ref{puregauge}), is compatible with many solutions. 
Therefore, the choice of a particular unit vector $\ket{a}$  complying with Eq.~(\ref{transpure}) implies a \emph{gauge selection}.

\begin{proposition}
\label{gaugegleasonpure}
Gauge transformations correspond to changing the phase of the Gleason's vector components.
\end{proposition}
\emph{Proof.} Since by definition, the working distribution is invariant, this results from Eq.~(\ref{transpure}). 
$\Box$
\paragraph{}

From Wigner's theorem these transformations, say  $\mathsf{\Theta}$, can be antiunitary or unitary.
In any case, this requires either to construct another Hilbert space or to consider another basis in the same Hilbert space. The first possibility will be noted \enquote{global gauge} and the second \enquote{local gauge}.
For definiteness, let us address the \emph{global gauge}%
\footnote{%
The model of \emph{local gauge} is left out of the current article.
}.
Construct a new Hilbert space $\mathcal{H}_{a'}$. 
\begin{equation}
\label{puretransition0}
\mathsf{\Theta} :  \quad \mathcal{H}_{a}\to\mathcal{H}_{a'}:\quad\ket{a}\mapsto \ket{a'} = \mathsf{\Theta}\ket{a}
\end{equation}

First, transcribing a real-valued problem into a complex-valued framework implies an initial gauge choice between $i$ and $-i$.
This choice is made \emph{once and for all} and is comparable to the initial choice of a discrete Boolean gauge, Definition~(\ref{sourcegauge}).
As a result the problem necessarily has two equivalent representations simply related by complex conjugation. 
Let $\mathsf{K}: \mathbb{C}\to\mathbb{C}: z\mapsto z^*$ denote the standard complex conjugation.
In the current basis, this change is expressed by  a \emph{antiunitary transformation},  $\mathsf{\Theta}=\mathsf{K}\times\mathds{1}_d$ as
\begin{equation}
\label{puretransitionantiunitary}
\ket{a}\mapsto \ket{a'} = \ket{a^*}, 
\end{equation}
where $a'_{\omega_a}=a^*_{\omega_a}$.
This transformation is involutive, that is, equal to its own inverse. This particular expression depends on the current basis of the Hilbert space and other possibilities exist. 
For the sake of generality, we will define later an intrinsic antiunitary gauge operator $\mathsf{C}$   instead of $\mathsf{K}\times\mathds{1}_d$ (see  Sec. (\ref{gtg}) below). This generates a discrete conjugation group $\mathscr{C}=\{\mathds{1}_d, \mathsf{C}\}$ acting on the Hilbert space $\mathcal{H}_a$.

Second,  there exists a continuous set of \emph{unitary matrices} $\mathsf{\Theta}$ complying with Proposition~(\ref{gaugegleasonpure}), for example in the current basis, the diagonal $d$-unitary matrix, $Diag(\exp\mathbf{i}\theta_i)$. 
These unitary solutions form a continuous unitary gauge group $\mathcal{G}$ acting on the Hilbert space $\mathcal{H}_a$ that we will also construct intrinsically in Sec.~(\ref{gtg}) below.

Finally, the full gauge group, say $\mathfrak{G}$,  will be constructed as a semi-direct product $\mathfrak{G} = \mathcal{G} \ltimes \mathscr{C}$.

\paragraph{}
Alternatively, a \emph{local gauge} could be built by keeping a single Hilbert space and assigning a specific basis to each gauge.

\subsubsection[Mixed state]{Transcription of a mixed state}
\label{computew}
A mixed simplicial state, $(\Sigma_\mu, \mathcal{W}_a)$, is defined by a simplex $\mathcal{W}_a$ composed of $r_a>1$ extreme points $w_i$ in  $\mathcal{P}_a$ and a set $\Sigma_\mu=\{\mu_i\}$ of simplicial coordinates. 

\begin{proposition}
\label{mixedgauge}
A mixed simplicial quantum state $(\Sigma_\mu, \mathcal{W}_a)$ can be transcribed as a density operator $\rho_a$.
Each extreme point $w_i$ of the simplex is transcribed independently as a pure state $\ket{a_i}\bra{a_i}$, where the vector $\ket{a_i}$ is the Gleason's vectors associated to $w_i$, while the simplicial coordinates $\mu_i$ are conserved. Then
\begin{equation}
\label{rhoa}
\rho_a=\sum_{i=1}^{r_a} \mu_i\ket{a_i}\bra{a_i}.
\end{equation}
The pure states  can be regarded as the extreme points of the transcribed simplex.
\end{proposition}
\emph{Proof.}
The working distribution $w_a$ can be viewed as a weighted combination of $r_a>1$ \emph{auxiliary pure states} of working distributions $w_i$ in $\mathcal{P}_a$ for $i\in\llbracket 1, r_a\rrbracket$. Since the weighting coefficients $\mu_i$ are independent of the simplex itself, the mixed state must be transcribed for consistency as the same weighted combination of the $r$ transcribed projectors $\ket{a_i}\bra{a_i}$ of the auxiliary pure states $w_i$. Then the mixed state in  $\mathcal{H}_a$ is also considered as a simplex, now  composed of the $r_a$ extreme points $\ket{a_i}\bra{a_i}\in\mathrm{D}(\mathcal{H}_a)$.
From Eq. (\ref{transpure}) , we obtain  Eq. (\ref{rhoa}).$\Box$
\paragraph{}
This construction can also be obtained by a \emph{purification} procedure.
\begin{proposition}
\label{purifymixted}
The transcription of a mixed simplicial state can be implemented by (1) \enquote{purifying} this mixed state, (2) transcribing the simplicial pure state to obtained a standard quantum pure state and (3) tracing out this pure state.
\end{proposition} 
\emph{Proof.}
We proceed in three steps. (1) \enquote{Purify} the simplicial quantum state $\{ w_a, \mathcal{W}_a \}$ of rank $r_a$ defined in the real probability space $\mathcal{P}_a$ into a pure state $w_c$ living in an auxiliary space $\mathcal{P}_c=\mathcal{P}_a\otimes\mathcal{P}_b$, as described in Sec. (\ref{purification}). (2) Transcribe the pure state $w_c$ into a projection operator $\ket{c}\bra{c}$ defined in a Hilbert space $\mathcal{H}_c=\mathcal{H}_a\otimes\mathcal{H}_b$. (3) Compute the partial trace over $\mathcal{H}_b$ of the projection operator $\ket{c}\bra{c}$ to obtain the relevant density operator $\rho_a$ in $\mathcal{H}_a$. 
Step (1) has been defined in Sec. (\ref{partialsystem}). Consider a real probability space $\mathcal{P}_b$ of dimension $d_b\ge r_a$. Assume that $d_b=r_a$ and select the set of $r_a$ basis vectors in $\mathcal{P}_b$, as described by Eq. (\ref{wc0}), 
\begin{align*}
\begin{aligned}
{w_c}\ident\sum_{\omega_b=1}^{r_a} \mu_{\omega_b}w_{\omega_b} \otimes \tilde{\omega}_b
\quad\mathrm{then}\quad
w_{c,(\omega_a; \omega_b)}= 
{\mu_{\omega_b}}w_{\omega_b,\omega_a}
\end{aligned}
\end{align*}
where  we changed the dummy subscripts  \enquote{$i$} into \enquote{$\omega_b$} for clarity. 

Step (2) has been constructed just above (Proposition \ref{puregauge}). 
%
Let us denote  $\ket{c}$  the Gleason's vector and 
$ c_{(\omega_a;\omega_b)}$ its entries.

Step (3) is a standard operation in quantum information with a unique solution. 
Resuming the subscripts \enquote{$\omega_b$} into \enquote{$i$}, we obtain 
\begin{equation*}
\rho_a=\tr_b (\ket{c}\bra{c})=\sum_{i=1}^{r_a} \mu_i\ket{a_i}\bra{a_i}
\end{equation*}
%
We have recovered Eq. (\ref{rhoa}) as required. 
$\Box$
%
%
\paragraph{Gauge selection.}
Any particular feasible Gleason's vector $\ket{c}$ constructed in Step (2) corresponds to a gauge selection, as described for pure states. 

\begin{proposition}
\label{gaugegleason}
Gauge transformations are the unitary or antiunitary operators that  modify the phase of the involved Gleason's vector components.
\end{proposition}
\emph{Proof.} This results from Proposition (\ref{gaugegleasonpure}) and the transcription method irrespective of the case. $\Box$
\paragraph{}
The complete set of gauge transformations will be addressed later in Sec. (\ref{gtg}) below. 

\paragraph{Standard density operator.}

The expansion of the density operator  $\rho_a$ as a weighted array of pure states,  Eq. (\ref{rhoa}), is not standard, albeit considered in detail by, among others, Jaynes \cite{jaynes2}.
Indeed, while of norm $1$, the Gleason's vectors $\ket{a_i}$ are \emph{not} orthogonal in general. Nevertheless, we can easily obtained an orthonormal set of vectors $\ket{e_j}\in\mathcal{H}_a$ by a standard diagonalization of $\rho_a$ as,
$$\rho_a =\sum_{i=1}^{r_a} \mu_i \ket{a_i}\bra{a_i}
\quad \Longrightarrow\quad
\rho_a=\sum_{j=1}^{r_a} \lambda_j\ket{e_j}\bra{e_j}\quad\mathrm{with}\quad \braket{e_j}{e_{j'}}=\delta_{jj'}$$
The computation of the eigenvalues $\lambda_j$ from the simplicial coefficients $\mu_i$ is then straightforward.   

Since there is a one-to-one correspondence between the vertexes $w_i$ and the Gleason's vectors $\ket{a_i}$, the source window is called \emph{regular}. The concept of \enquote{regular window} as opposed to \enquote{blind window} will be clarified in the next section (Definition \ref{defregular} below).
On the other hand, retrieving the simplicial coefficients $\mu_j$ or the  $r$ vertices $w_j$ from $\rho_a$ is not that trivial and will be detailed below in Sec. (\ref{reversetrans}).

Let us define the spectrum of the density operator as $$\mathrm{spec}(\rho_a)=\Sigma_a=\{\lambda_j\}.$$
We obtain the final result:
\begin{proposition}
The simplicial quantum state $\{ w_a, \mathcal{W}_a \}$ in $\mathcal{P}_a$ is transcribed in $\mathcal{H}_a$ as a density operator $\rho_a$, depending on a transcription gauge. 
Starting from the simplicial representation,
$$ w_a = \sum_{i=1}^{r_a} \mu_i\ w_i\quad;\quad\mathcal{W}_a=\mathrm{conv}(w_i)\quad;\quad\mu_i\in\Sigma_\mu$$
the transcribed density operator is
\begin{equation}
\label{densityop}
\rho_a  \ident\sum_{i=1}^{r_a} \mu_i\ \ket{a_i}\bra{a_i}=\sum_{i=1}^{r_a} \lambda_i \ket{e_{i}}\bra{{e}_{i}}
\quad\mathrm{with}\quad
\lambda_i\in\Sigma_{\scriptscriptstyle \Lambda}=\mathrm{spec}(\rho_a)
\end{equation}
where $\ket{e_i}$ are a set of $r_a=d_a-m_a+1$ orthonormal vectors.
In particular there is a real gauge with $a_{i,\omega_a}=\sqrt{w}_{i,\omega_a}$.
\end{proposition}
\paragraph{Working distribution versus density operator.}
Irrespective of the gauge, it is straightforward to recover the working distribution $w_a$ from the density operator $\rho_a$. 
\begin{proposition}
 The working distribution $w_a=(w_{a,\omega_a})$ in $\mathcal{P}_a$ is the diagonal probability  distribution of the density operator $\rho_a$ and can be recovered  as
\begin{equation}
\label{waa}
\forall \omega_a\in\Omega_a:\quad w_{a, \omega_a} = \bra{{\omega}_a}\rho_a\ket{{\omega}_a}
\end{equation}
\end{proposition}
\emph{Proof.}
From $\rho_a= \sum_i \mu_i\ \ket{a_i}\bra{a_i}$ we have
$$ \bra{{\omega}_a}\rho_a\ket{{\omega}_a}=  \bra{{\omega}_a} \sum_{i=1}^{r_a}( \mu_i\ \ket{a_i}\bra{a_i})\ket{{\omega}_a}= \sum_{i=1}^{r_a} \mu_i\ |\braket{{\omega}_a}{a_i}|^2=\sum_{i=1}^{r_a} \mu_i w_{i,\omega_a} = w_{a, \omega_a}\quad\Box.$$
%

\paragraph{Simplicial entropy versus von Neumann entropy.}
The simplicial entropy is closely related to the von Neumann entropy of the density operator $\rho_a$. Start from the standard theorem
\begin{theorem}
The von Neumann entropy $S(\rho_a)$ of the quantum state, $\rho_a$, is 
$$ S(\rho_a)=\mathbb{H}(\Sigma_a)=\sum_{i=1}^{r_a} -\lambda_i \log \lambda_i=S_a. $$
\end{theorem}

\emph{Proof.} This is a standard result of quantum information. Since $\lambda_i$ are the eigenvalues of the density operator $\rho_a$, we have $S_a=-\tr\rho_a\log\rho_a$. $\Box$
\paragraph{}We have the additional result:
\begin{proposition}[Jaynes' inequality]
\label{bound}
The von Neumann entropy $S(\rho_a)=\mathbb{H}(\Sigma_a)$ is bounded above by the simplicial entropy in any window $S_\mu=\mathbb{H}(\Sigma_\mu)$.
\begin{equation}
\label{eqbound}
 \mathbb{H}(\Sigma_a)\le \mathbb{H}(\Sigma_\mu)
\end{equation}
\end{proposition}
\emph{Proof.} 
In another wording, the inequality is due to Jaynes (Ref.~\cite{jaynes2}, Appendix A). The proof works as follows.
Basically, in Eq. (\ref{densityop}), we have 
$\sqrt{\mu_i}\ket{a_i}=\sum_{j=1}^{r_a} U_{ij}\sqrt{\lambda_j}\ket{e_j}$
where $(U_{ij})$ is some $r_a\times r_a$ unitary matrix. From this and the orthogonality of $\ket{e_j}$, it follows that
$ \mu_i = \sum_{j=1}^{r_a} u_{ij} \lambda_j$
where $u_{ij} =|U_{ij}|^2$ with
$\sum_i  u_{ij} = \sum_j  u_{ij} =1$.
Given the well-known inequality $x\log x\ge x-1$ based on convexity, we obtain,
%
$$\sum_{i=1}^{r_a} -\mu_i \log\mu_i\ge\sum_{i=1}^{r_a} -\lambda_i \log\lambda_i\quad\mathrm{or}\quad S_\mu=\mathbb{H}(\Sigma_\mu)\ge\mathbb{H}(\Sigma_a) = S_a.$$
In addition, we will see that the inequality is saturated in a principal window (Proposition \ref{satsimple} below). $\Box$.
\paragraph{Window entropy from the density operator.}
Recall that the entropy of the working distribution is the window entropy (Definition \ref{defwindowentropy}).
It can be immediately computed from the density operator. 
\begin{proposition}
The window entropy of a quantum state $\rho_a=(\rho_{ij})$ is the entropy of the diagonal probability distribution $\rho_{ii}$.
\label{propwindowentropy}
\begin{equation}
\label{eqwindowentropy}
\mathbb{H}(\Omega_a)= \sum_{\omega_a\in\Omega_a} -\bra{{\omega}_a}\rho_a\ket{{\omega}_a}\log_2\bra{{\omega}_a}\rho_a\ket{{\omega}_a}
\end{equation}
\end{proposition}
\emph{Proof.} Obvious from Eq. (\ref{waa}). $\Box$

\subsection{Transcription of observables} 
\label{transcrob}

 Consider a probability space $\mathcal{P}_a$ and the Hilbert space $\mathcal{H}_a$. By construction, the covectors of the dual space $\mathcal{P}_a^*$ are transcribed into $\mathcal{H}_a$ so as to ensure the consistency of the dual forms.
 As a result, the transcription does not depend on the gauge.
Let $w_a\in\mathcal{W}_a$ denote the working distribution of a quantum state.
Consider an arbitrary observable $Q_a(\omega_a)=\mathrm{q}_{\omega_a}$ and let $\mathrm{q}_a=(\mathrm{q}_{a,\omega_a}) \in \mathcal{P}_a^*$.
 
 \begin{proposition}[Transcription of observables]
Irrespective of the gauge,  a covector $\mathrm{q}_a$ in $\mathcal{P}_a^*$ is transcribed into a diagonal operator acting on $\mathcal{H}_a$:
\begin{equation}
\label{hq}
\mathrm{q}=(\mathrm{q}_{a,\omega}) \in \mathcal{P}_a^* \mathrm{~is~transcribed~as~} \mathsf{Q}_a = \diag{\omega_a\in\Omega_a}(\mathrm{q}_{a,\omega_a}).
\end{equation}
\end{proposition}
\emph{Proof.} 
Define a diagonal operator acting on $\mathcal{H}_a$ as
$\mathsf{Q}_a = \mathrm{Diag}(\mathrm{q}_{a,\omega_a}).$ Computing the trace, we have identically from Eq. (\ref{densityop}) in a particular gauge,
$$\langle Q_a\rangle_a=\langle  \mathrm{q}_a w_a\rangle = \sum_{i=1}^{r_a} \mu_i\langle  \mathrm{q} w_i\rangle= \sum_{i=1}^{r_a} \mu_i\tr ( \mathsf{Q}_a \ket{a_i}\bra{a_i})= \tr(\mathsf{Q}_a \rho_a)$$
$\Box$
\paragraph{}
By anticipation, note that since this transcription leads to a Hermitian diagonal operator, its uniqueness whatever the gauge will only hold in that window where the operator is diagonal, i.e., in the proper window of the observable (Definition \ref{observableproperwindow}). By contrast, in other windows, the Hermitian operator remains a Hermitian operator but depends generally on the gauge and  the observable can no longer be reverse-transcribed within that window.

The transcription of a Boolean formula is noteworthy.

\begin{proposition}[Boolean formulas]
Irrespective of the gauge, a Boolean formula is transcribed into an orthogonal projection operator.
\end{proposition}
\emph{Proof.} From Proposition (\ref{basicexpectation}), a Boolean formula is represented by a particular observable, namely, an indicator function composed only of 0 and 1 entries.
$\Box$

\subsection{Expectation and Born rule}
\label{Bornrule}
Let us resume our usual notation, i.e., leave the subscript \enquote{$a$} or replace \enquote{$a$} by \enquote{$\Lambda$} where appropriate.
A simplicial quantum state $\{w_{\scriptscriptstyle \Lambda}, \mathcal{W}_{\scriptscriptstyle \Lambda} \}$ is transcribed as a density operator $\rho_{\scriptscriptstyle \Lambda}$ depending on the gauge. An observable $Q$ is transcribed as a diagonal operator $\mathsf{Q}$ independent of the gauge.
Then, irrespective of the gauge, the dual forms, $\langle \mathrm{q} w_{\scriptscriptstyle \Lambda}\rangle$ with $\mathrm{q}\in\mathcal{P}^*$  are transcribed as $\langle \mathrm{q} w_{\scriptscriptstyle \Lambda}\rangle=\tr (\mathsf{Q}\rho_{\scriptscriptstyle \Lambda})$.
The expectation of an observable $Q(\omega)=\mathrm{q}_\omega$ with respect to the probability distribution $\Prob(\omega)=w_{{\scriptscriptstyle \Lambda},\omega}\in\mathcal{W}_{\scriptscriptstyle \Lambda}$ is then, 
\begin{equation}
\label{hdual}
\langle Q\rangle= \langle \mathrm{q} w_{\scriptscriptstyle \Lambda}\rangle =\tr (\mathsf{Q}\rho_{\scriptscriptstyle \Lambda})
\end{equation}

\begin{proposition}
\label{staticexpectation}
In the transcription of a source system into a Hilbert space the expectation value of an observable is computed by the Born rule.
\end{proposition}
\emph{Proof.}
From Eq. (\ref{hq}) all observables are transcribed as Hermitian operators. From Definition (\ref{qexpectation}) the Born rule Eq. (\ref{hdual}) is obvious. Note that for pure states, this is the very content of  Gleason's theorem. $\Box$
\paragraph{}

 More generally, a resolution of the tautology described by a set $\Gamma$ of non-negative forms,  q$_\gamma\in\mathcal{P}^*, \gamma\in\Gamma$, is translated as a commutative POVM $\{\mathsf{Q}_\gamma\}$ acting on $\mathcal{H}$ and
  $$\prob(\gamma) = \tr (\rho \mathsf{Q}_\gamma ),$$
so that general commutative measurements can be performed. 

We will show later (Theorem \ref{generalexpectation}) that beyond the source system,   the Born rule holds as well in general systems, i.e., for observables depicted by arbitrary Hermitian operators $\mathsf{Q}$, not necessarily diagonal. Let us name \enquote{proper window} the window where the Hermitian operator is diagonal.
\begin{definition}[Proper window of an observable]
\label{observableproperwindow}
The proper window of an observable $\mathsf{Q}$ in a Hilbert space $\mathcal{H}$ is a window where the Hermitian operator $\mathsf{Q}$ is diagonal.
\end{definition}

When the observable is an orthogonal projection operator onto a subspace $\mathcal{H}_\ell\subseteq\mathcal{H}$ of the Hilbert space, this definition applies to this  subspace.
\begin{definition}[Proper window of a subspace]
A proper window of a subspace $\mathcal{H}_\ell\subseteq\mathcal{H}$ in a Hilbert space $\mathcal{H}$ is a window where the subspace is spanned by basis vectors.
\end{definition}

 \subsection{Bayesian theater and observation windows}
 \label{defwindow}
 Until now we have used the concepts of \enquote{observation window}  and \enquote{Bayesian theater} informally.
At this stage, it is already possible to formalize our terminology by anticipating the notion of reverse transcription (Sec. \ref{reversetrans} below).

 
The problem is initially formulated with a particular Boolean variable batch of sample set $\Omega_0$ as a Bayesian prior $(\Lambda_0)$ in  a particular probability space $ \mathcal{P}_0$. The constraints can be completely captured by a simplicial quantum state $(w_0, \mathcal{W}_0)$.
However, the technique of Bayesian inference makes it possible to reformulate the same problem with other batches of related Boolean variables. An intermediate step is required, namely, transcribe the probability system into a Hilbert space $\mathcal{H}$. The initial sample set  $\Omega_0$ is transcribed as a basis, still called for simplicity $\Omega_0=\{\ket{\omega_0}\}$ in  $\mathcal{H}$ and the simplicial quantum state $(w_0, \mathcal{W}_0)$ is transcribed as a density operator $\rho^{(0)}$ acting on   $\mathcal{H}$.
The complete system of  related Boolean variable batches is then obtained by changing the basis in  $\mathcal{H}$ from $\Omega_0$ to new bases $\Omega_i$ leading to new expressions of the density operator from $\rho^{(0)}$ to $\rho^{(i)}$. Next, the density operators $\rho^{(i)}$ are reverse-transcribed as new  simplicial quantum states $(w_i, \mathcal{W}_i)$ defined in new  probability spaces $ \mathcal{P}_i$.
The \enquote{Bayesian theater} is the overall system while each particular variable batch defines an \enquote{observation window} $\Omega_i$.
\begin{center}
\begin{minipage}{0.9\linewidth}
\begin{tikzpicture}
  \matrix (m) [matrix of math nodes,row sep=2em,column sep=8em,minimum width=1em] 
  {
    \mathrm{windows}\rightarrow&  \Omega_0 & \Omega_i   \\
   \mathrm{Hilbert~space}:   &  \rho^{(0)} & \rho^{(i)}   \\
    \mathrm{Probability~spaces}: &  (w_0, \mathcal{W}_0) & (w_i, \mathcal{W}_i) \\
    }
     ;
  \path[-stealth] 
    (m-2-2) edge [<->, thick]  (m-3-2)
            edge [double,->] node [below] {$\mathrm{Unitary~operator}$} (m-2-3)
    (m-2-3) edge [->, thick] (m-3-3)
    ;
\end{tikzpicture}
\end{minipage} 
\end{center}
It turns out that the complex part of the Bayesian theater corresponds identically to the standard model of quantum information.
In addition, based on the saturation of the entropic inequalities (see below Eq. (\ref{POVMinequality3}) and Sec. \ref{entropicinequalities}), the union of all windows represents the complete set of related Boolean variable batches. 

\begin{definition}[Bayesian theater and observation window]
A Bayesian theater is the representation by Bayesian inference of a logical problem with multiple discrete degrees of freedom, regardless of the particular Boolean variable batch.
An observation window is a particular implementation of a Bayesian theater with a specific variable batch, which requires the allocation of a  distinct Boolean variable to each degree of freedom. 
The Bayesian theater can be depicted either by the complete set of windows or equivalently by their transcription into a single Hilbert space. 
\end{definition}
 
 \begin{proposition}[Individual  window $\Omega_i$]
 In the Hilbert space $\mathcal{H}$ every individual  window $\Omega_i$ corresponds to a specific basis, also noted $\Omega_i$ and the probability distribution is expressed by a standard \enquote{quantum state}, i.e., a density operator  expressed in this basis.  
Equivalently, the individual  window $\Omega_i$ is depicted by a  Bayesian LP system on a real-valued probability space $\mathcal{P}_i=\mathbb{R}^{\Omega_i}$ and the probability distribution is expressed by a \enquote{simplicial quantum state}.  
 \end{proposition}
In Sec. (\ref{statetranscription}), we have seen that any source window in $\mathcal{P}$ can be transcribed into  $\mathcal{H}$ using a particular transcription gauge.  
In Sec. (\ref{reversetrans}), we will show that conversely any window in $\mathcal{H}$ can be regarded as a source window in $\mathcal{P}$ except for some exceptional cases that will be referred to as \enquote{blind windows}.

\section{General systems}
\label{dynamic}
Let us first recall the concept of quantum channel, which is the tool to explore the complete set of Boolean variable batches.

\subsection{Quantum channels}

In standard quantum information, quantum channels represent operations that transform the states of one register into states of another register~\citep{holevo}.
Here, we will use  quantum channels to explore a unique Hilbert space $\mathcal{H}$. The various windows represent the same logical problem formulated with different batches of Boolean variables. A channel $\Phi:\mathrm{D}(\mathcal{H})\to \mathrm{D}(\mathcal{H})$ transforms a state $\rho$ in the initial basis into a new state $\rho'$ in a second basis. 
Technically, $\Phi$ must be trace-preserving and completely positive, so that any probability remains a probability while being compatible with a concatenation of registers.

\paragraph{Kraus representation.}
We characterize a quantum channel,  $\Phi:\mathrm{D}(\mathcal{H})\to \mathrm{D}(\mathcal{H})$, by the so-called \enquote{Kraus representation}. Let $\rho=\sum_i\lambda_i \ket{e_i}\bra{e_i}$ be a density operator of rank $r$. Let $\Gamma=\{\gamma\}$ denote a finite set and $\mathsf{M}_\gamma$ a set of linear operators in $\mathcal{H}$ such that $\mathsf{M}_\gamma^\dagger\mathsf{M}_\gamma$ is a resolution of the identity  for $\gamma\in\Gamma$. We have,
\begin{align}
\label{Kraus}
\begin{aligned}
&\rho'\ident\Phi(\rho)=\sum_{\gamma\in\Gamma} \mathsf{M}_\gamma\rho\mathsf{M}_\gamma^\dagger = \sum_{i=1}^{r}\sum_{\gamma\in\Gamma}\lambda_i \mathsf{M}_\gamma\ket{e_i}\bra{e_i}\mathsf{M}_\gamma^\dagger  \\
&\mathrm{with~} \sum_{\gamma\in\Gamma} \mathsf{M}_\gamma^\dagger\mathsf{M}_\gamma=\mathds{1}_d
\end{aligned}
\end{align}
The operators $\mathsf{M}_\gamma$ are the \enquote{Kraus operators}.

\paragraph{Unitary channels.}

The most basic channels are those that only change the batch of binary variables, i.e., change the observation window. They are reversible and trivially specified by a single Kraus operator. As a result, they are simply the unitary operators acting on the Hilbert space and form the unitary group $\mathrm{U}(d)$. It is convenient to call this group the \enquote{window group}.

\begin{definition}[Window group]
\label{windowgroup}
The window group is the transformation group of the different bases in the  Hilbert space $\mathcal{H}$.
\end{definition}

Unitatary channels conserve the von Neumann entropy of the density operator.
By contrast, general channels  are usually irreversible, leading to an increase of the von Neumann entropy~\cite{petz}.

\paragraph{Probability induced by a channel.}
By reverse transcription a window means a probability distribution $\Prob$ over the classical states $\omega\in\Omega$ of a batch of Boolean variables. This distribution will be computed in the following section.
Assume that the density operator $\rho$ is mapped to a new state $\rho'$ by a unitary quantum channel $\Phi$. In the new basis, the reverse transcription of $\rho'$  defines a new specific simplex $\mathcal{W}'$, a new sample set $\Omega'$, and a new working  distribution $w'$.

\subsection{Reverse transcription into a source system}
\label{reversetrans}
Reverse transcription is always possible, so that any window can be regarded as a source window with the exception of some exceptional windows that we will call \enquote{blind}.

 To this end, the simplex $\mathcal{W}_{\scriptscriptstyle \Lambda}$ is defined by a specific set of extreme points $\{w_i\}$ while the working distribution corresponds to a set of simplicial coefficients $\Sigma_\mu= \{\mu_i\}$. 
$$\{(\lambda_j,\ket{e_j}\bra{e_j})\} \mapsto \{ (\mu_i,w_i)\}.$$
Informally, this mapping transforms a convex ensemble in the set of density operators $\mathrm{D}(\mathcal{H})$, namely, $\mathrm{conv}_j(\ket{e_j}\bra{e_j})$ into another convex set in the tautological simplex $\mathcal{W}_{\scriptscriptstyle I}$, namely,  $\mathrm{conv}_i(w_i)$. The pure states  are transformed into the extreme points of the simplex and the working distribution $w_{\scriptscriptstyle \Lambda}$ is directly displayed by the diagonal of the density operator $\rho_{\scriptscriptstyle \Lambda}$ in accordance with Eq. (\ref{waa}).

\subsubsection[Pure state]{Reverse transcription of a pure state}
\label{LPtranspure}
Reverse transcription of a pure state is straightforward.
Let $\rho_{\scriptscriptstyle \Lambda}= \ket{e}\bra{e}$ denote a pure density matrix in $\mathcal{H}$. From Eq. (\ref{transpure}), the working distribution is $w_{\scriptscriptstyle \Lambda}= |e|^2\in\mathcal{P}$, i.e., $w_{{\scriptscriptstyle \Lambda},\omega}= |e_\omega|^2$. The simplex $\mathcal{W}_{\scriptscriptstyle \Lambda}$ is reduced to the isolated vertex $\{w_{\scriptscriptstyle \Lambda}\}$.
\begin{proposition}
\label{reversepure}
A density operator $\rho_{\scriptscriptstyle \Lambda}= \ket{e}\bra{e}$ of rank 1 is reverse-transcribed as a simplex $\mathcal{W}_{\scriptscriptstyle \Lambda}=\{w_{\scriptscriptstyle \Lambda}\}$ composed of an isolated vertex $w_{{\scriptscriptstyle \Lambda}}=(w_{{\scriptscriptstyle \Lambda},\omega})$ with $w_{{\scriptscriptstyle \Lambda},\omega}= |e_\omega|^2$.
\end{proposition}

\paragraph{LP system.} The vector $w_{\scriptscriptstyle \Lambda}$ is trivially the solution  of the linear system $p=|e|^2$ of rank $m=d$
\begin{align}
\label{reversepureLP}
\begin{aligned}
p_{\omega}=&\ |e_\omega|^2\quad (\forall\omega\in\Omega)
\end{aligned}
\end{align}
Alternatively,  the system can be formulated as
\begin{align*}
\mathrm{Assign~}\Prob \mathrm{~subject~to~} \langle \tilde{\omega}^*\rangle= |e_\omega|^2\quad (\forall\omega\in\Omega)
\end{align*}
where $\tilde{\omega}^*$ is the indicator function corresponding to the classical state $\omega$. The normalization arises from the normalization of $e$.

\subsubsection[Mixed state]{Reverse transcription of a mixed state}

Start from a density operator $\rho_{\scriptscriptstyle \Lambda}$ of rank $r$ acting on a standard Hilbert space $\mathcal{H}$ as
$$\rho_{\scriptscriptstyle \Lambda}= \sum_{i=1}^r \lambda_i\ket{e_i}\bra{e_i},$$
where the $r$ vectors $\ket{e_i}$ form an orthonormal array in $\mathcal{H}$. 
Let $\mathcal{P}$ denote the real probability space associated  with $\mathcal{H}$ and $\mathcal{W}_{\scriptscriptstyle I}$ the tautological simplex in $\mathcal{P}$, Definition (\ref{tautosimplex}).
Construct the vectors $v_i= |e_i|^2=(v_{i,\omega})\in\mathcal{W}_{\scriptscriptstyle I}$ as $v_{i,\omega} =|e_{i,\omega}|^2$ and $w_{\scriptscriptstyle \Lambda}= \sum_{i=1}^r \lambda_i v_i$. Clearly, $w_{\scriptscriptstyle \Lambda}\in\mathcal{P}$ is a probability distribution.

\paragraph{Regular windows.}
Define a \enquote{regular} window as a window in which the rank of the set of vectors $\{v_i\}$ in $\mathcal{P}$ is also $r$. 
\begin{definition}[Regular window, blind window]
\label{defregular}
A  window of rank $r$ is \enquote{regular} when the $r$ extreme orthonormal vectors $\ket{e_i}$ in the Hilbert space are reverse transcribed as a system $v_i = |e_i|^2$ of same rank $r$ in the probability space.
Otherwise, the window is called \enquote{blind}.
\end{definition}
In particular, a pure window is trivially regular.

\paragraph{Reverse transcription by purification of the window.}
Let $\mathcal{H}_b$ be an auxiliary Hilbert space of dimension $r$.
It is always possible to purify the mixed state into a Hilbert space $\mathcal{H}_c=\mathcal{H}\otimes\mathcal{H}_b$ of dimension $d\times r$, and next to reverse transcribe the pure state into a probability space  $\mathcal{P}_c=\mathcal{P}\otimes\mathcal{P}_b$ as in Sec. (\ref{LPtranspure}). The quantum state $(w_{\scriptscriptstyle \Lambda}, \mathcal{W}_{\scriptscriptstyle \Lambda})$ is then computed by applying Proposition (\ref{partialLPpure}). 

Alternatively, it is possible to reverse transcribe a regular window by extending the method used in pure windows as follows.

\paragraph{Reverse transcription of a regular window.}
Construct the $r$-dimensional subspace $\mathbb{W}_r=\mathrm{Span}_i(v_i)\subseteq\mathcal{P}$ and the tautological simplex $\mathcal{W}_{\scriptscriptstyle I}$ in $\mathcal{P}$. 
Identify $\mathbb{W}_r$ with an effective probability space and define the polytope
$$\mathcal{W}_{\scriptscriptstyle \Lambda} = \mathcal{W}_{\scriptscriptstyle I}\cap\mathbb{W}_r$$
From Proposition (\ref{propspecificpolytope}), $\mathcal{W}_{\scriptscriptstyle \Lambda}$ is a simplex with $r$ equivalent vertices, say  $w_j$.
Since $w_{\scriptscriptstyle \Lambda}$ is a probability distribution and $w_{\scriptscriptstyle \Lambda}\in \mathbb{W}_r$,  then $w_{\scriptscriptstyle \Lambda}\in\mathcal{W}_{\scriptscriptstyle \Lambda}$ so that
$$ w_{\scriptscriptstyle \Lambda} =\sum_{j=1}^r \mu_j w_j,$$
for a specific set of simplicial coefficients $\mu_j$.

Finally, the reverse transcribed simplicial quantum state is  $(w_{\scriptscriptstyle \Lambda}, \mathcal{W}_{\scriptscriptstyle \Lambda})$.
From the demonstration in Sec. (\ref{computew}), this explicit method is consistent with the purification procedure and provides the same result.
\paragraph{}
On the other hand, in \emph{blind windows}, the rank of the set $\{v_{i}\}$ is less than $r$ and may even be reduced to 1.  This occurs specifically when the window carries no information. For instance this happens when the current window is complementary of a principal window (see Sec. \ref{MUB} below) because in that case all information is concentrated in the principal window and then the current window is devoid of any information or rather the only information is the rank $r$ of the state.
As a result, the window is unable to serve as a \enquote{source window}. 
%
However, reverse transcription is still possible by purifying the window as we saw just above.

\paragraph{Recovering the LP system.} The LP system of rank $m=d-r+1$ can be specified by the pair of the linear system of rank $d-r$ describing the $r$-dimensional subspace $\mathbb{W}_r=Span(w_i)$ and an additional constraint of normalization, namely, the LP system of rank $1$ describing the tautological simplex $\mathcal{W}_{\scriptscriptstyle I}$, Eq. (\ref{tautoLP}).

\paragraph{} Finally, we reach the final result,

\begin{theorem}[Quantum state]
\label{concQu}
A quantum state can be represented either by a standard density operator $\rho_{\scriptscriptstyle \Lambda}$ in a Hilbert space $\mathcal{H}$ or by a simplicial quantum state, i.e., a working distribution $w_{\scriptscriptstyle \Lambda}$ within a simplex $\mathcal{W}_{\scriptscriptstyle \Lambda}$ in a real probability space $\mathcal{P}$. For a definite simplicial state  ($w_{\scriptscriptstyle \Lambda}, \mathcal{W}_{\scriptscriptstyle \Lambda}$) in $\mathcal{P}$, the corresponding density operator $\rho_{\scriptscriptstyle \Lambda}$ in $\mathcal{H}$ is defined up to a gauge selection. 
\end{theorem}

\subsubsection[Observable]{Reverse transcription of an observable} 
We are given an observable $\mathsf{Q}$, i.e., an Hermitian operator acting on a Hilbert space.  Recall from Definition (\ref{defobservable}) that an observable is a real-valued function on a sample set $\Omega$. 
\begin{proposition}
\label{domainobservable}
An observable $\mathsf{Q}$ acting on a Hilbert space $\mathcal{H}$ depicts a function $Q : \Omega\to\mathbb{R}$ whose domain is the sample set $\Omega$ of its proper window.
 \end{proposition} 
 \emph{Proof.} The Hermitian operator is constructed in a source window as a diagonal operator, that is in a proper window of the operator itself. $\Box$
 \paragraph{} 
 
The interpretation of an observable requires moving to its proper window, say $\Omega$.
In that window, the Hermitian operator $\mathsf{Q}$ is converted into a covector $\mathrm{q}=(q_{\omega_i} )$ in $\mathcal{P}$ such that $q_{\omega_i} $ is the eigenvalue of $\mathsf{Q}$ belonging to the eigenvector $\ket{i}$ in $\mathcal{H}$.
$$\langle Q\rangle= \tr(\mathsf{Q}\rho_{\scriptscriptstyle \Lambda})=\langle \mathrm{q}w_{\scriptscriptstyle \Lambda}\rangle.$$
By construction, this definition does not depend on the gauge.

\begin{theorem}
\label{reverseobservable}
Any Hermitian operator $\mathsf{Q}$ acting on a Hilbert space $\mathcal{H}$ can be considered as an observable defined in the real-valued probability space $\mathcal{P}$ obtained by reverse transcription into the proper window $\Omega$ of $\mathsf{Q}$. The covector components $\mathrm{q}_\omega$ in the dual space $\mathcal{P}^*$ are the eigenvalues of the Hermitian operator $\mathsf{Q}$.
\end{theorem}

Let $\ket{e_i}$ with $i\in\llbracket 1, d\rrbracket$ denote the proper basis of the observable $\mathsf{Q}$.  From Proposition (\ref{propwindowentropy}) the proper window entropy (characterizing only the proper basis and not the observable as such) is 
$$\mathbb{H}(\Omega) = \sum_{k=1}^d -\bra{e_i} \rho_{\scriptscriptstyle \Lambda}\ket{e_i}\log_2\bra{e_i} \rho_{\scriptscriptstyle \Lambda}\ket{e_i}/$$

\subsection{Principal window}
\label{canoniwindow}

The logical problem was initially expressed using any batch of Boolean variables but a specific window plays a central role.
Indeed, it is possible to \emph{diagonalize} the density matrix $\rho_{\scriptscriptstyle \Lambda}$ in $\mathcal{H}$ by means of a unitary channel. This particular window in the Hilbert space will be called  \enquote{principal window} because it contains on its own all the Shannon information of the Bayesian theater, although in fact the principal basis is not unique when the eigenvalues are not all distinct.
\begin{definition}[Principal window]
\label{defcanoniwindow}
A principal window is a window in which the density operator is diagonal. 
\end{definition}
It is convenient to describe the other windows as twisted, as they produce entangled states.

\begin{definition}[Twisted window]
\label{twistedwindow}
A twisted window is a window in which the density operator is non-diagonal. 
\end{definition}

Let $\ket{\omega_i}$ be the $d$ basis vectors in the Hilbert space $\mathcal{H}$ in a principal observation window.
Let $\ket{e_i}$ denote the eigenvectors normalized to unity and $\lambda_i$ the non-negative eigenvalues of the density operator. 
 Since $\rho_{\scriptscriptstyle \Lambda}$ is diagonal, we have   $\ket{e_i}=\ket{\omega_i}$ up to arbitrary phase factors.

After reordering the basis vectors if necessary, we can assume that the eigenvalues $\lambda_i$ are sorted in descending order. The density operator reads
\begin{equation}
\label{principaldensity}
\rho_{\scriptscriptstyle \Lambda}=Diag(\lambda_1,\dots, \lambda_{r}, 0,\dots,0).
\end{equation}
Then
$$\rho_{\scriptscriptstyle \Lambda}=\sum_{i=1}^r \lambda_i\ \ket{e_i}\bra{e_i}, $$ 
We have $\sum_i\lambda_i=\tr(\rho_{\scriptscriptstyle \Lambda})=1$.
Complement the set $\Sigma_{\scriptscriptstyle \Lambda}=\{\lambda_i\}$ as an ensemble of $d\ge r$ coefficients with $\lambda_i=0$ for $i>r$ so that $\Sigma_{\scriptscriptstyle \Lambda}$ is the spectrum of $\rho_{\scriptscriptstyle \Lambda}$.
%
\begin{proposition}
\label{canonicgauge}
In a principal window, the expression of the density operator $\rho_{\scriptscriptstyle \Lambda}$ is independent of the gauge.
\end{proposition}
\emph{Proof.}
From Proposition (\ref{gaugegleason}) gauge transformations just change the phases of the Gleason's vector $\ket{e_i}$ with respect to the basic vectors $\ket{\omega_i}$. The diagonal matrices are not affected.$\Box$ 
\paragraph{}
The  Hilbert space  $\mathcal{H}$ is the direct sum of the eigensubspaces $\mathsf{h}_k$ of the density operator $\rho_{\scriptscriptstyle \Lambda}$ as  $\mathcal{H}= \bigoplus_k \mathsf{h}_k$.
Let $\mathsf{A}_k$ denote the orthogonal projector on  $\mathsf{h}_k$, $\mathcal{H}\to\mathsf{h}_k\subseteq\mathcal{H}$ and let $n_e$ be the number of distinct values of multiplicity $d_k$, ending with zero. Let $\alpha_k$ be the common eigenvalues $\lambda_i$ in  $\mathsf{h}_k$. For ease of exposition,  set yet $\alpha_{n_e}= 0$ with $d_{n_e}= 0$ if zero is not an eigenvalue.
Then, irrespective of the gauge,
\begin{equation}
\label{decomprhog1}
\rho_{\scriptscriptstyle \Lambda}= \sum_{k=1}^{n_e} \alpha_k\mathsf{A}_k.
\end{equation}
The  observables $\mathsf{A}_k$ are  diagonal with entries $0$ or $1$ and $\tr( \mathsf{A}_k)=d_k$. 
By reverse transcription,  $\mathsf{A}_k$ is the indicator function of some Boolean formula in a principal window.

\subsubsection{Reverse transcription of a principal window}

The reverse transcription of a principal  window is straightforward and leads to a strictly conventional joint probability problem on the principal sample set $\Omega$, with the distribution $\mathbb{P}(\omega_i) = \lambda_i$.
As a result, the principal window can immediately be interpreted in terms of standard probability distribution on the Boolean classical states.
\begin{proposition}[Principal probability distribution]
\label{principaldensitysysteme}
A principal window is always regular.
By reverse transcription into a probability space $\mathcal{P}$, the diagonal density operator $\rho_{\scriptscriptstyle \Lambda}$ acting on the Hilbert space $\mathcal{H}$ leads to a completely divisible simplicial quantum state $(w_{\scriptscriptstyle \Lambda}, \mathcal{W}_{\scriptscriptstyle \Lambda})$, Definition~(\ref{divisiblestate}), describing a strictly classical distribution. The vertices $w_i$ of the  simplex $\mathcal{W}_{\scriptscriptstyle \Lambda}$ are basic vectors in  $\mathcal{P}$, i.e. deterministic states,  $w_i=\tilde{\omega}_i$, $\forall i\in\llbracket 1, r\rrbracket$ and the probability distribution is $\mathbb{P}(\omega_i) = \lambda_i$, $\forall i\in\llbracket 1, d\rrbracket$.
\end{proposition}
\emph{Proof.}
The proof consists in checking that it is possible to construct from scratch a relevant source window in a real-valued $d$-dimensional probability space $\mathcal{P}$ with basis $\{\tilde{\omega}_i\}$. Set $w_i=\tilde{\omega}_i\in\mathcal{P}$ for  $i\in\llbracket 1, r\rrbracket$, so that the rank of the set $\{w_i\}$ is $r$.
Define $\mathcal{W}_{\scriptscriptstyle \Lambda}=\mathrm{conv}(w_i)$ and $w_{{\scriptscriptstyle \Lambda},\omega_i} = \lambda_i$, so that the working distribution is
$$w_{\scriptscriptstyle \Lambda}=(w_{{\scriptscriptstyle \Lambda},\omega_i})=\sum_{i=1}^r \lambda_i\ \tilde{\omega}_i$$
By inspection, from Eq. (\ref{rhoa}), the direct transcription of the quantum state $(w_{{\scriptscriptstyle \Lambda}},\mathcal{W}_{\scriptscriptstyle \Lambda})$ is indeed the diagonal operator $\rho_{\scriptscriptstyle \Lambda}$. In addition, the rank $r$ of the density operator $\rho_{\scriptscriptstyle \Lambda}$ is equal to the number of vertices of the simplex $\mathcal{W}_{\scriptscriptstyle \Lambda}$, which proves that the system is regular.
At last, since the vertices are deterministic the simplex is separable (Definition \ref{separablesimplex}) with respect to any Kronecker factorization of the Hilbert space , i.e., any split of the principal register and therefore completely divisible (Definition \ref{divisiblestate}).
$\Box$

\paragraph{} In standard quantum information, the  property for a state of being separable or entangled is regarded as intrinsic. 
This is because, implicitly, there is only a unique batch of variables, which is therefore considered intrinsic.
By contrast, in the present model, each window corresponds to a specific variable batch and Proposition (\ref{principaldensitysysteme}) shows that every state is always separable in its principal window. Therefore, in other window, \emph{entanglement reflects the departure of the current window from the principal window} and is by no means specific of the state. 

\begin{proposition}[Independent binary variables]
\label{principalwindow}
A principal window specifies a batch of mutually independent Boolean variables.
\end{proposition}
\emph{Proof.}
In a principal window, all basic vectors of the probability space are deterministic solutions of the LP problem. Therefore, from Proposition (\ref{mutualindependence}) and Sec. \ref{separability}, the binary variables are mutually independent.
$\Box$
 
We have previously defined completely  divisible states, Definition (\ref{divisiblestate}). It turns out that this property is not intrinsic but depends on the window. For clarity define thus the notion completely divisible window.

\begin{definition}[Completely divisible window]
\label{divisiblewindow}
A  completely divisible window is a window in which the density operator is completely divisible.
\end{definition}  

\begin{proposition}
A completely divisible window defines a batch of mutually independent Boolean variables.
\end{proposition}
\emph{Proof.}
Diagonalization of the $N$ individual $2\times2$ elementary density operators leads to a principal window, which in turn specifies a batch of independent Boolean variables from Proposition (\ref{principalwindow}). This batch is uniquely defined by the window.
 $\Box$

\begin{theorem}
\label{paldivisible}
All Bayesian theaters are completely divisible.
\end{theorem}
\emph{Proof.} The density operator is always diagonalizable in a principal window. As a result, Theorem (\ref{paldivisible}) follows from Proposition (\ref{principalwindow}) and Definition (\ref{divisiblestate}). $\Box$
\paragraph{}
\emph{Remark.} This fundamental theorem is at odds of the common belief. It states that all paradoxes of quantum information only result from ill-tuned batches of binary variables. $\Box$

\begin{proposition}[Mixed distribution]
\label{mixedistribution}
The mixed distribution of a standard \enquote{mixed quantum state} is the working distribution $w_{\scriptscriptstyle \Lambda}$ in a principal window.
\end{proposition}
\emph{Proof.}
In a principal window, the set $\{\lambda_i\}$ represents all at once the set of simplicial coefficients, the components of the working distribution and the spectrum of the density operator. 
It corresponds also to the mixed distribution of the standard mixed quantum states.
$\Box$
\paragraph{The principal Bayesian LP problem.}
Now, we aim to recover the Bayesian system. Again, it is straightforward to construct the relevant LP problem in $\mathcal{P}$.
\begin{proposition}
\label{principalP}
When $r<d$, the principal LP problem can be formulated as
\begin{align}
\label{expecsimplic}
(\Lambda):\ \mathrm{Given~} d-r \mathrm{~classical~states~} \omega_{i'} \mathrm{~assign~} \Prob \mathrm{~subject~to~}\langle \tilde{\omega}_{i'}^* \rangle=0. 
\end{align}
 When $r=d$, the prior $(\Lambda)$ is simply the statement that $d=2^N$.  
\end{proposition}
\emph{Proof.}
The $r$ basis vectors $\tilde{\omega}_{i }$ span the effective probability space $\mathbb{W}_r\subseteq\mathcal{P}$ and the specific simplex $\mathcal{W}_{\scriptscriptstyle \Lambda}$ is the tautological simplex $\mathcal{W}_{r}$ in  $\mathbb{W}_r$.
Complement the $r$ basis vectors $\tilde{\omega}_{i }$ by  $d-r$ other basis vectors $\tilde{\omega}_{i'}$ in $\mathcal{P}$.
In Eq. (\ref{expecsimplic}),  $\tilde{\omega}_{i'}^*$ denote the $d-r$ indicator functions corresponding to the classical states $\omega_{i'}$. 
$\Box$
\paragraph{}
Alternatively, the sum $A_{n_e}\ident \sum_{i'=r+1}^d \tilde{\omega}_{i'}^* $ is the indicator function of a Boolean formula. With a relevant order of the indexes, its corresponding covector is $(0,\dots,0, 1,1,\dots,1)$.
Since a sum of positive terms is zero if and only if each individual term is zero, a more compact formulation is
\begin{align}
\label{intrinobser1}
(\Lambda):\ \mathrm{Given~} \mathrm{the~indicator~function~} A_{n_e} \mathrm{~assign~} \Prob \mathrm{~subject~to~}\langle A_{n_e} \rangle=0. 
\end{align}
The $r$ vertices of the simplex in the probability space $\mathcal{P}$ are the basic vectors $\tilde{\omega}_{i}$ for  $i\in\llbracket 1, r\rrbracket$.

\paragraph{}

Surprisingly, it follows from proposition (\ref{principalP}) that the core of any Bayesian system is simply limited to its order $r$. Consequently, the main actual input is the mixed contextual distribution.

\begin{theorem}
\label{fundamental0}
Any Bayesian theater can be specified in a principal window by the pair of a specific Boolean formula 
 $(\omega_1,\dots,\omega_r)=1$
and a mixed distribution $\Sigma_\lambda=\{\lambda_1,\dots,\lambda_r,0,\dots,0\}$.
\end{theorem}
\emph{Proof.}
For simplicity, let $A_{n_e}$ denote also the Boolean \emph{formula} of  indicator function $A_{n_e}$. 
Then, the logical assertion $A_{n_e}$ is compelled to be false, or equivalently, its negation $\overline{A}_{n_e}$ is compelled to be true. Clearly, we have $\overline{A}_{n_e}= \bigvee_{i=1}^r \tilde{\omega}_{i}^*$. When the rank $r$ is equal to the dimension $d$, $A_{n_e}=\varnothing$ and $\overline{A}_{n_e}$ is the tautology.
This encompasses the most general logical problem subject to constraints. 
To get a complete description, we need to assign an exogenous contextual distribution $\Sigma_\lambda=\{\lambda_i\}$.
$\Box$
\paragraph{}
The indicator function ${A}_{n_e}\in\mathcal{P}^*$ depicts the $d_{n_e}=d-r$ vertices $\tilde{\omega}'_{i}$ of zero probability, $\lambda_{n_e}=0$.
Taking into account the other contextual multiplicities, let  ${A}_k\in\mathcal{P}^*$ denote the indicator function of the union of all $d_{k}$ vertices $\tilde{\omega}_{i}$ corresponding to the same probability $\alpha_k$.
Since the eigenvalues are sorted in descending order,  $A_k$ is the indicator function of a set of basic vectors with contiguous indexes, say $k_1$ to $k_2$, with $d_k$ non zero entries, for instance, $A_k$ may be the covector $(0,0,0,1,1,1,1, 0,0,0)\in\mathcal{P}^*$. 

Now, for all $k\in\llbracket1,n_e\rrbracket$, the dual form $ \langle A_k\,p\rangle$ with 
$p\in\mathcal{P}$ is $\langle A_k\,p\rangle=\sum_{i=k_1}^{k_2} p_i$ while the expectation $\langle A_k\,w_{\scriptscriptstyle \Lambda}\rangle$ is $\langle A_k \rangle =d_k\alpha_k$.
Clearly, the system is invariant under arbitrary permutation of the $d_k$ indexes of same mixed probability $\alpha_k$. This defines a \emph{contextual symmetry}.

\begin{definition}[Contextual symmetry]
\label{defintrinsicsym}
A contextual symmetry is a transformation of the sample set $\Omega$ in a principal window, leaving invariant the mixed probability distribution.
\end{definition}

\begin{proposition}
\label{intrinsicsym}
The contextual symmetry group is the direct product $\mathrm{S}_{d_{1}}\times\mathrm{S}_{d_{2}}\times\dots\mathrm{S}_{d_{n_e}}$ of the permutation symmetric groups of degree $d_k$. 
\end{proposition}
\emph{Proof.} Any product of vertex permutations of same mixed probability $\alpha_k$ is a contextual symmetry by definition. $\Box$
\paragraph{}
Note that from Proposition (\ref{principalP}), strictly speaking, the symmetric group $\mathrm{S}_{d_{n_e}}$ does not depend on the context but on the core LP problem.

\subsubsection{Fundamental theorem}
\label{fundamental}
A principal window  depicts a very conventional probability problem, composed of $d$ \emph{deterministic outcomes} mutually exclusive, namely $\omega_i\in\Omega$ with $i\in\llbracket 1,d\rrbracket$, and a standard probability distribution, $\Sigma_{\scriptscriptstyle \Lambda}=\{\lambda_i\}$, on the sample set $\Omega$. Only $r\le d$ probability masses $\lambda_i$ are non-zero.

\begin{theorem}[Fundamental theorem]
\label{fondamental}
Any density operator $\rho_{\scriptscriptstyle \Lambda}$ of  spectrum $\Sigma_{\scriptscriptstyle \Lambda}=\{\lambda_i\}$ in a Hilbert space $\mathcal{H}$ is the image by a unitary channel of a strictly conventional probability problem consisting in drawing one object among $d$ deterministic classical states ${\omega}_i\in\Omega$ with respect to the contextual probability distribution $\Sigma_{\scriptscriptstyle \Lambda}$.
\end{theorem}
\emph{Proof.} This is a trivial corollary of Theorem (\ref{fundamental0}). $\Box$
\paragraph{}
In fact, much of this result is known since von Neumann~\cite{vonneumann1}. The only novelty lies in the interpretation at odds of the common belief: Now, from Theorem (\ref{fondamental}), \emph{a Bayesian theater represents a quite classical logical system}.  In other words, every quantum system can always be expressed as a classical random system provided it is expressed with a relevant batch of variables. Entanglement is a property of the variable batch and not of the problem itself.

\begin{proposition}
\label{puredeterministic}
A pure quantum state depicts a deterministic distribution expressed in the principal sample set.
\end{proposition}
\emph{Proof.}
By definition, a pure state is of rank 1 and thus deterministic in a principal window.
$\Box$
\paragraph{}
This can be expressed in striking form: With a relevant discrete Boolean gauge, a pure state represents just a reset register.
\begin{proposition}
\label{reset}
 It is always possible to choose a discrete Boolean gauge so that the deterministic distribution of a pure state coincides with the empty atom $\varpi_0$, that is a reset register composed of $N$ zeros, $(0,0,\dots,0)$.
\end{proposition}
\emph{Proof.}
This is a straightforward consequence of the discrete Boolean gauge definition, Definition~(\ref{sourcegauge}).
$\Box$
\paragraph{}
In other words, a pure quantum state, deterministic in a principal window, is simply genuinely deterministic. In another window, it still represents a deterministic state but evaluated from a maladjusted viewpoint. The probabilities thus involved are only Bayesian estimations, that is primarily technical coefficients indicating that the window is ill-matched. Again, this interpretation is at odds of the common belief.

\begin{theorem}[Information stored in the Bayesian theater]
\label{fondamentalVonNeumann}
A Bayesian theater described a classical memory with a storage capacity of $N$ bits. The current information stored in the system is 
equal to  $N-S(\rho_{\scriptscriptstyle \Lambda})$.
\end{theorem}
\emph{Proof.}
From Theorem (\ref{fondamental}), a principal window is explicitly classical and depicts a conventional memory space with a storage capacity of $N$ bits.  Accordingly, the information $N-S(\rho_{\scriptscriptstyle \Lambda})$ effectively stored in the memory is  characterized by a genuine Shannon entropy which is simply the von Neumann entropy $S(\rho_{\scriptscriptstyle \Lambda})$ of the density operator. 
$\Box$
\paragraph{} 
In standard quantum information theory, the amount of information stored in a system is not that clear and even challenging, since quantum information is generally believed to be essentially different from strict Shannon information (see e.g., Ref.~\cite{caves2}). In the present model, there is no difference at all. For instance a pure state, with $S(\rho_{\scriptscriptstyle \Lambda})=0$, carries an information of exactly $N$ bits, meaning that a wave vector in an infinite dimensional Hilbert space would convey an infinite amount of information.

\subsubsection{Information expressions}
 In a principal window, three probability distributions are identical: (1) the working distribution $w_{\scriptscriptstyle \Lambda}$ in the sample set $\Omega$, (2) the simplicial distribution $\mu_i$ of the contextual distribution in $\Sigma_\mu$ and  (3) the distribution $\lambda_i$ in the spectrum $\Sigma_{\scriptscriptstyle \Lambda}$ of the density operator $\rho_{\scriptscriptstyle \Lambda}$.
\paragraph{Entropy.}
 Let us recall the definition of the entropy of these different distributions in general.

\begin{definition}[Forms of entropy]~\\
- The entropy of the working distribution $w_{\scriptscriptstyle \Lambda}$ in a particular window is the \emph{window entropy} $S_w=\mathbb{H}(\Omega)=\mathbb{H}(w)$.\\ 
- The entropy of the contextual distribution (or simplicial distribution) in a particular window is the \emph{simplicial entropy} $S_\mu=\mathbb{H}(\Sigma_\mu)$.
We will use interchangeably the terms \enquote{simplicial entropy} and \enquote{contextual entropy}.\\
- The entropy of the Bayesian theater is the \emph{von Neumann entropy} $S_{\scriptscriptstyle \Lambda}=S(\rho_{\scriptscriptstyle \Lambda})=\mathbb{H}(\Sigma_{\scriptscriptstyle \Lambda})$.  We will use interchangeably the terms \enquote{von Neumann entropy} and \enquote{mixed entropy}.
\end{definition}
The von Neumann entropy $S(\rho_{\scriptscriptstyle \Lambda})$ is invariant under a  unitary channel and can be regarded as the global \enquote{theater entropy} while the window entropy $S_w$ and the simplicial entropy $S_\mu$ are window-dependent by definition. 
\begin{proposition}
\label{commonentropy}
In a principal window, we have
$$S_{\scriptscriptstyle \Lambda}= S_\mu =S_w.$$
\end{proposition}
\emph{Proof.} In a principal window, the three distributions are identical and therefore the entropies are identical as well.
$\Box$
\begin{proposition}
\label{satsimple}
The von Neumann entropy is the lower bound of the simplicial entropy over all possible windows.
$$S_{\scriptscriptstyle \Lambda}= \min_{\mathrm{windows}} (S_\mu).$$
\end{proposition}
\emph{Proof.}
From Jaynes' inequality, Proposition (\ref{bound}), $S_{\scriptscriptstyle \Lambda}\le S_\mu$. From Proposition (\ref{commonentropy}), the inequality is saturated in a principal window. 
$\Box$.
\paragraph{}
The upper bound of the simplicial entropy is trivially $\log r$ when the working distribution coincides with the center of mass of the specific simplex.

At last, it is convenient to define also the overall information, or  von Neumann negentropy, as  $\mathbb{I}(\rho_{\scriptscriptstyle \Lambda})=N-S(\rho_{\scriptscriptstyle \Lambda})$.
\begin{definition}[von Neumann information] 
\label{defVonNeumannegentropy}
The von Neumann information, or von Neumann negentropy of a density operator $\rho_{\scriptscriptstyle \Lambda}$ acting on a $d$-dimensional Hilbert space is $\mathbb{I}(\rho_{\scriptscriptstyle \Lambda}]=N-S(\rho_{\scriptscriptstyle \Lambda})$, where $d=2^N$ and $S(\rho_{\scriptscriptstyle \Lambda})= -\tr (\rho_{\scriptscriptstyle \Lambda}\log_2\rho_{\scriptscriptstyle \Lambda})$.
\end{definition} 

\paragraph{Other expressions.}
\label{otherentropy}

Now, any probability expression in conventional information theory, whether function or inequality, is \emph{ipso facto} valid in the very conventional principal distribution $(\Omega,\Sigma_{\scriptscriptstyle \Lambda})$. Therefore \emph{in a principal window} the same expression is valid by formally replacing the eigenvalues $\lambda_i$ by the operator $\rho_{\scriptscriptstyle \Lambda}$ in the Hilbert space, on the model of $S(\rho_{\scriptscriptstyle \Lambda})=\mathbb{H}(\Omega)$ with implicitly $\Prob(\omega_i)=\lambda_i\in\Sigma_{\scriptscriptstyle \Lambda}$ and $\omega_i\in\Omega$. 
\begin{proposition}
\label{entropequation}
Any valid probability expression in the principal sample set $\Omega$ with the probability distribution $\Prob(\omega_i)= \lambda_i$ is also valid in any window by replacing $\lambda_i$ by $\rho_{\scriptscriptstyle \Lambda}$ and then formally $\mathbb{H}$ by $S$ and $\Omega$ by $\rho$.
\end{proposition}
 In particular, since the principal distribution is actually a joint distribution, this applies to any entropy measure in a pair of register, e.g. for conditional or partial entropy.
We will give examples in Sec. (\ref{pairH}).

\subsection{Gauge transcription group}
\label{gtg}

We constructed a Hilbert space $\mathcal{H}$ from a simplicial quantum state $(w, \mathcal{W})$ transcribed into a density operator $\rho$.
We found that the transcription implies necessarily a gauge choice.
Conversely, the consistency of the model demands that the simplicial states $(w, \mathcal{W})$ reverse-transcribed from the density operator $\rho$ be independent of the gauge, which in turn entails a particular gauge structure.
In this section, we will investigate this gauge structure.

 
The direct approach is to link each particular Boolean batch to a specific basis, regardless of the gauge.
As a result, both the initial gauge and the gauge changes are generated from a single source window.
This leads to construct a particular Hilbert space for each gauge and therefore the gauge is termed \emph{global}. This is addressed from Sec. (\ref{globalgauge}). 

By contrast,  one can demand that the gauge could be changed locally, i.e., independently in each particular window, within a unique Hilbert space. 
This requires to transcribe every Boolean batch into a different basis for each gauge. 
This second option is expressed in physics of particles by the so-called \enquote{gauge principle}. This is beyond the scope of the present paper.

In any case, the gauge transformations form a group that we will naturally call the gauge group, say, $\mathfrak{G}$. Since the probability distribution is conserved, gauge operators are either \emph{unitary} or \emph{antiunitary}.

At last, infinitesimal gauge transformations open up a different approach, namely the use of differential analysis. This is beyond the scope of the present paper.

\paragraph{}
\emph{Notation.} We use three closely related but distinct concepts, \enquote{basis}, \enquote{window} and \enquote{frame}. A basis is the standard basis of Hilbert spaces. A frame is a particular set of ordered basis vectors. An observation window is associated with a particular batch of Boolean variables.  If the gauge is global, the source window determines a unique basis  in a particular Hilbert spaces for each gauge. By contrast, if the gauge is local, the same window is represented by a particular basis for each gauge in a unique Hilbert space. 

Let us first address the direct transcription, that is to say, global gauges.

\subsubsection{Global gauges}
\label{globalgauge}
 The initial transcription of a simplicial quantum state is performed in the source window by  fixing a particular gauge, say $g$.  However,  the particular source window itself is widely indifferent because it is straightforward to perform the transcription from any other regular window.
\begin{proposition}
\label{indifference}
For any gauge $g$, it is possible to construct a unique Hilbert space $\mathcal{H}_{g}$ irrespective of the regular source window used for the transcription. 
\end{proposition}
\emph{Proof.}
Transcribe the simplicial quantum from a source window. This defines a gauge $g$ and  determines both a particular  density operator and, by reverse transcription, a particular simplicial quantum state in every particular window. 
Now just decide by convention that this particular density operator in any regular window is precisely the result of the direct transcription with the same gauge $g$ of the corresponding particular simplicial quantum state when this regular window is regarded as the source window.
$\Box$
\paragraph{}

 This convention can be regarded as a definition of a global gauge over the Bayesian theater.
  Irrespective of the source window, we will refer to this unique Hilbert space as  $\mathcal{H}_{g}$ and denote $\rho_g$ the density operator.

 \begin{definition}[Global gauge]
 \label{defglobalgauge}
 A global gauge representation $g$ is the specific transcription of the logical system into a specific Hilbert space $\mathcal{H}_g$.
 \end{definition}

\subsubsection{Changing the global gauge}
\label{globalchanging}
Consider a second gauge, $g'$ and therefore a new Hilbert space $\mathcal{H}_{g'}$. Let $\rho_{g}$ and $\rho_{g'}$ denote the density operators acting on $\mathcal{H}_{g}$ and $\mathcal{H}_{g'}$ respectively.  First, make sure that as far as $g'\ne g$,  $\mathcal{H}_{g}$ and $\mathcal{H}_{g'}$ must indeed be distinct.
\begin{proposition}
\label{distinct}
When the gauge is global, distinct gauges require distinct Hilbert spaces.
\end{proposition}
\emph{Proof.}
From Proposition (\ref{canonicgauge}), irrespective of the gauge, the density operators are identical in a principal window. If the Hilbert spaces were the same for every gauge, the density operators would be also identical in every windows and the gauges would not be distinct. 
$\Box$
 \begin{proposition}
 \label{gaugeigen}
 Any change from a gauge $g$ to a gauge $g'$ maps the eigensubspaces of $\rho_g$ onto  the eigensubspaces of $\rho_{g'}$. 
 \end{proposition}
 
\emph{Proof.}
Since the expressions of the density operators are identical in both principal windows, the eigensubspaces are transformed into eigensubspaces.
$\Box$
\paragraph{}

Now, from Wigner's theorem the gauge operators $\Theta: \mathcal{H}_{g}\to\mathcal{H}_{g'}$ expressed in the source window $\Omega$ are either unitary or antiunitary. By definition, since the gauges are global, for every Boolean variable batch, that is for every window, the bases in the two Hilbert spaces are henceforth identical, irrespective of the gauge. As a result,  when changing the source window itself, the operator $\Theta$ changes accordingly.

\begin{proposition}
Using another source window $\Omega'$ obtained from the initial source window $\Omega$ by a unitary transition matrix $\mathsf{U}\in\mathrm{U}(d)$, the gauge operator $\Theta\in\mathfrak{G}$, whether unitary or antiunitary  is expressed as
\begin{equation}
\label{class}
\Theta' = \mathsf{U}\Theta\mathsf{U}^{-1}
\end{equation}
\end{proposition}
\emph{Proof.} 
Since the gauge is global, the two bases $\Omega$ and $\Omega'$ are by hypothesis identical in the two distinct Hilbert spaces $ \mathcal{H}_{g}$ and $ \mathcal{H}_{g'}$. As a result, the transition unitary matrices $\mathsf{U}: \ket{\psi}\to\ket{\psi'}$ are also  identical, where $\ket{\psi}$ and $\ket{\psi'}$ denote the expression of a current vector in the bases $\Omega$ and $\Omega'$ respectively.
\begin{center}
\begin{minipage}{0.4\linewidth}
\begin{tikzpicture}
  \matrix (m) [matrix of math nodes,row sep=2em,column sep=4em,minimum width=1em] 
  {
  \mathcal{H}_{g}:              &  \ket{\psi_g}   &  \ket{\psi'_g}   \\ 
    \mathcal{H}_{g'}:        &   \ket{\psi_{g'}} &  \ket{\psi'_{g'}}   \\ 
    } ;
  \path[-stealth] 
    (m-1-2)  edge [->, thick]   node [left] {$\Theta$} (m-2-2)
                 edge [double,->] node [below] {$\mathsf{U}$} (m-1-3)
     (m-2-2) edge [double,->] node [above] {$\mathsf{U}$} (m-2-3)
     (m-1-3) edge [->, thick]   node [right] {$\Theta'$} (m-2-3)
    ;
\end{tikzpicture}
\end{minipage}
\qquad\quad~
\begin{minipage}{0.5\linewidth}
From simple inspection of the commutative diagram  we have
$\ket{\psi'_{g}} =\mathsf{U}\ket{\psi_g}$ and $\ket{\psi'_{g'}} =\mathsf{U}\ket{\psi_{g'}}$
so that, irrespective of $\ket{\psi_g}$, $\ket{\psi'_{g'}} =\Theta'\mathsf{U}\ket{\psi_g}= \mathsf{U}\Theta\ket{\psi_g} $ 
and thus $\Theta' = \mathsf{U}\Theta\mathsf{U}^{-1}$. $\Box$
\end{minipage}
\end{center}

  Since from Proposition (\ref{indifference}) the source window is indifferent, it is convenient to select henceforth the source window as a \emph{principal window} corresponding to a  batch of mutually independent Boolean variables. 
  
  The gauge operators  $\Theta$ can be unitary or antiunitary.
 Let us start by investigating the unitary gauge group.

\subsubsection{The unitary gauge group $\mathcal{G}$}
Obviously, the unitary transformations of a global gauge into another global gauge form a unitary group.
By construction, the groups operators are expressed on a common source window, that is a common basis of the Hilbert spaces $\mathcal{H}_g$.

\begin{definition}[Unitary gauge group $\mathcal{G}$]
\label{defgp}
The unitary gauge group  $\mathcal{G}$ is the unitary transformation group of the global gauges. 
\end{definition}

The unitary gauge group can be precisely characterized by its action on the eigensubspaces of the density operator.
\begin{proposition}
\label{setcanon}
The unitary group $\mathcal{G}$ is the group of unitary operators leaving invariant the eigensubspaces of the density operator expressed  in any particular gauge.
\end{proposition}
\emph{Proof.}
From  Proposition (\ref{canonicgauge}), the eigensubspaces of the density operator are invariant under every gauge transformation and conversely, any unitary transformation leaving invariant these eigensubspaces leaves invariant the density operator in any principal window and thus defines a gauge change. $\Box$
\paragraph{Constructing the unitary gauge group $\mathcal{G}$.}
We will hereafter regard the unitary gauge group $\mathcal{G}$ as realized by unitary matrices acting on the $d$-dimensional Hilbert space $\mathcal{H}_{g_0}$ for an arbitrary but fixed gauge $g_0$ and expressed in a common principal basis, so that the group is isomorphic to a subgroup of the standard unitary matrix group $\mathrm{U}(d)$.

In the principal window, after reordering the basis vectors if necessary, suppose that the eigenvalues $\lambda_i$ of the density operator $\rho_{g_0}$ are sorted in descending order.  Let $\ket{\omega_i}\in\mathcal{H}_{g_0}$  for  $i\in\llbracket 1, d\rrbracket$ denote the basis vectors. 
The  Hilbert space  $\mathcal{H}_{g_0}$ is the direct sum of the eigensubspaces $\mathsf{h}_k$ of the density operator $\rho_{g_0}$ as  $\mathcal{H}_{g_0}= \bigoplus_k \mathsf{h}_k$. 
Let $\mathsf{A}_k$ denote the orthogonal projectors $\mathcal{H}_{g_0}\to\mathsf{h}_k\subseteq\mathcal{H}_{g_0}$ and let $n_e$ be the number of distinct eigenvalues $\alpha_k$ of multiplicity $d_k$, including possibly zero.
Then, from Eq. (\ref{decomprhog1}),
\begin{equation*}
\rho_{g_0}= \sum_{k=1}^{n_e} \alpha_k\mathsf{A}_k
\end{equation*}
\begin{proposition}
\label{directproduct}
The unitary gauge group  $\mathcal{G}$ is a Lie group of dimension $\sum_k d_k^2$ isomorphic to the direct product $\mathrm{U}(d_1)\times \mathrm{U}(d_2)\times \mathrm{U}(d_3)\dots\times \mathrm{U}(d_{n_e})$, where $\mathrm{U}(d_k)$ are respectively the unitary groups acting on the $d_k$-dimensional eigensubspaces  $\mathsf{h}_k$ of the density operator. 
\end{proposition}
\emph{Proof.}
By construction, the Hilbert space $\mathcal{H}_{g_0}$ is a linear representation of dimension $d$ of  the gauge group $\mathcal{G}$.
On each subspace $\mathsf{h}_k$ of dimension $d_k$, ($k\in\llbracket 1,n_e\rrbracket$),  $\mathcal{G}$ acts as the full unitary group $\mathrm{U}(d_k)$ so that any subspace $\mathsf{h}_k$ is a linear representation of dimension $d_k$.
Finally $\mathcal{H}_g$ is a completely decomposable representation of $\mathcal{G}$. As a result,  each subgroup $\mathrm{U}(d_k)$ is normal in $\mathcal{G}$ and $\mathcal{G}$ is the direct product $\mathrm{U}(d_1)\times \mathrm{U}(d_2)\times \mathrm{U}(d_3)\dots\times \mathrm{U}(d_{n_e})$.
The dimension of a unitary Lie group $\mathrm{U}(d_k)$ is $d_k^2$, so that the dimension of the $n_e$-tuple is $\sum_{k=1}^{n_e} d_k^2$.
$\Box$
\paragraph{}\noindent
Conversely,  the set of eigensubspaces $\{\mathsf{h}_k\}$  determines the density operator,  up to a possible rescaling of the mixed distribution $\{\alpha_k\}$ leaving the multiplicities unchanged, allowing just a modification of the source contextuality.
By contrast, a  complete rescaling of the mixed distribution $\{\lambda_i\}$ can e.g. increase the number of eigensubspaces, which would express a break of symmetry.

\begin{proposition}
There is a one-to-one correspondence between the unitary gauge subgroups $\mathrm{U}(d_k)$ and the intrinsic symmetry subgroups $\mathrm{S}_k$, Definition (\ref{defintrinsicsym}). Moreover, the intrinsic symmetry group is a discrete subgroup of the Lie gauge group.
\end{proposition}
\emph{Proof.}
The unitary gauge group and the intrinsic symmetry group are both determined by the same set $(d_k)$ of the $n_e$ multiplicities.
Moreover, from Proposition (\ref{intrinsicsym}), the gauge group contains any permutation of the basis vectors in a principal window, leaving invariant the eigensubspaces, that is the intrinsic symmetry group.
$\Box$
\paragraph{}
 
Especially, the Lie gauge group of \emph{any pure state} is always $\mathcal{G}=\mathrm{U}(1)\times\mathrm{U}(d-1)$, but the converse is false in general because the eigenspaces are not necessarily affected by a rescaling of the mixed distribution. 
 It is useful to define an \enquote{effective} subgroup of the gauge group by ignoring $\mathrm{U}(d_{n_e})$ when the eigenvalue  $\lambda_{n_e}$ is zero because this last subgroup most often has no effect. 
 \begin{definition}[Effective unitary gauge group $\mathcal{G}_\mathrm{eff}$]
 \label{eff}
The effective gauge group $\mathcal{G}_\mathrm{eff}$ is the direct product of the groups $\mathrm{U}(d_k)$ associated to all non-zero eigenvalues $\lambda_{k}$ of the density operator.
 \end{definition}
\noindent
Now,  the effective gauge group is $\mathcal{G}_\mathrm{eff}=\mathrm{U}(1)$ \emph{if and only if}  the state is pure.
Obviously, the effective gauge group determines the gauge group proper as $\mathcal{G}=\mathcal{G}_\mathrm{eff}\times\mathrm{U}(d_{n_e})$ where $d_{n_e} = d-\sum d_{k}$ (with $\alpha_k>0$).
\paragraph{}
Reversing the logic, the unitary gauge group $\mathcal{G}$ determines to some extent the density operator.
In fact, the group of gauges does not specify the contextual distribution and is equivalent to simply giving the specific simplex, that is the LP system.
\begin{theorem}[Correspondence between the unitary gauge group and the quantum state]
\label{reversePb}
The unitary gauge group $\mathcal{G}$ determines the quantum state up to  a rescaling of the mixed distribution $\{\alpha_k\}$. 
Conversely, the quantum state is specified by the set $\{d_k, \alpha_k\}$ with  $\sum d_k=d$  and $\sum d_k\alpha_k=1$ for $k\in\llbracket1,n_e\rrbracket$.
\end{theorem} 
\emph{Proof.}
The only feasible gauge groups are direct products of subgroups $\mathrm{U}(d_k)$. Therefore the set $\{d_k\}$ is completely determined by   $\mathcal{G}$. The eigenvalues $\{\alpha_k\}$ of the density operator can be arbitrary chosen provided they be positive, distinct and sum to $1$ when accounting for the multiplicity. Therefore the quantum state is determined by the set $\{d_k, \alpha_k\}, k\in\llbracket1,n_e\rrbracket$. 
$\Box$

In particular, for a pure state, the gauge group is  $\mathcal{G}=\mathrm{U}(1)\times\mathrm{U}(d-1)$ with $d_1=1$, $d_2=d-1$, $\alpha_1=1$ and $\alpha_2=0$.

\subsubsection{Invariant observables and Noether constants}

By definition, the eigenprojectors $ \mathsf{A}_k$ are invariant under the gauge group action. Consequently, they play a role similar to that of the Hamiltonian in standard physics and the eigenvalues are therefore just Noether constants of the gauge group.
\begin{proposition}
The eigenprojectors $ \mathsf{A}_k$ are invariant under the gauge group and commute with any group operator. They form a commutative POVM of mutually orthogonal observables.
By reverse-transcription into any principal window, they are depicted by $n_e$ indicator functions $A_k$ corresponding to the  union of the $d_k$ classical states of same mixed probability $\alpha_k$ so that $\langle{A}_k\rangle=\langle\mathsf{A}_k\rangle=d_k\alpha_k$. 
\end{proposition}
\emph{Proof.} By construction, the group operators leaves invariant the subspaces $\mathsf{h}_k$. The projectors $\mathsf{A}_k$ on $\mathsf{h}_k$ commute with any group operator and therefore are invariant under the gauge group. They commute and have a common proper window, namely, any principal window.
They sum to the identity, $\sum_{k=1}^{n_e} \mathsf{A}_k=\mathds{1}_d$. Therefore, they form a commutative POVM of orthogonal observables.
In a principal window, they are reverse-transcribed as  indicator functions $A_k$. Finally $\langle{A}_k\rangle=\langle\mathsf{A}_k\rangle=d_k\alpha_k$. 
$\Box$

\begin{definition}[Invariant observables and Noether constants]
\label{currents}
The eigenprojectors  $\mathsf{A}_k$ constitute a set of invariant observables.
The Noether constants  $\langle \mathsf{A}_k\rangle=d_k\alpha_k$ are the expectation values of these observables.
\end{definition}
Now it is possible to reformulate the correspondence between the gauge group and the quantum state, Theorem (\ref{reversePb}), in terms of these entities. 
%
\begin{proposition}
\label{noether} 
The Bayesian theater is completely determined by the $n_e$ invariant observables $\mathsf{A}_k$ and the corresponding Noether constants, namely, the   $n_e$ expectations $\langle \mathsf{A}_k\rangle=d_k\alpha_k$.
\end{proposition}
The unitary gauge group $\mathcal{G}$ does not exhaust all gauge transformations because the antiunitary operators have been omitted. Let us now investigate these antiunitary gauge changes, obtained by complex conjugation $\mathcal{H}_{g}\to\mathcal{H}_{g^*}$

\subsubsection{The conjugation gauge group $\mathscr{C}$}  

Let $\mathsf{K} : z\mapsto z^*$ denote the standard complex conjugation in $\mathbb{C}$. 
Consider the global conjugation gauge $\mathcal{H}_{g}\to\mathcal{H}_{g^*}$, obtained by changing each vector $\ket{\psi_g}$ into its complex conjugate $\ket{\psi_{g^*}}=\ket{\psi_{g}}^*$ in the source window. 
Let $\mathds{1}_d\times\mathsf{K}$, or simply $\mathsf{K}$ when no confusion can occur, denote the diagonal matrix $\mathrm{Diag}(\mathsf{K}, \mathsf{K},\dots,\mathsf{K})$.
Now from a theorem by E. Wigner \cite{wigner3}, any antiunitary operator is of the form $\mathsf{U}\mathsf{K}$ where $\mathsf{U}$ is unitary.

\begin{proposition}
 \label{gaugeantiunitary}
 In a principal window, any antiunitary gauge operator $\Theta$ is the product $\mathsf{G}\mathsf{K}$ of a unitary gauge operator $\mathsf{G}\in\mathcal{G}$ by the matrix $\mathds{1}_d\times\mathsf{K}$.
 \end{proposition}
 
\emph{Proof.}
Let $\mathsf{C}$ denote a conjugation gauge operator. As antiunitary operator $\mathsf{C}= \mathsf{G}\mathsf{K}$ where $\mathsf{G}$ is unitary \cite{wigner3}.
In a principal window the density operator $\rho$ is real and invariant by any gauge operator. Therefore $\mathsf{G}$  is a unitary gauge operator.
$\Box$
\paragraph{}
Since $\mathsf{C}^2=\mathds{1}_d$, for definiteness, it is possible to select the initial conjugation operator in the principal source window as  $\mathsf{C}=\mathds{1}_d\times\mathsf{K}$. Let us term this matrix \enquote{conjugation gauge operator}.
\begin{definition}[Conjugation gauge operator $\mathsf{C}$]
\label{charge}
The conjugation operator $\mathsf{C}$ is expressed in a principal source window $\Omega$ by the matrix $\mathds{1}_d\times\mathsf{K}$  so that in this window
$$\mathsf{C}:\quad\mathcal{H}_g\to\mathcal{H}_{g^*}\quad:\quad  \ket{\psi_g}\mapsto \ket{\psi_{g^*}}= \mathsf{K}\ket{\psi_g}= \ket{\psi_g}^* $$
\end{definition}

\begin{definition}[Conjugation gauge group $\mathscr{C}$]
\label{chargegroup}
The conjugation gauge group is the involutive group $\mathscr{C}=\{\mathds{1}_d,\mathsf{C} \}$
\end{definition}

From Eq. (\ref{class}) the expression $\mathsf{C}'$ of the group generator $\mathsf{C}$ in another window derived from the principal window by a transition matrix $\mathsf{U}$ is

$$\mathsf{C}' = \mathsf{U}\mathsf{C}\mathsf{U}^{-1}=\mathsf{U}\mathsf{U}^T\times\mathsf{K}$$
because $\mathsf{U}$ is unitary and then $\mathsf{U}^{-1}=\mathsf{U}^\dagger= \mathsf{U}^{T*}$ so that $\mathsf{K}\mathsf{U}^{-1}= \mathsf{U}^T\mathsf{K}$.

\subsubsection{The full gauge group $\mathfrak{G}=\mathscr{C}\rtimes\mathcal{G}$}

We have defined two gauge groups, the discrete conjugation group $\mathscr{C}$ and the continuous unitary group $\mathcal{G}$.

\begin{proposition}
 \label{fullgaugegroup}
 The full gauge group $\mathfrak{G}$ is the semi-direct product $\mathscr{C}\rtimes\mathcal{G}$.
 \end{proposition}
 
\emph{Proof.}
By construction, the two groups operators are expressed in a common principal source window, that is a common basis of the four Hilbert spaces $\mathcal{H}_g$, $\mathcal{H}_{g^*}$, $\mathcal{H}_{g'}$ or $\mathcal{H}_{{g^*}'}$.
\begin{center}
\begin{minipage}{0.35\linewidth}
\begin{tikzpicture}
  \matrix (m) [matrix of math nodes,row sep=2em,column sep=4em,minimum width=0em] 
  {
                             &  \ket{\psi_g}              &  \ket{\psi_{g'}}   \\ 
                            &   \ket{\psi_{g^*}}           &  \ket{\psi_{{g^*}'}}   \\ 
    } ;
  \path[-stealth] 
    (m-1-2)  edge [->, thick]   node [left] {$\mathsf{C}$} (m-2-2)
                 edge [thick,->] node [below] {$\mathsf{G}$} (m-1-3)
     (m-2-2) edge [thick,->] node [above] {$\mathsf{G}$} (m-2-3)
     (m-1-3) edge [->, thick]   node [right] {$\mathsf{GCG}^{-1}$} (m-2-3)
    ;
\end{tikzpicture}
\end{minipage}
\qquad\quad~
\begin{minipage}{0.55\linewidth}
 Applying complex conjugation $\mathsf{C}\in\mathscr{C}$ and then a unitary transformation $\mathsf{G}\in\mathcal{G}$ has the same effect as applying the unitary transformation $\mathsf{G}$ first and then the group-conjugate  $ \mathsf{GCG}^{-1}=\mathsf{G}\mathsf{G}^T\times\mathsf{K}$ of the complex conjugation $\mathsf{C}$. 
\end{minipage}
\end{center}
As a result, the complex conjugation group $\mathscr{C}$ is a normal subgroup of the full gauge group $\mathfrak{G}$, that is to say that the full gauge group $\mathfrak{G}$ is the semi-direct product $\mathscr{C}\rtimes\mathcal{G}$.

The conjugation gauge operator $\mathsf{C}$ is specifically expressed by $\mathsf{K}$ in the initial unitary gauge, that is for  $\mathsf{G}=\mathds{1}_d$ or more generally when $\mathsf{G}$ is real. But of course in any cases, 
$$\mathsf{C}^2=(\mathsf{G}\mathsf{G}^T\mathsf{K})(\mathsf{G}\mathsf{G}^T\mathsf{K})=\mathsf{G}\mathsf{G}^T(\mathsf{G}^*\mathsf{G}^{T*})\mathsf{K}^2=\mathsf{G}\mathsf{G}^T (\mathsf{G}^T)^{-1}\mathsf{G}^{-1})\mathds{1}_d=\mathds{1}_d.$$


\subsection{Measurement and uncertainty}
\label{measurement}
Let $\mathcal{H}$ denote a Hilbert space.
In a general window, consider  a density operator $\rho$,  i.e., a positive Hermitian operator of unit trace acting on $\mathcal{H}$ and a set of observables, i.e., Hermitian operators $\mathsf{Q}$ acting on $\mathcal{H}$.

\subsubsection{Born rule}
We need first to verify that the Born rule, valid in the source window, is also valid in full generality in the Bayesian theater.
\begin{theorem}[Born rule]
\label{generalBorn}
In a Hilbert space the Born rule  applies in full generality regardless of the density matrix $\rho$ and whatever the observable $\mathsf{Q}$,
\begin{equation}
\label{eqgborn}
\langle \mathsf{Q} \rangle = \tr (\rho\mathsf{Q}).
\end{equation}
\end{theorem}
\emph{Proof.}
Any observable is described by a Hermitian operator. 
First, diagonalize the Hermitian operator, i.e.,  map the initial window to a proper window of the observable.
By reverse transcription, it is possible to regard the proper window as a source window.
By Proposition (\ref{staticexpectation}),  the Born rule holds in the source window and therefore in the current window as well because the computation of a tensor does not depend on the basis. 
$\Box$

\subsubsection{General measurement} 
\label{generalmeasurement}
Again, we still need to verify that the POVMs, valid for commutative diagonal observables, are also valid in full generality in the Bayesian theater. Actually this is a direct consequence of Theorem (\ref{generalBorn}).
Let $\rho$ denote an arbitrary density operator in a $d$-dimensional Hilbert space $\mathcal{H}$. Let $\Gamma$ be a finite set. Consider a  resolution of the tautology in $\mathcal{H}$ described by a set  of positive Hermitian operators $\{\mathsf{Q}_\gamma\}_{\gamma\in\Gamma}$, not necessarily commutative nor diagonal  in the current window, such that
$$\mathsf{Q}_\gamma\ge 0;\quad\sum_{\gamma\in\Gamma}\mathsf{Q}_\gamma =\mathds{1}_d$$
From the Born rule, Theorem (\ref{generalBorn}), define
  $$\prob(\gamma) = \tr (\rho \mathsf{Q}_\gamma ).\quad\mathrm{By~linearity,~we~have:}\sum_{\gamma\in\Gamma}\prob(\gamma) =1.$$
As a result, general positive-operator valued measurements (POVM) can be performed exactly like in conventional quantum information theory.
We obtain the important result:

\begin{theorem}[General measurement]
\label{generalexpectation} General POVMs can be performed regardless of the density matrix and whatever the positive  observables.
\end{theorem}

\emph{Interpretation.}
For commutative observables the measurement  estimates the probability of  outcomes collected from a unique viewpoint on the register.
By contrast, for non commutative observables $\mathsf{Q}_\gamma$, 
 the measurement estimates the probability of outcomes collected from different viewpoints.
Far from being exceptional, such measurements are also performed in classical physics (see Sec. \ref{qversusc}).
\paragraph{}
\emph{Generalization to weak POVMs.} The standard concept of POVM can be extended to that of \enquote{weak POVM} defined only with respect of a particular density operator.
\begin{definition}[Weak POVM]
\label{weakPOVM}
A weak POV measurement is defined by a set  of Hermitian operators $\{\mathsf{Q}_\gamma\}_{\gamma\in\Gamma}$, such that with a particular density operator
$$\langle\mathsf{Q}_\gamma\rangle\ge 0;\quad\sum_{\gamma\in\Gamma}\langle\mathsf{Q}_\gamma\rangle=1.$$
\end{definition}

\paragraph{}
\emph{Measurement operators.}
 Instead of $\mathsf{Q}_\gamma$, it is possible to introduce the so-called \enquote{measurement operators} $\mathsf{M}_\gamma$  acting on $\mathcal{H}$ such that $\mathsf{Q}_\gamma=\mathsf{M}_\gamma^{\dagger}\mathsf{M}_\gamma$~\cite{luders}.
 Then $\sum_\gamma \mathsf{M}_\gamma^{\dagger}\mathsf{M}_\gamma= \mathds{1}_d$ and  $\prob(\gamma)= \tr(\mathsf{M}_\gamma\rho\mathsf{M}_\gamma^{\dagger})$.
 
 In standard quantum information, following a general measurement, the state still can be viewed as a quantum state defined by a residual density operator $\rho'$ composed of an array of individual density operators $\rho_\gamma$ (when $\prob(\gamma)\not=0$) defined from the measurement operators as,
 \begin{equation}
 \label{geneM}
 \rho \mapsto \rho'= \sum_{\gamma\in\Gamma}\mathsf{M}_\gamma\rho\mathsf{M}_\gamma^{\dagger}= \sum_{\gamma\in\Gamma} \prob(\gamma)\times \rho_\gamma
 \qquad
\mathrm{where}\quad\rho_\gamma=\frac{\mathsf{M}_\gamma\rho\mathsf{M}_\gamma^{\dagger}}{\prob(\gamma)}
\end{equation}
In the present model,  we can take this concept as a definition.

\subsubsection{POVM entropy}
From Theorem (\ref{fondamentalVonNeumann}),  a Bayesian theater in a state $\rho$ contains $N-S(\rho)$ information bits. This raises the question of how to extract this information.
Actually, a POVM $\{\mathsf{Q}_\gamma\}_{\gamma\in\Gamma}$ extracts a fraction of this information depicted by the probability distribution $p=(\prob(\gamma))_{\gamma\in\Gamma}$. 

Consider first a completely random state, $\rho_0=(1/d)\times\mathds{1}_d$ corresponding to an absence of information. Define $q_\gamma= \tr(\mathsf{Q}_\gamma) $. Then the distribution $p_0=(\prob_0(\gamma))_{\gamma\in\Gamma}$ is
$$ \prob_0(\gamma) = \tr(\rho_0\mathsf{Q}_\gamma)= \frac{q_\gamma}{d}$$
In the current state $\rho$, the information gain $\mathbb{I}(\rho\Vert\Gamma)$ provided by the POVM probability distribution $p=(\prob(\gamma))_{\gamma\in\Gamma}$ is measured with respect to the state $\rho_0$ of no information as the relative entropy $\mathbb{H}(\prob\Vert\prob_0)$.

\begin{definition}[POVM information gain]
\label{defPOVMentropy}
The information $\mathbb{I}(\rho\Vert\Gamma)$ is the maximum information that can be extracted by a POVM $(\Gamma):\ \{\mathsf{Q}_\gamma\}_{\gamma\in\Gamma}$ as 
\begin{equation}
\label{eqPOVMentropy}
\mathbb{I}(\rho\Vert\Gamma)\ident\mathbb{H}(\prob\Vert\prob_0)=\sum_{\gamma\in\Gamma}\prob(\gamma)\log_2\frac{ \prob(\gamma)}{\prob_0(\gamma)} = N+ \sum_{\gamma\in\Gamma}\prob(\gamma)\log_2\frac{ \prob(\gamma)}{q_\gamma}
\end{equation}
\end{definition}
This information gain $\mathbb{I}(\rho\Vert\Gamma)$ is trivially less than the storage capacity $N$ of the register and even of the total information $N-S(\rho)$ currently stored in the Bayesian theater. This conception is not conventional. In standard quantum information theory, this bound, called \emph{Holevo bound}~\cite{holevo} is regarded as paradoxical and provided from the so-called \enquote{Holevo $\chi$-quantity} defined in the context of quantum channels (Eq. \ref{geneM}) as
\begin{equation}
\label{holevochi}
\chi(\Gamma) \ident S(\rho')-\sum_{\gamma\in\Gamma} \prob(\gamma)\times S(\rho_\gamma) =S( \sum_{\gamma\in\Gamma} \prob(\gamma)\times \rho_\gamma) -  \sum_{\gamma\in\Gamma} \prob(\gamma)\times S(\rho_\gamma)
\end{equation}

It is convenient to define the POVM entropy as $\mathbb{H}(\Gamma)= N-\mathbb{I}(\rho\Vert\Gamma)$. From Eq. (\ref{eqPOVMentropy}) we have
\begin{equation}
\label{eqPOVMinequality}
\mathbb{H}(\Gamma)=\sum_{\gamma\in\Gamma}-\prob(\gamma)\log_2\frac{ \prob(\gamma)}{q_\gamma}\ge S(\rho)\ge 0
\end{equation}
\begin{definition}[POVM entropy]
The POVM entropy $\mathbb{H}(\Gamma)$, Eq. (\ref{eqPOVMinequality}), is the entropy $N-\mathbb{I}(\rho\Vert\Gamma)$ of the maximum information $\mathbb{I}(\rho\Vert\Gamma)$ that can be extracted by a POVM.
\end{definition}
In particular, assume that the POVM corresponds to a von Neumann measurement in a particular window of sample set $\Omega= \{\omega\}$. Let $\ket{\omega}$ be the basis in this window. Then, $\Gamma=\Omega$ and $\mathsf{Q}_\omega = \ket{\omega}\bra{\omega}$ so that $q_\omega=1$. As a result, the POVM entropy $\mathbb{H}(\Omega)$ is just the window entropy, Definition (\ref{defwindowentropy}).
\begin{proposition}[Window entropy]
\label{substancewindowEntropy}
The window entropy $\mathbb{H}(\Omega)$ represents the entropy of the maximum information $N-\mathbb{H}(\Omega)$ that can be extracted by a von Neumann measurement in the window.
\end{proposition}

In standard quantum information, a POVM is called  \enquote{information-complete} when the operators $\mathsf{Q}_\gamma$, $\gamma\in\Gamma$ span the complete space $\mathcal{L}(\mathcal{H})$. 
Indeed, such a measurement provides $|\Gamma|\ge d^2-1$ coefficients $\prob(\gamma)$ that allow the unique reconstruction of the density operator $\rho$ and then the Bayesian probability distribution.  This does not necessarily mean that the POVM entropy is equal to $S(\rho)$ because this information is encoded in a particular way, which can cause a bias not taken into account in Eq. (\ref{eqPOVMentropy}) and then a loss of information (or an increase of entropy).
When there is no bias, the POVM can be called \enquote{centered} on the density operator.

\begin{definition}[Centered POVM]
A {information-complete} POVM is centered with respect to a density operator when its POVM entropy is equal to the von Neumann entropy of the density operator.
\end{definition}

In general, a particular measurement  is not information-complete and therefore the determination of the density operator requires independent measurements from additional POVMs.

\subsubsection{Independent POVMs}
\label{POVMinequality}
Suppose that a POVM $\{\mathsf{Q}_{\gamma}\}_{\gamma\in\Gamma}$, that we will refer to as ($\Gamma$), is information-\emph{incomplete} and consider the possibility to complement this POVM by another POVM. 

The set of density operators $\mathrm{D}(\mathcal{H})\subset\mathcal{L}(\mathcal{H})=\{\rho\}$ is a convex ensemble located in an \emph{affine} subspace of real dimension $d^2-1$. Motivated by  Ref. \cite{durt}, it is helpful to consider rather the set of traceless Hermitian operators, $\{\mathbf{e}\}$ defined as
$$ \mathbf{e}=\rho-\frac{1}{d}\mathds{1}_d,$$
because this ensemble is located in a \emph{linear} vector space $\mathcal{E}\subset\mathcal{L}(\mathcal{H}) $ still of dimension $d^2-1$. 
This mapping $\mathrm{D}(\mathcal{H})\to\mathcal{E}$ can be extended to all operators of a POVM as follows.
Consider the POVM $(\Gamma)$, $\{\mathsf{Q}_{\gamma}\}_{\gamma\in\Gamma}$ and define $\mathsf{Q}_{\gamma}\mapsto \mathbf{e}_\gamma$ as
\begin{equation}
\label{QE}
q_\gamma= \tr(\mathsf{Q}_{\gamma})>0 \quad;\quad\mathsf{E}_\gamma =  \frac{1}{q_\gamma} \mathsf{Q}_{\gamma}\in\mathrm{D}(\mathcal{H})\quad;\quad  \mathbf{e}_\gamma=\mathsf{E}_\gamma-\frac{1}{d}\mathds{1}_d\in\mathcal{E}
\end{equation}
The POVM is then characterized by  
\begin{equation}
\label{POVMcarac}
\sum_{\gamma\in\Gamma} q_\gamma=d\quad;\quad\sum_{\gamma\in\Gamma} q_\gamma \mathsf{E}_\gamma=\mathds{1}_d\quad;\quad\sum_{\gamma\in\Gamma} q_\gamma \mathbf{e}_\gamma=0
\end{equation}
At last, define a Hermitian inner product in  $\mathcal{E}$ as
\begin{equation}
\label{innerproduct}
\langle\mathbf{e_1}\cdot \mathbf{e_2}\rangle\ident \tr (\mathbf{e}_1^\dagger\ \mathbf{e}_2).
\end{equation}

 Let $\mathsf{Q}<\mathds{1}_d$ be an additional Hermitian positive operator.
Let $q= \tr(\mathsf{Q})>0$, $\mathsf{E}_{\scriptscriptstyle{\mathrm{Q}}} =  ({1}/{q}) \mathsf{Q}\in\mathrm{D}(\mathcal{H})$ and $\mathbf{e}_{\scriptscriptstyle{\mathrm{Q}}}=\mathsf{E}_{\scriptscriptstyle{\mathrm{Q}}}-({1}/{d})\mathds{1}_d\in\mathcal{E}$.
It turns out that $\mathsf{Q}$ is independent of the POVM if and only if $\mathbf{e}_{\scriptscriptstyle{\mathrm{Q}}}$ is orthogonal to every $\mathbf{e}_\gamma$.
Indeed, assume  that $\mathbf{e}_{\scriptscriptstyle{\mathrm{Q}}}$ is orthogonal to the subspace $ Span\{\mathbf{e}_\gamma\}_{\gamma\in\Gamma}\subseteq\mathcal{E}$. We compute easily from Eqs. (\ref{QE}-\ref{innerproduct})
$$ \forall \gamma\in\Gamma: \quad \langle \mathbf{e}_{\scriptscriptstyle{\mathrm{Q}}}\cdot \mathbf{e}_{\gamma}\rangle =0\quad\Longleftrightarrow\quad\frac{1}{qq_{\scriptscriptstyle \Lambda}}\tr(\mathsf{Q}\mathsf{Q}_{\gamma})-\frac{1}{d}=0$$
We have then
\begin{equation}
\label{eqPOVMindependence}
\forall\gamma\in\Gamma\quad  \tr (\mathsf{Q}\mathsf{Q}_{\gamma}) = \frac{\tr(\mathsf{Q})\tr(\mathsf{Q}_\gamma)}{d}
\end{equation}
Conversely, if Eq. (\ref{eqPOVMindependence}) holds, then $\mathbf{e}_{\scriptscriptstyle{\mathrm{Q}}}$ is orthogonal to every $\mathbf{e}_\gamma$.

To check the independence of the additional operator $\mathsf{Q}$, construct a second POVM with two operators, $\{\mathsf{Q}, \mathds{1}_d-\mathsf{Q}\}$.
Assume that the system \enquote{lives} in the first POVM set, meaning that $\rho=\rho_{\scriptscriptstyle{\Gamma}}\in Span(Q_\gamma)_{\gamma\in\Gamma}$.
Then, from linearity, Eq. (\ref{eqPOVMindependence}) and $\tr(\rho_{\scriptscriptstyle{\Gamma}})=1$,  the second measurement yields
$$\prob(\mathsf{Q}) = \tr(\rho_{\scriptscriptstyle{\Gamma}} \mathsf{Q})=  \frac{\tr(\mathsf{Q})\tr(\rho_{\scriptscriptstyle{\Gamma}})}{d}= \frac{\tr(\mathsf{Q})}{d}=\tr\Big(\frac{\mathds{1}_d}{d}\times\mathsf{Q}\Big)
\quad;\quad 
\prob(\mathds{1}_d-\mathsf{Q})) =1-\prob(\mathsf{Q})$$
exhibiting the effective  density operator $\rho_\mathrm{void}=\mathds{1}_d/d$ of a completely random system.
Therefore $\prob(\mathsf{Q})$ is totally independent of the density matrix $\rho_{\scriptscriptstyle{\Gamma}}\in Span(Q_\gamma)_{\gamma\in\Gamma}$.
Similarly, if the system lives in the second POVM set, $\rho=\rho_{\scriptscriptstyle{\mathrm{Q}}}\in Span(Q, \mathds{1}_d-\mathsf{Q})$ then the first POV-measurement yields
$$\prob(\mathsf{Q}_\gamma) = \tr(\rho_{\scriptscriptstyle{\mathrm{Q}}}\mathsf{Q}_\gamma)=\tr\Big(\frac{\mathds{1}_d}{d}\times\mathsf{Q}_\gamma\Big)$$
and again the coefficients $\prob(\mathsf{Q}_\gamma)$ are totally independent of the density matrix $\rho_{\scriptscriptstyle{\mathrm{Q}}}$
We will refer to the two POVMs as mutually \enquote{independent}. More generally, consider two distinct POVMs, $\{\mathsf{Q}_{\gamma_1}\}_{\gamma_1\in\Gamma_1}$ and $\{\mathsf{Q}_{\gamma_2}\}_{\gamma_2\in\Gamma_2}$.
For brevity, we say that a system defined by a density operator $\rho\in\mathcal{L}(\mathcal{H})$ \enquote{lives} in a POVM  $\{\mathsf{Q}_{\gamma}\}_{\gamma\in\Gamma}$ when $\rho\in Span\{\mathsf{Q}_{\gamma}\}_{\gamma\in\Gamma}$.

\begin{definition}[Independent POVMs]
\label{defPOVMindependence}
Two distinct POVMs, $\{\mathsf{Q}_{\gamma_1}\}_{\gamma_1\in\Gamma_1}$ and $\{\mathsf{Q}_{\gamma_2}\}_{\gamma_2\in\Gamma_2}$ are mutually independent if the measurement with one POVM when the system \enquote{lives} in the other POVM is identical to a measurement in a completely random state $\rho_\mathrm{void}=\mathds{1}_d/d$.
\end{definition}

\begin{proposition}
\label{propPOVMindependence}
Two distinct POVMs, $\{\mathsf{Q}_{\gamma_1}\}_{\gamma_1\in\Gamma_1}$ and $\{\mathsf{Q}_{\gamma_2}\}_{\gamma_2\in\Gamma_2}$ are mutually independent if and only if
\begin{equation}
\label{eq2POVMindependence}
\forall\gamma_1\in\Gamma_1,\quad\forall\gamma_2\in\Gamma_2:\quad  \tr (\mathsf{Q}_{\gamma_1}\mathsf{Q}_{\gamma_2}) = \frac{\tr(\mathsf{Q}_{\gamma_1})\tr(\mathsf{Q}_{\gamma_2})}{d}
\end{equation}
\end{proposition}
\emph{Poof.} From Eq. (\ref{eqPOVMindependence})  each $\mathbf{e}_{\gamma_{i}}$ is orthogonal to every $\mathbf{e}_{\gamma_{3-i}}$ ($i=1,2$).
$\Box$
\paragraph{}
Now, given that the two POVMs are independent, the information gains provided by the two measurements do not overlap.  As a result the sum of the two information gains is still bounded by the total information, $N-S(\rho)$, stored in the system.
\begin{proposition}[POVM entropic inequality]
Let $\Gamma_1: \{\mathsf{Q}_{\gamma_1}\}_{\gamma_1\in\Gamma_1}$ and $\Gamma_2: \{\mathsf{Q}_{\gamma_2}\}_{\gamma_2\in\Gamma_2}$ be two independent POVMs acting on a system in the state $\rho$. Then
\begin{equation}
\label{POVMinequality2}
\mathbb{H}(\Gamma_1)+\mathbb{H}(\Gamma_2) \ge N+S(\rho)\ge N
\end{equation}
\end{proposition}
\emph{Proof.}
Proceed to the transformations  $\mathbf{e}=\rho-\mathds{1}_d/d$, $q_{\gamma_i}= \tr(\mathsf{Q}_{\gamma_i})$, $\mathsf{E}_{\gamma_i} =  ({1}/{q_{\gamma_i}}) \mathsf{Q}_{\gamma_i}\in\mathrm{D}(\mathcal{H})$ and $\mathbf{e}_{\gamma_i}=\mathsf{E}_{\gamma_i}-({1}/{d_{\gamma_i}})\mathds{1}_d\in\mathcal{E}$, where $i\in\llbracket 1,2\rrbracket$ and $\gamma_i\in\Gamma_i$. Let $\mathcal{E}_i =Span_{\gamma_i\in\Gamma_i} (\mathbf{e}_{\gamma_i})$.
The space $\mathcal{E}$ splits into three mutually orthogonal subspaces, $\mathcal{E}=\mathcal{E}_1\oplus\mathcal{E}_2\oplus\mathcal{E}_0$. 
As a result, we have a unique decomposition $\mathbf{e} = \mathbf{e}_1+\mathbf{e}_2+\mathbf{e}_0$. 
Define $\rho_i =\mathbf{e}_i +\mathds{1}_d/d$.
Then,  still for $i\in\llbracket 1,2\rrbracket$ and $\forall \gamma_i\in\Gamma_i $ we obtain successively by a straightforward computation
\begin{align*}
\langle\mathbf{e}\cdot\mathbf{e}_{\gamma_i}\rangle=
\langle(\mathbf{e}_{0}+\mathbf{e}_{1}+\mathbf{e}_{2})\cdot\mathbf{e}_{\gamma_i}\rangle &=\langle \mathbf{e}_{i} \cdot\mathbf{e}_{\gamma_i}\rangle\\
\tr\Big[\big(\rho-\frac{\mathds{1}_d}{d} \big)\big(\frac{\mathsf{Q}_{\gamma_i}}{q_{\gamma_i}}-\frac{\mathds{1}_d}{d} \big)\Big] &=\tr\Big[\big(\rho_i-\frac{\mathds{1}_d}{d} \big)\big(\frac{\mathsf{Q}_{\gamma_i}}{q_{\gamma_i}}-\frac{\mathds{1}_d}{d} \big)\Big]\\
\tr(\rho\mathsf{Q}_{\gamma_i}) &=\tr(\rho_i\mathsf{Q}_{\gamma_i}).
\end{align*}
so that $\prob(\gamma_i)=\tr(\rho\mathsf{Q}_{\gamma_i})$ depends only on $\rho_i$. Therefore, the two information gains $\mathbb{I}_1=\mathbb{I}(\rho\Vert\Gamma_1)$ and $\mathbb{I}_2=\mathbb{I}(\rho\Vert\Gamma_2)$ are independent and the total information extracted by the two POVMs is the sum of the two information gains. This sum is trivially bounded by the storage capacity $N$ of the register, and even by the actual information stored in the register $N-S(\rho)$, i.e., $\mathbb{I}_1+\mathbb{I}_2\le N-S(\rho)\le N$. In terms of entropy, $\mathbb{H}(\Gamma_i)= N-\mathbb{I}_i$, we obtain Eq. (\ref{POVMinequality2}).
$\Box$
\paragraph{}
To our knowledge, the POVM inequality, Eq. (\ref{POVMinequality2}), is new but the concept of \enquote{unbiased POVM} was previously defined by  Kalev and Gour~\cite{kalev}.
In standard quantum information, the inequality is rather expressed for von Neumann measurements. Independent POVMs are then particularized by independent von Neumann measurements in the so called \enquote{mutually unbiased bases}.

\subsubsection{Mutually unbiased bases (MUB)}
\label{MUB}
Mutually unbiased bases, first introduced by J. Swinger in 1960~\cite{schwinger} are extensively used in standard quantum information~\cite{durt}.  
Let us first define precisely a pair of mutually unbiased bases  $\Omega_1$ and $\Omega_2$ in the present model. Each basis $\Omega_i$, of basic vectors $\ket{\omega_i}$, $(\omega_i\in\Omega_i)$, ($i\in\llbracket 1,2\rrbracket$), defines a von Neumann measurement i.e., a particular POVM, namely $\{\ket{\omega_i}\bra{\omega_i}\}_{\omega_i\in\Omega_i}$
\begin{definition}[Mutually unbiased bases (MUB) or mutually unbiased windows]
\label{defMUBpovm}
A pair of bases are mutually unbiased when they determine two independent von Neumann measurements.
\end{definition}
Let us recover the standard definition by the following proposition:
\begin{proposition}[MUB]
\label{propMUB1}
In a $d$-dimensional Hilbert space, two distinct orthonormal windows of index set $\Omega_1$ and $\Omega_2$ and  of basic vectors $\ket{\omega_1}$, $(\omega_1\in\Omega_1)$ and  $\ket{\omega_2}$, $(\omega_2\in\Omega_2)$ are mutually unbiased if and only if
\begin{equation}
\label{defMUBstand}
\forall \omega_1\in\Omega_1,\ \forall\omega_2\in\Omega_2\quad: \quad |\braket{\omega_1}{\omega_2}|^2 =\frac{1}{d}.
\end{equation}
\end{proposition}
\emph{Proof.}
From Eq. (\ref{eq2POVMindependence}) two von Neumann measurements are independent if and only if Eq. (\ref{defMUBstand}) holds.
$\Box$
\paragraph{}

Consider a pair of mutually unbiased bases,  defining two independent von Neumann measurements. Then, Eq. (\ref{POVMinequality2}) holds, with $\Omega_i$ standing for $\Gamma_i$, as 
\begin{equation}
\label{POVMinequality3}
\mathbb{H}(\Omega_1)+\mathbb{H}(\Omega_2) \ge N+S(\rho)\ge N
\end{equation}
We recover the well known entropic relations of standard quantum information theory that will be considered more generally in Sec. (\ref{setobservables}) below.
The first bound, $N+S(\rho)$, corresponds to a special case of the Frank-Lieb's inequality~\cite{frank} and the second bound, $N$, to the less tight Massen-Uffink's inequality~\cite{maassen}. 
Note the the present model provides an intuitive basis to these inequalities, usually regarded as somewhat esoteric technical results.

Beyond a single pair of bases, starting from an initial basis, it is possible to construct the set of all bases mutually unbiased, i.e., containing independent information. Indeed, it turns out that there are always $d$ additional bases, i.e., a cluster of $d+1$ distinct MUBs, $\Omega_i$, when the dimension $d$  of the Hilbert space is a power of a prime integer and then specifically when $d=2^N$~\cite{bandyopadhyay}. 
This set is both maximum and information-complete, meaning that there is no additional unbiased basis and that the full ensemble of $d(d+1)$  projectors $\ket{\omega_i}\bra{\omega_i}$, while not linearly independent, spans the space  $\mathcal{L}(\mathcal{H})$.  This allows the unique reconstruction of an arbitrary positive operator in $\mathrm{D}(\mathcal{H})$~\cite{durt}. Indeed, due to normalization, each basis provides $d-1$ independent probability $\prob(\gamma_i)$ and the whole $d+1$ bases provide $(d+1)\times(d-1)= d^2-1$ parameters.

By iterating, the inequality Eq. (\ref{POVMinequality3}) can be generalized to $K$ distinct MUBs, meaning that a maximum of $N-S(\rho)$ bits of information and no more can be distributed among the $K$ windows, i.e., $\sum \mathbb{I}_k\le  N-S(\rho)$, or in terms of window entropies
$$\sum_{k=1}^{K}\mathbb{H}(\Omega_k)\ge N  (K-1) + S(\rho),$$
where $\Omega_k$ are the sample sets of the $K\le d+1$ different MUBs. At last, for $K=d+1$ we have
\begin{equation}
\label{infoMUB}
\sum_{k=1}^{d+1}\mathbb{H}(\Omega_k)\ge   N d + S(\rho),
\end{equation}
One might expect that the inequality Eq. (\ref{infoMUB}) be saturated. However, this is not the case in general because the probability distributions in the cluster are encoded in a particular way which causes a bias not taken into account even in Eq. (\ref{eqPOVMentropy}), i.e., an excess of entropy, say $\Delta_\mathbb{H}$.

\begin{proposition}
The totality of the information stored in the Bayesian theater can be recovered from a principal window. 
\end{proposition}
\emph{Proof.}
The bound $ N d+ S(\rho)$ in Eq. (\ref{infoMUB}) is attained when one of the $d+1$ windows is principal because the density operator $\rho$ is diagonal in this window. Then its window entropy is equal to $S(\rho)$ and the $d$ others window are completely devoid of information with a window entropy of $N$ bits.
Such a cluster can be called \enquote{centered} on the state $\rho$. 
Reversing the logic, we can assess the lack of centering of a general cluster from the excess of entropy $\Delta_\mathbb{H}$ in Eq. (\ref{infoMUB}). 
$\Box$

\begin{proposition}
\label{completude}
The set of all windows in a Bayesian theater covers the complete set of relevant Boolean variable batches up to a discrete Boolean gauge change.
\end{proposition}
\emph{Proof} Since the totality of the information of the Bayesian theater can always be recovered, there is no additional window, that is, there is no additional relevant Boolean variable batch up to a discrete Boolean gauge change (Definition \ref{sourcegauge}).
$\Box$

\subsubsection{Effects}
Consider just one non-negative observable $\mathsf{Q}\le\mathds{1}_d$. \emph{Irrespective of the window}, such an operator, also called \enquote{effect}  \cite{busch}  describes an autonomous object with a specific probability, a specific entropy and an internal probability distribution.

The concept of \enquote{effect} can be extended to cases of Hermitian operators that are not necessarily positive but whose expectation with respect to the current density operator $\rho$ is positive and less than or equal to 1 and that we propose to call \enquote{\emph{weak effects}}.

\begin{definition}[Effect, weak effect]
An effect is an autonomous object specified by a non-negative observable $\mathsf{Q}\le\mathds{1}_d$. A weak effect is defined with respect to a particular density operator $\rho$ by an observable $\mathsf{Q}$ whose expectation is positive and less than 1, $0\le\langle\mathsf{Q}\rangle\le{1}$.
\end{definition}

\paragraph{Specific probability.}
The probability of the effect is trivially its expectation. In particular, we can recover some standard instances of the Born rule.
\begin{proposition} 
The specific probability of a bounded positive observable $\mathsf{Q}\le\mathds{1}$ is its expectation.
\begin{equation}
\label{probaeffect}
\prob(\mathsf{Q})=\tr( \rho\mathsf{Q}).
\end{equation}
In particular, the probability  of a rank 1 projection operator, $\mathsf{Q}=\ket{u}\bra{u}$ is $\prob(u)=\bra{u}\rho\ket{u}$. If the density operator depicts a pure state $\rho =\ket{v}\bra{v}$, the \emph{conditional probability} of $\ket{u}$ given $\ket{v}$ is $\prob(u|v)=|\braket{u}{v}|^2$. 
\end{proposition}
\emph{Proof.} Include $\mathsf{Q}$ into any POVM, e.g. $\{\mathsf{Q}, \mathds{1}-\mathsf{Q}\}$. $\Box$
\paragraph{}

For a weak effect, we use similarly weak POVM (Definition \ref{weakPOVM}).
\paragraph{Induced probability distribution.}
It is also possible to define a probability distribution \emph{inside} the effect. 
\begin{proposition}
 In the proper window of a bounded positive observable, the density operator $\rho$ induces by reverse-transcription a probability distribution inside the effect as
 \begin{equation}
\label{inducedproba}
h_\omega \ident \frac{\mathrm{q}_\omega w_{{\scriptscriptstyle \Lambda},\omega}}{\langle\mathrm{q} w_{\scriptscriptstyle \Lambda} \rangle}.
\end{equation}
\end{proposition}
\emph{Proof.}
Proceed to the reverse transcription of  the  system in the  proper window of the effect (in which the observable is diagonal), that is $\mathsf{Q}=Diag(\mathrm{q}_\omega)$. Let $\mathcal{P}$ be the real-valued probability space of this window so that $\mathrm{q}=(\mathrm{q}_\omega)\in\mathcal{P}^*$  is the covector of the observable $Q$. 
In the proper window, the pair of the working distribution $w_{\scriptscriptstyle \Lambda}$ and the observable $Q$ induces trivially a probability distribution $h_\omega$  given by Eq. (\ref{inducedproba}).

In particular, when $\mathsf{Q}$ is an orthogonal projection operator, $q_\omega\in\llbracket0,1\rrbracket$ and the observable $Q$ in $\mathcal{P}$ depicts a Boolean function, so that the probability distribution $h_\omega$ is just the restriction of the working distribution $w_{{\scriptscriptstyle \Lambda},\omega}$ to the support of this Boolean function.
$\Box$

\begin{definition}[Induced probability distribution inside an effect]
The induced probability distribution inside an effect is the distribution Eq. (\ref{inducedproba}).
\end{definition}

As a result, it is also possible to define an induced entropy.
\begin{definition}[Induced entropy of an effect]
\label{effectentropy}
The induced entropy  $\mathbb{H}(h)$ of an effect is the entropy of its induced probability distribution $h$.
\end{definition}

For instance, the projection operator on an eigensubspace of multiplicity $d_k$ of the current density operator $\rho$ is an effect characterized by a completely random induced probability distribution and thus an induced entropy of $\log_2 d_k$  bits. This simply expresses the equivalence of the $d_k$ eigenvalues.

\paragraph{}
We will define later a \enquote{window entropy} of general observables, Definition (\ref{windowobservablentropy}) below, that has nothing to do with this \enquote{induced entropy} .

\subsection{Set of observables}
\label{setobservables}

In standard physics, observables are defined by Hermitian operators acting on the Hilbert space. This is of course valid in the present model, but the basic definition of an observable is primarily found in the probability space $\mathcal{P}$  (Definition \ref{defobservable}).
\begin{align*}
Q:\quad \Omega \to \mathbb{R} : \quad \omega\mapsto Q(\omega)=\mathrm{q}_\omega.
\end{align*}
Naturally,  these observables with the same proper window $\Omega$ commute.

Let us address the general case of non commutative observables.

\subsubsection[Entropic inequalities]{Entropic inequalities between non commutative observables}
\label{entropicinequalities}
When two observables $\mathsf{Q}_1$ and $\mathsf{Q}_2$ in a Hilbert space $\mathcal{H}$ have no common proper window they describe information from two distinct sample sets, $\Omega_1$ and $\Omega_2$.
In general, they are  non-commutative.
In standard physics and in infinite dimension, this information is estimated with respect to a pure quantum state  by a formulation of the Heisenberg uncertainty principle due to E. H. Kennard~\cite{kennard} and generalized by H. P. Robertson~\cite{robertson}.

In the present model the Hilbert space is finite dimensional. The Robertson's inequality is ineffective but \emph{entropic inequalities} are appropriate with the same meaning.
We already computed the entropic relations in the case of independent POVMs in Secs. (\ref{POVMinequality}, \ref{MUB}) above. Now we address again this question but for non necessarily independent measurements.

The entropic inequalities were defined by I. Bialynicki-Birula \emph{et al}~\cite{bialynicki} and computed by H. Maassen and J. B. M. Uffink~\cite{maassen} with respect to a pure quantum state. The Maassen-Uffink bound was  extended to general quantum states and significantly improved in 2011 by R. Frank and  E. Lieb~\cite{frank}. 
These relations concern the proper windows of a set of observable and specifically their entropy.

Let us define the \enquote{window entropy of an observable}. 
This entropy characterizes only the \emph{proper basis} in contrast with the induced entropy (Definition \ref{effectentropy}). All  regular commutative observables have the same window entropy. 

\begin{definition}[Window entropy of an observable]
\label{windowobservablentropy}
The window entropy  of an observable with distinct eigenvalues is the window entropy $\mathbb{H}(\Omega)$ of its proper window.
\end{definition}

Let $\Omega_1$ and $\Omega_2$ respectively denote the proper windows of a pair of non-commutative observables $\mathsf{Q}_1$ and $\mathsf{Q}_2$.
We need to define  the so called \emph{window-overlap},  $\delta$,  between two windows. For generality, define this window-overlap as a special case of a \enquote{POVM-overlap} between two POVMs.

\begin{definition}[POVM-overlap]
\label{defPOVMoverlap}
The overlap $\delta$ of two distinct POVMs, $\{\mathsf{Q}_{\gamma_1}\}_{\gamma\in\Gamma_1}$ and $\{\mathsf{Q}_{\gamma_2}\}_{\gamma\in\Gamma_2}$ is the square-root of the maximum absolute value of $\tr (\mathsf{Q}_{\gamma_1}\mathsf{Q}_{\gamma_2})$
\begin{equation}
\label{eqPOVMoverlap}
\delta =\max_{\gamma_1\in\Gamma_1, \gamma_2\in\Gamma_2} |\tr (\mathsf{Q}_{\gamma_1}\mathsf{Q}_{\gamma_2})|^{1/2}
\end{equation}
\end{definition}
\noindent From the Cauchy-Schwarz inequality, $\delta\le 1$.

\begin{definition}[Window-overlap]
\label{overlap}
The overlap $\delta$ of two distinct windows $\Omega_1$ and $\Omega_2$  is the POVM-overlap of the two von Neumann measurements in the windows.
\end{definition} 
\noindent Let $\ket{\omega_1}$ and  $\ket{\omega_2}$ denote the basis vectors in $\Omega_1$ and $\Omega_2$ respectively. Then, the two POVMs are $\{\ket{\omega_1}\bra{\omega_1}\}_{\omega_1\in\Omega_1}$ and  $\{\ket{\omega_2}\bra{\omega_2}\}_{\omega_1\in\Omega_2}$ respectively and therefore
$$\delta = \max_{\omega_1,\omega_2} |\braket{\omega_1}{\omega_2}| \quad\mathrm{for~} \omega_1\in\Omega_1,\  \omega_2\in\Omega_2$$

Let $\mathbb{H}(\Omega_1)$ and $\mathbb{H}(\Omega_2)$ denote the window entropies of $\mathsf{Q}_1$ and $\mathsf{Q}_2$  respectively and $\delta$ their overlap.
The Maassen-Uffink entropic inequality~\cite{maassen} reads 
\begin{equation}
\label{entropicinequality}
\mathbb{H}(\Omega_1)+ \mathbb{H}(\Omega_2)\ge  \log(1/\delta^2).
\end{equation}
A more precise bound taking into account the von Neumann entropy $S(\rho)$ was established  by R. Frank and  E. Lieb~\cite{frank} as

\begin{equation}
\label{frank}
\mathbb{H}(\Omega_1)+ \mathbb{H}(\Omega_2)\ge  \log(1/\delta^2) + S(\rho).
\end{equation}
The two inequalities Eqs. (\ref{entropicinequality}) and (\ref{frank}) are identical for deterministic states ($S(\rho)=0$). For mutually unbiased bases, we already saw that $\delta=1/\sqrt{d}$ and $\log_2(1/\delta^2)=N$ bits (Sec. \ref{MUB}).
At last for completely random state, $S(\rho)=N$ bits.

 
\subsubsection{Complementary observables} 
In a Hilbert space of infinite dimension, the Fourier transform provides a \emph{complementary viewpoints  to a given observable}.
In a Hilbert space of finite dimension, the discrete Fourier transform and more generally \enquote{complex Hadamard matrices} ~\cite{tadej} conveniently rescaled, say  $\mathsf{U}$, play the same role. They transform the initial basis into a new basis, so that the two windows are \enquote{mutually unbiased} (MUB).

\begin{proposition}~\\
\label{propMUB}
Let $\mathsf{U}$ be the unitary operator mapping an initial basis $\ket{\omega_1}$ onto a second basis $\ket{\omega_2}$ in a $d$-dimensional Hilbert space. The two bases are mutually unbiased if the norm $|U_{\omega_1\omega_2}|^2$ of the $d^2$ entries  expressed in the initial window is constant. The transition operator $\mathsf{U}$ is then a rescaled complex Hadamard matrix and $|{U_{\omega_1\omega_2}}|^2=1/d$.
\end{proposition}
Consider a particular observable and its proper window.  Define a new window by a complex Hadamard matrix so that the two bases are mutually unbiased. Then the new observable is complementary of the initial observable.

\begin{definition}[Complementary observables]
A pair of  observables is complementary when the two proper windows are mutually unbiased.
\end{definition}

With some mathematical precautions, the limit when $N\to\infty$ leads to the complementary pairs of quantum observables like position and momentum in Hilbert space of infinite dimension. In fact, such a pair of complementary observables describes a continuous degree of freedom. Interestingly there is no additional mutually unbiased base beyond each pair in infinite dimension \cite{weigert}.

\subsection{Pair of systems}
\label{pairH}
In this section, we shortly review the results of Sec. (\ref{pairofregisters}) but in the full Hilbert space. Actually, we recover identically the standard quantum information theory, e.g..  conditional entropy or \enquote{entanglement entropy}.

Consider two Hilbert spaces, $\mathcal{H}_a$ and $\mathcal{H}_b$, and let  $\mathcal{H}_c= \mathcal{H}_a\otimes\mathcal{H}_b$.
In addition, consider a global density operator $\rho_c$ of rank $r_c$ acting on $\mathcal{H}_c$. 
Define the partial traces, $\rho_a=\tr_b(\rho_c)$ acting on $\mathcal{H}_a$ and $\rho_b=\tr_a(\rho_c)$ acting on $\mathcal{H}_b$.

\paragraph{Reverse transcription.}
The reverse transcription of the system is composed of three probability spaces, $\mathcal{P}_a$, $\mathcal{P}_b$ and  $\mathcal{P}_c= \mathcal{P}_a\otimes\mathcal{P}_b$. Let $(w_c,\mathcal{W}_c)$ denote the quantum state in $\mathcal{P}_c$.

Now, the results of Sec. (\ref{pairofregisters}) hold.
Construct the two partial systems derived from  the working distribution $w_c$ in $\mathcal{P}_c$, namely, $(w_a, \mathcal{W}_a)$ and $(w_b, \mathcal{W}_b)$. Let  $\Prob_a=w_a$  and $\Prob_b=w_b$ denote the marginal probability distributions,  in $\mathcal{P}_a$ and $\mathcal{P}_b$ respectively. By construction,  $(w_a, \mathcal{W}_a)$ and $(w_b, \mathcal{W}_b)$ are consistently transcribed in $\mathcal{H}_a$ and $\mathcal{H}_b$ respectively as $\rho_a$ and $\rho_b$.

\paragraph{Entanglement entropy.}
Usually, the entanglement of a \emph{pure state} $\rho_c$ with respect to the factorization $\mathcal{H}_c= \mathcal{H}_a\otimes\mathcal{H}_b$ is identified with the von Neumann entropy $S'_2(\rho)$ of either of the two reduced states $\rho_a$ or $\rho_b$ in $\mathcal{H}_a$ and $\mathcal{H}_b$ respectively. 
%
$$S'_2(\rho_c)\ident S(\rho_a)=S(\rho_b)$$
However, this definition is irrelevant for a mixed state $\rho_c$ because it does not grasp the correlation between the two factor spaces~\cite{mf2}. An alternative formulation was proposed by V. Vedral \emph{et al}~\cite{vedral1} as the minimum of the relative entropy of the state $\rho_c$ with respect to all disentangled states, $\sigma_c$ as
$$S_2(\rho_c)\ident\min_{\sigma_c\in\mathrm{D}(\mathcal{H}_a)\otimes\mathrm{D}(\mathcal{H}_b)}S(\rho_c\Vert\sigma_c) $$
where $\mathrm{D}(\mathcal{H}_a) =\{\sigma_a\}$ and $\mathrm{D}(\mathcal{H}_b) =\{\sigma_b\}$ are the sets of density operators acting on $\mathcal{H}_a$ or $\mathcal{H}_b$ respectively. From Sec. (\ref{pairofregisters}), the minimum is attained for  $\sigma_a=\rho_a$ and $\sigma_b=\rho_b$. Finally, we adopt the following definition

\begin{definition}[Entanglement entropy]
The entanglement entropy $S_2(\rho_c)$ of a quantum state $\rho_c$ with respect to the factorization $\mathcal{H}_c= \mathcal{H}_a\otimes\mathcal{H}_b$ is the relative entropy of $\rho_c$ with respect to the separable state $\rho_a\otimes\rho_b$ as
\begin{equation}
\label{entanglemententropy}
S_2(\rho_c)\ident S(\rho_c\Vert\rho_a\otimes\rho_b)
\end{equation}
where $\rho_a =\tr_b(\rho_c)$ and $\rho_b =\tr_a(\rho_c)$ are the two reduced states of $\rho_c$  in $\mathcal{H}_a$ and $\mathcal{H}_b$ respectively. 
\end{definition}

Consider a principal window $\Omega_c$ of $\rho_c$. Let  $\Omega_a$ and $\Omega_b$ denote the reduced windows of of $\rho_c$ in $\mathcal{H}_a$ and $\mathcal{H}_b$ respectively. 
 
 \begin{proposition}
 \label{entangtropy}
 The entanglement entropy, Eq. (\ref{entanglemententropy}), of a bipartite quantum state is the mutual information of the corresponding principal distributions.
 \begin{align}
\begin{aligned}
S_2(\rho_c)=\mathbb{H}(\Omega_a;\Omega_b) &= \mathbb{H}(\Omega_a) - \mathbb{H}(\Omega_a|\Omega_b) =\mathbb{H}(\Omega_b)-\mathbb{H}(\Omega_b|\Omega_a) \\
 &=\mathbb{H}(\Omega_a)+\mathbb{H}(\Omega_b)-\mathbb{H}(\Omega_a,\Omega_b)
 \end{aligned}
\end{align} 
 \end{proposition}
 \emph{Proof.} The global quantum state $\rho_c$ and the two partial states $\rho_a$ and $\rho_b$ are simultaneously diagonal in a common principal window. Then, from Proposition (\ref{entropequation}), the computation similar to Eq. (\ref{relativeentropy}) is performed in a conventional probability distribution as in  Eq. (\ref{mutualinformation}). $\Box$

%
\paragraph{Conditional entropy.}
Consider the entropy $S(\rho_a|\rho_b)$ of the state $\rho_a$ in $\mathcal{H}_a$ conditional on the state $\rho_b$ in $\mathcal{H}_b$.  In conventional quantum information, this expression is considered problematic~\cite{guerra}. In the present model, it makes sense by switching to the principal  window of $\rho_c$, as stated by Proposition (\ref{entropequation}) in Sec. (\ref{otherentropy}). In this window, $\rho_c$ is diagonal in $\mathcal{H}_c$, and so are the partial traces $\rho_a$ and $\rho_b$ in $\mathcal{H}_a$ and in $\mathcal{H}_b$ respectively.   By reverse transcription, let  $\Omega_a$, $\Omega_b$ and the Cartesian product $\Omega_c=(\Omega_a, \Omega_b)$ denote the sample sets respectively. 
From Proposition (\ref{entropequation}) we have the formal correspondence
\begin{align*}
\mathbb{H}(\Omega_a) ; \mathbb{H}(\Omega_b) ; \mathbb{H}(\Omega_c)  &\implies S(\rho_a)=\mathbb{H}(\Omega_a) ; S(\rho_b)=\mathbb{H}(\Omega_b) ; S(\rho_c)=\mathbb{H}(\Omega_c)\\
\mathbb{H}(\Omega_a|\Omega_b) =\mathbb{H}(\Omega_c)-\mathbb{H}(\Omega_a) &\implies S(\rho_a|\rho_b) =S(\rho_c) -S(\rho_b) 
\end{align*}
where $\mathbb{H}(.)$ only refers to a principal window while $S(.)$ is valid irrespective of the window.
Therefore, in the present model,  $S(\rho_a|\rho_b)$ is a well-defined function.


\section[Illustrative examples]{Examples} 
\label{examples}
To illustrate the present theory, we propose to review some examples. We begin with a system with only one bit. 
It is remarkable that this simple instance is already a real Bayesian theater.
The model describes both a classical bit, that is a state of rank 2, and a genuine qubit of rank 1. Next, a 2-bit system allows the description of the singlet and the triplet states. In passing, we turn briefly to the problem of the EPR pair and the non-signaling property. Finally, we propose to demystify some paradoxes of the non-local PR-box in the framework of the present theory.

\subsection{One-bit system}
\subsubsection[Mixed state]{Mixed one-bit system} 
\label{onebit}

Consider a register of only one Boolean variable $\mathsf{X}_1$ without any constraint.  
The Bayesian prior $(\Lambda)$ is simply 
$$(\Lambda)\ident\{ N=1\}.$$

\paragraph{Source window.}

In a Bayesian framework, we leave indeterminate the truth value of the Boolean variable and describe this uncertainty by the formalism of random variables.
The sample set $\Omega =\{ \omega_1, \omega_2\}$ comprises two classical states, say $\omega_1=\overline{\mathsf{X}}_1$ and $\omega_2={\mathsf{X}}_1$. This choice is of course arbitrary and defined up to a swap of the two states. While trivial in this example, this corresponds to the discrete Boolean gauge group (Definition \ref{sourcegauge}), whose operators are here simply  the identity and the swap operator.
The formulation of the problem by the logical states of a particular Boolean variable amounts to defining an \emph{observation window} and, as it is the initial description, it is called \enquote{source window}.

It is possible to construct a real-valued probability space based on this source window, say  $\mathcal{P}\ident Span(\omega_1,\omega_2)$, of dimension $d=2^N=2$. 
Define $p=(p_1,p_2)$ where $p_1=\Prob(-1)\ident\Prob({\mathsf{X}_1}=0|\Lambda)$ and $p_2=\Prob(1)\ident\Prob({{\mathsf{X}}_1}=1|\Lambda)$.
The LP system Eq. (\ref{lpeq}) is just composed of the relevant universal equations, Eqs. (\ref{simplet}, \ref{doublet}, \ref{triplet}, etc.),  limited here to the sole normalization equation, 
\begin{align}
\label{lp1bit}
\begin{aligned}
&p_1 + p_2=1\\
&\mathrm{subject~to~}
p \ge 0 
\end{aligned}
\end{align}
so that the rank of the LP system is $m=1$. 
Each solution  is a particular probability distribution $\Prob$ on the sample set $\Omega$. 
The Bayesian formulation Eq. (\ref{expectation}) is reduced to its simplest expression without any explicit constraint as
\begin{align*}
(\Lambda):\quad\mathrm{Assign~a~probability~distribution~}\Prob\mathrm{~on~}\Omega.
\end{align*}
Let $\tilde{\omega}_1=(1,0)$ and $\tilde{\omega}_2=(0,1)$ denote the two deterministic solutions in $\mathcal{P}$. ~The LP system,
 
\begin{minipage}{0.25\linewidth}
\begin{tikzpicture}[scale=2]
\draw[->, >=latex,very thick] (0, 0) -- (1.2, 0);
\draw [->, >=latex,very thick](0, 0) -- (0, 1.2);
\draw [very thick](0,1) -- (1, 0);
\draw (0, 0) node[below]{O};
\draw (0, 1) node{$\bullet$};
\draw (0, 1) node[right]{$\ \tilde{\omega}_2$};
\draw (1, 0) node{$\bullet$};
\draw (1, 0) node[above]{$\ \tilde{\omega}_1$};
\draw (0.5, 0.5) node{o};
\draw (0.5, 0.5) node[above,right]{$\tilde{c}$};
\draw (0.3, 0.7) node{o};
\draw (0.3, 0.7) node[above,right]{$w_{\scriptscriptstyle \Lambda}$};
\draw (1, 0) node[below]{1};
\draw (0, 1) node[left]{1};
\draw (0, 1.2) node[above]{$p_2$};
\draw (1.2, 0) node[right]{$p_1$};
\end{tikzpicture}
\end{minipage}
\begin{minipage}{0.7\linewidth}
 Eq. (\ref{lp1bit}), accepts not only the two classical deterministic distributions $\tilde{\omega}_1$ and $\tilde{\omega}_2$ but also a continuous set of solutions on their convex hull. The feasible solutions are located on a specific polytope 
 $\mathcal{W}_{\scriptscriptstyle \Lambda}$, that is the line segment $[\tilde{\omega}_1,\tilde{\omega}_2]$ identical to the tautological simplex of one variable $\mathcal{W}_{\scriptscriptstyle I}$. The line itself is an affine 1-dimensional subspace $P_{\scriptscriptstyle \Lambda}$. The simplex vertices are  $w_1=\tilde{\omega}_1$ and $w_2=\tilde{\omega}_2$. Therefore, the system is simplicial (Definition \ref{defsimplicial}) and $\mathcal{W}_{\scriptscriptstyle \Lambda}=\mathcal{W}_{\scriptscriptstyle I}=\mathrm{conv}(\tilde{\omega}_1,\tilde{\omega}_2)$.
\end{minipage}
\paragraph{Simplicial quantum state.}
 The system, Eq. (\ref{lp1bit}) defines a \enquote{mixed state} of rank $r=2$. 
 The specific polytope $\mathcal{W}_{\scriptscriptstyle \Lambda}$ is the tautological simplex. It is possible to single up a particular solution, $w_{\scriptscriptstyle \Lambda}$, called \enquote{working distribution}, by assigning a weight to each vertex of the simplex that is a discrete \emph{contextual probability distribution}. Define
 $$\Sigma_\lambda=\{\lambda_1,\lambda_2 \}\quad\mathrm{where~} \lambda_1,\lambda_2\ge 0\quad\mathrm{and~} \lambda_1+\lambda_2=1 ,$$
so that $w_{\scriptscriptstyle \Lambda} = \lambda_1\tilde{\omega}_1+\lambda_2\tilde{\omega}_2 \in\mathcal{W}_{\scriptscriptstyle \Lambda}$.
 By default, the working distribution $w_{\scriptscriptstyle \Lambda}$ is the center of mass of the polytope, i.e.,  $\tilde{c} = (1/2)(\tilde{\omega}_1+\tilde{\omega}_2) $. It is also the mean point with respect to an auxiliary uniform density, say $\sigma$, on the line segment $[\tilde{\omega}_1,\tilde{\omega}_2]$. 
 The pair, $(w_{\scriptscriptstyle \Lambda},\mathcal{W}_{\scriptscriptstyle \Lambda})$, is termed \enquote{simplicial quantum state}.  
 The default simplicial quantum state is $(\tilde{c},\mathcal{W}_{\scriptscriptstyle \Lambda})$.
  If $\lambda_1=0$ or $1$ we have a conventional deterministic bit. Otherwise, we have a random bit, still conventional described by the simplicial quantum state $(w_{\scriptscriptstyle \Lambda}, \mathcal{W}_{\scriptscriptstyle \Lambda})$.
The window entropy $\mathbb{H}(\Omega)$ and the  simplicial entropy $\mathbb{H}(\Sigma_\lambda)$ are equal and 
$$\mathbb{H}(\Omega)=\mathbb{H}(\Sigma_\lambda)=-\lambda_1\log\lambda_1-\lambda_2\log\lambda_2.$$

\paragraph{Observable.}
In the source window, consider an observable $Q:\Omega\to\mathbb{R}$ and let  $Q(\omega)=\mathrm{q}_\omega$. The expectation is defined as
\begin{equation*}
\langle Q \rangle=\langle \mathrm{q} p \rangle|_{p=w_{\scriptscriptstyle \Lambda}} =\langle \mathrm{q} w_{\scriptscriptstyle \Lambda} \rangle = \lambda_1\mathrm{q}_{\omega_1} + \lambda_2\mathrm{q}_{\omega_2}.
\end{equation*}
%
For instance, consider  the particular observable $S_Z(\omega)=\mathrm{s}_\omega$ defined as
$$S_Z(\omega_1)=1\quad;\quad S_Z(\omega_2)=-1\quad\mathrm{i.e.}\quad \mathrm{s}=(1,-1)\in\mathcal{P}^*$$
We have
$$\langle S_Z \rangle=\langle \mathrm{s} w_{\scriptscriptstyle \Lambda} \rangle = \lambda_1 - \lambda_2$$ 

\paragraph{Other observation windows.}
We started from a unique Boolean variable, $\mathsf{X}_1$.
Surprisingly enough, in a Bayesian framework, it is possible to pose the problem by using \emph{other alternatives than $\mathsf{X}_1$ and  $\overline{\mathsf{X}}_1$}, that is to consider \emph{other observation windows}. 
These new alternatives are necessary in order to compute the expectation value of every relevant observable.
For instance, a Boolean variable, e.g. the spin of a particle in physics, points in a specific direction compatible with particular observation windows. However, the current observation window has no reason to coincide with one of these directions. Nevertheless, Bayesian inference always provides a probabilistic estimation for any direction. It can be viewed as a form of artifact and is a major novelty of Bayesian inference technique.

To change the observation window,  we make use of a new tool.

\paragraph{Transcription into $\mathcal{H}$.}
Indeed, to construct these new observation windows, the fundamental innovation of quantum information is to transcribe the source window into a Hilbert space $\mathcal{H}$ defined as the \emph{complex} span of $(\omega_1, \omega_2)$.  Let  $(\ket{1}, \ket{2})$ denote its basis vectors.  Afterward, the new alternatives will be simply computed by changing this initial basis. The initial simplicial quantum  state  is transcribed as a density operator $\rho_{\scriptscriptstyle \Lambda}=\lambda_1\ket{1}\bra{1} + \lambda_2\ket{2}\bra{2}$, or
$$
\rho_{\scriptscriptstyle \Lambda}= \lambda_1 
\begin{bmatrix}
  1  & 0 \\
   0 & 0
\end{bmatrix}
+\lambda_2 
\begin{bmatrix}
  0  & 0 \\
   0 & 1
\end{bmatrix}
=
\begin{bmatrix}
  \lambda_1  & 0 \\
   0 & \lambda_2
\end{bmatrix}.
$$
The density operator is diagonal. As a result the source window is called \emph{principal}.
In general, a transcription is not unique and depends on a gauge selection but here, the operator is diagonal  and its transcription is unique likewise. 
There is nevertheless a gauge group composed of a unitary and antiunitary subgroups that leaves the density operator invariant.
The unitary gauge subgroup is $\mathrm{U}(1)\times\mathrm{U}(1)$ when $\lambda_1\ne\lambda_2$ and $\mathrm{U}(2)$ when $\lambda_1=\lambda_2=1/2$. The antiunitary subgroup corresponds to the standard complex conjugation.

The contextual distribution $\{\lambda_1,\lambda_2 \}$ is identical to the spectrum of $\rho_{\scriptscriptstyle \Lambda}$, $\Sigma_{\scriptscriptstyle \Lambda}=\{\lambda_1,\lambda_2 \}$. 
The simplicial entropy and also  the von Neumann entropy are both equal to $S = -\lambda_1\log\lambda_1-\lambda_2\log\lambda_2$.

Irrespective of the gauge, an observable in the source window $Q:\Omega\to\mathbb{R}$ is transcribed as the following diagonal operator
$$\mathsf{Q}= 
\begin{bmatrix}
  \mathrm{q}_{\omega_1}  & 0 \\
   0 & \mathrm{q}_{\omega_2}
\end{bmatrix}
$$
For instance, the observable $S_Z$ is transcribed as
\begin{equation}
\label{spin2}
\mathsf{S}_Z=
\sigma_3 =  
\begin{bmatrix}
  1  & 0 \\
   0 & -1
\end{bmatrix}
\end{equation}
where $\sigma_3$ is a Pauli matrix.

\paragraph{Changing the window.}To obtain new alternatives, we simply have to change the basis in $\mathcal{H}$. 
It turns out that the new corresponding probability problem can be simply retrieved by reverse transcription in the new basis.
In general, the new density operator is no longer diagonal in the new basis, so that the new observation window is not principal but \emph{twisted}.
Let $\ket{e_{i'}}= (\alpha_{i',1}, \alpha_{i',2})^T$ for $i'=1,2$ be the expression of its eigenvectors in the new basis. The new expression $\rho'_{\scriptscriptstyle \Lambda}=\lambda_1\ket{e_{1'}}\bra{e_{1'}} + \lambda_2\ket{e_{2'}}\bra{e_{2'}}$ of the density operator is thus
$$
\rho'_{\scriptscriptstyle \Lambda}= \lambda_1 
\begin{bmatrix}
  \alpha_{1',1}\, \alpha_{1',1}^*  & \alpha_{1',1}\, \alpha_{1',2}^* \\
   \alpha_{1',2}\, \alpha_{1',1}^* & \alpha_{1',2}\, \alpha_{1',2}^*
\end{bmatrix}
+\lambda_2 
\begin{bmatrix}
  \alpha_{2',1}\, \alpha_{2',1}^*  & \alpha_{2',1}\, \alpha_{2',2}^* \\
   \alpha_{2',2}\, \alpha_{2',1}^* & \alpha_{2',2}\, \alpha_{2',2}^*
\end{bmatrix}
=
\begin{bmatrix}
  w'_1  & \rho'_{12} \\
   \rho'_{21} & w'_2
\end{bmatrix}
$$
and we have $\tr(\rho'_{\scriptscriptstyle \Lambda})=\tr(\rho_{\scriptscriptstyle \Lambda})=w'_1+w'_2=1$. 
For example for $\ket{e_{1'}}= (\cos\theta, \sin\theta)^T$ and $\ket{e_{2'}}= (-\sin\theta, \cos\theta)^T$, we obtain
$$\rho'_{\scriptscriptstyle \Lambda}=\begin{bmatrix}
  \lambda_1\cos^2\theta +\lambda_2\sin^2\theta  & (\lambda_1-\lambda_2)\sin\theta\cos\theta \\
   (\lambda_1-\lambda_2)\sin\theta\cos\theta & \lambda_1\sin^2\theta+ \lambda_2\cos^2\theta
\end{bmatrix}.$$

To reverse transcribe into a new real-valued probability space $\mathcal{P}'$, use the eigenvectors $\ket{e_{i'}}\in\mathcal{H}$ to define the  vectors $v_i'=(|\alpha_{i',1}|^2, |\alpha_{i',2}|^2)^T$ in $\mathcal{P}'$. In the example, $w'_1= \lambda_1\cos^2\theta +\lambda_2\sin^2\theta$, $w'_2=\lambda_1\sin^2\theta+ \lambda_2\cos^2\theta $, $v_1'= (\cos^2\theta, \sin^2\theta)^T$ and $v_2'=(\sin^2\theta, \cos^2\theta)^T$


By exception, when $v'_1=v'_2$, the new window is \enquote{blind} and $w'_1=w'_2=1/2$. For example, this occurs when the new alternative describes the balance or not of the new truth table, which is obtained for e.g., $\theta=\pi/4$, $\ket{e_{1'}}= (1/\sqrt{2})(1, 1)$ and $\ket{e_{2'}}=  (1/\sqrt{2})(-1, 1)$.

Otherwise, the new simplex is the affine segment $[v'_1,v'_2]$ and the new working distribution is $w'=(w'_1, w'_2)^T$.
Finally, this defines a new sampling set $\Omega'$. Although the system is basically classical,  Bayesian inference leads to a \emph{twisted} observation window because the basis vectors are correlated and no longer independent. 

Obviously, the old observables $\Omega\to\mathbb{R}$ will change accordingly in $\mathcal{H}$ and will no longer be diagonal. Therefore, they \emph{cannot} be reverse-transcribed in the new window because they are still defined on $\Omega\ne\Omega'$. By contrast, the new window matches different observables, inaccessible from the old window, $\Omega'\to\mathbb{R}$ which became diagonal in the new window. Nevertheless, all observables can always be computed in the Hilbert space in any observation window because each observable is expressed as an operator, whether diagonal or not, acting on the Hilbert space $\mathcal{H}$. 
\paragraph{Purification.}
From Sec. (\ref{purification}), it is possible to regard the 1-bit mixed state as the partial subsystem of a pure 2-bit quantum state. Define a second 1-bit LP space $\mathcal{P}_b$ and let $\mathcal{P}_c=\mathcal{P}\otimes\mathcal{P}_b$. From Eq. (\ref{wc0}), construct the 2-bit working distribution $w_{c}=(w_{c,(\omega_i; \omega_b)})\in \mathcal{P}_c$ as
$w_{c,11}=\mu_1 ; w_{c,12}= 0; w_{c,21}=0 ; w_{c,22}=\mu_2 .$
Then $w_{\scriptscriptstyle \Lambda}$ in $\mathcal{P}$ is the marginal of $w_c$ in $\mathcal{P}_c$. Similarly, $\rho_{\scriptscriptstyle \Lambda}$ can be purified in a 4-dimensional Hilbert space as a projection operator $\ket{c}\bra{c}$ where $\ket{c}$ is defined up to a phase factor as
$$\ket{c}= \sqrt{\mu}_1 e^{i\phi}\ket{11}+\sqrt{\mu}_2 e^{-i\phi} \ket{22}$$
and where the gauge phase $\phi$ is arbitrary. Finally $\rho_{\scriptscriptstyle \Lambda}=\tr_b(\ket{c}\bra{c})$.

\subsubsection[Qubit]{Qubit, pure 1-bit state} We define a qubit as a pure state in a 1-bit LP system. For the sake of generality, assume that the source window is not necessarily principal. Define a covector $\mathrm{a}_\theta =(\mathrm{a}_{\theta, \omega_1},\mathrm{a}_{\theta, \omega_2})$  in $\mathcal{P}^*$ depending on a setting $\theta$ associated with an observable, $A_\theta$, so that $A_\theta(p)=\mathrm{a}_{\theta, \omega_1}\times p_1+\mathrm{a}_{\theta, \omega_2}\times p_2$. Without loss in generality for feasible LP problems, we can choose the following formulation of $\mathrm{a}_\theta$
 $$\mathrm{a}_\theta =(\mathrm{a}_{\theta, \omega_1},\mathrm{a}_{\theta, \omega_2})  =(\sin^2\theta/2, -\cos^2\theta/2),$$
The qubit is the unique solution of the Bayesian problem Eq. (\ref{expectation})
\begin{align*}
(\theta):\quad\mathrm{Assign~}\Prob \mathrm{~subject~to~} \langle A_\theta\rangle=0
\end{align*}
The rank of the LP system is $m=d=2$ and the solution is $w_\theta=(\cos^2\theta/2,\sin^2\theta/2)$.
The quantum state $(w_\theta,\mathcal{W}_\theta)$ is thus characterized by the isolated vertex $w_\theta$ and $\mathcal{W}_\theta=\{w_\theta\}$.
\paragraph{Observable.}
Consider an observable $Q(\omega)=\mathrm{q}_{\omega}$. The quantum expectation is defined as,
$$\langle Q \rangle=\langle \mathrm{q} w_\theta \rangle = \mathrm{q}_{\omega_1} w_{\theta,1}+ \mathrm{q}_{\omega_2}w_{\theta,2}= \mathrm{q}_{\omega_1}\cos^2\theta/2 + \mathrm{q}_{\omega_2}\sin^2\theta/2$$
Specifically, the expectation of the observable $S_Z=\sigma_3=(1,-1)$, Eq. (\ref{spin2}), is $\langle S_Z \rangle=\cos^2\theta/2-\sin^2\theta/2 = \cos\theta$.

\paragraph{Transcription into $\mathcal{H}$.}
The Hilbert space is still the complex span of $(\omega_1, \omega_2)$. 
As a pure state, the effective unitary gauge subgroup is $\mathcal{G}_{\mathrm{eff}}=\mathrm{U}(1)$ (Definition \ref{eff}).
Consider a gauge labelled $\phi$  defined by the gauge operator $\mathsf{G}=Diag(e^{i\phi/2}, e^{-i\phi/2})$. 
With this gauge, the quantum state is transcribed as the rank 1 density operator $\rho_{\theta,\phi}=\ket{a}\bra{a}$ with the  Gleason's vector  (Definition \ref{defgleason}),
$$\ket{a}=D_{11}\sqrt{w}_{\theta,1} \cdot\ket{1}+ D_{22}\sqrt{w}_{\theta,2}\cdot\ket{2}=e^{i\phi/2}\cos\theta/2\cdot\ket{1} +e^{-i\phi/2}\sin\theta/2\cdot\ket{2}$$
 as
$$
\rho_{\theta,\phi} =\ket{a}\bra{a}=  
\frac{1}{2}
\begin{bmatrix}
  1+\cos\theta  & e^{-i\phi}\sin\theta \\
   e^{i\phi}\sin\theta & 1-\cos\theta
\end{bmatrix}
$$
There is also a antiunitary gauge subgroup generated by complex conjugation, that is simply here $\phi\mapsto-\phi$.
\paragraph{Mutually unbiased bases.}
Consider the three unitary matrices
$$ 
\mathsf{U}_1=
\begin{bmatrix}
  1  & 0 \\
   0 & 1
\end{bmatrix} 
;\quad\mathsf{U}_2=\frac{1}{\sqrt{2}}
\begin{bmatrix}
  1  & ~~1 \\
   1 & -1
\end{bmatrix} 
;\quad\mathsf{U}_3=\frac{1}{\sqrt{2}}
\begin{bmatrix}
  1  & ~~1 \\
   i & -i
\end{bmatrix}
. 
$$
The two column vectors of each matrix define a basis.
The identity matrix $\mathsf{U}_1$ depicts the initial basis and  $\mathsf{U}_2$,  $\mathsf{U}_3$ are two rescaled complex Hadamard matrices~\cite{tadej}.  Therefore, the three bases are mutually unbiased (MUB). In each basis $\mathsf{U}_i$, the Gleason's vector of the density operator is $\ket{a_i}=\mathsf{U}_i^{-1}\ket{a}$. Select the natural gauge ($\phi=0$) for simplicity. Then
$$ 
\ket{a_1}=
\begin{bmatrix}
  \cos \theta/2 \\
  \sin \theta/2
\end{bmatrix}
;\quad\ket{a_2}=\frac{1}{\sqrt{2}}
\begin{bmatrix}
  \cos \theta/2 +\sin \theta/2\\
  \cos \theta/2 -\sin \theta/2
\end{bmatrix}
;\quad\ket{a_3}=\frac{1}{\sqrt{2}}
\begin{bmatrix}
  e^{-\mathbf{i}\theta/2}\\
  e^{+\mathbf{i}\theta/2}
\end{bmatrix}
$$
By reverse transcription, the working distributions $w_i$ read (naturally irrespective of the gauge)
$$
w_1=\frac{1}{2}(1+\cos \theta,1-\cos\theta)
;\quad w_2=\frac{1}{2}(1+\sin \theta,1-\sin\theta)
;\quad w_3=\frac{1}{2}(1,1)
$$

The window entropies are respectively 
\begin{align*}
&\mathbb{H}_1=-\frac{1+\cos \theta}{2}\log \frac{1+\cos \theta}{2}-\frac{1-\cos \theta}{2}\log \frac{1-\cos \theta}{2}\\
&\mathbb{H}_2=-\frac{1+\sin \theta}{2}\log \frac{1+\sin \theta}{2}-\frac{1-\sin \theta}{2}\log \frac{1-\sin \theta}{2}\\
&\mathbb{H}_3=1~\mathrm{bit}
\end{align*}

At last, as a pure state the von Neumann entropy $S(\rho_{\theta,\phi})=0$ is zero and we have in accordance with Eq. (\ref{infoMUB}) (where $N=1$ and $d=2^N=2$) 
$$\mathbb{H}_1+\mathbb{H}_2+\mathbb{H}_3\ge2$$
For instance, for $\theta=\pi/4$, we have $\mathbb{H}_1+\mathbb{H}_2+\mathbb{H}_3=2.125$ bits. In other words, the excess entropy of the MUB cluster is $\Delta_\mathbb{H} = 0.125$ bit. By contrast, if $\theta=0$, the first window is principal. Then $\mathbb{H}_1=0$, $\mathbb{H}_2=\mathbb{H}_3=1$ bit and therefore $\Delta_\mathbb{H}=0$. The cluster is centered.

Still when $\theta=0$, the full information is concentrated in the first window. In the two unbiased windows, the density operators are $\rho_2=\rho_3 = (1/2)\times\mathds{1}_2$. They depict completely random distributions.

\paragraph{Principal window.}
A principal window is obtained from the initial basis by diagonalization with a unitary operator, $\mathsf{U}$.
Irrespective of the gauge, the principal density operator,  $\rho_Z$ reads
$$
\rho_Z =  \mathsf{U}\ \rho_{\theta,\phi}\ \mathsf{U}^\dagger = 
\begin{bmatrix}
  1  & 0 \\
   0 & 0
\end{bmatrix}
$$
A pure state is explicitly a deterministic state in its principal window.
In the Bloch representation, the gauge group describes an axial symmetry around the axis $Z=(0,0,1)$.

\subsection{Two-bit system} 
\label{bipartite}
We come to a two-bit system.

\subsubsection{Mixed states} 
Consider a system of two bits $\mathsf{X}_1$ and $\mathsf{X}_2$ without any other constraint describing the simplest  LP problem. 
The prior is reduced to $$(\Lambda)\ident\{ N=2\}.$$

There are 8 unknowns, namely $\Prob(\pm 1)$, $\Prob(\pm 2)$  $\Prob(\pm 1; \pm 2)$. In order to describe a probability distributions, these unknowns are subject to the relevant universal equations, Eqs. (\ref{simplet}, \ref{doublet}, \ref{triplet}, etc.). Here, we have
\begin{align}
\label{norm1} &\Prob(1) + \Prob(-1)=1\\
\label{norm2} &\Prob(2) + \Prob(-2)=1\\
\label{consis12} &\Prob(\pm 1)=\Prob(\pm 1;2)+\Prob(\pm 1;-2)\\
\label{consis21} &\Prob(\pm 2)=\Prob(1;\pm 2)+\Prob(-1;\pm 2)\\
\mathrm{subject~to~}&
\Prob(\pm1) \ge 0 ; \Prob(\pm 2)\ge 0;\Prob(\pm 1;\pm 2)\ge 0.
\end{align}
Eqs. (\ref{norm1}, \ref{norm2}) provide normalization while Eqs. (\ref{consis12}, \ref{consis21}) ensure the overall consistency.

It is easy to eliminate the unknowns $\Prob(1), \Prob(-1)$, $\Prob(2), \Prob(-2)$  involving only one literal. 
The sample set $\Omega =\{ \omega_i |\  i\in\llbracket 1, 4\rrbracket\}$ comprises four classical states, $\omega_1=(\overline{\mathsf{X}}_1;\overline{\mathsf{X}}_2)$, $\omega_2=(\overline{\mathsf{X}}_1;{\mathsf{X}}_2)$, $\omega_3=({\mathsf{X}}_1;\overline{\mathsf{X}}_2)$ and $\omega_4=({\mathsf{X}}_1;{\mathsf{X}}_2)$.
Let $\mathcal{P}\ident Span( \omega_i\  |\  i\in\llbracket 1, 4\rrbracket)$ denote the real-valued probability space of dimension $d=2^N=4$ and let $p_i=\Prob(\omega_i)$.
The LP system of rank $m=1$, Eq. (\ref{lpeq}), reads
\begin{align}
\label{twobitstate}
\begin{aligned}
p_1+p_2+p_3+p_4&=1\\
\mathrm{subject~to~}p&\ge 0.
\end{aligned}
\end{align}
In $\mathcal{P}$, there is a continuous set of feasible distributions located on the tautological simplex of two variables. 
From LP theory~\cite{murty}, there are trivially $r=d-m+1=4$ deterministic extreme points, namely, $w_i=\tilde{\omega}_i$ with $p_i = 1$ (for $i=1, 2, 3, 4)$ corresponding to the basic vectors of $\mathcal{P}$.  Deterministic states are \emph{separable}~\cite{mf3}, that is $\Prob(\pm 1; \pm2) = \Prob(\pm 1)\times\Prob(\pm 2)$.
Other solutions, depending on 3 independent parameters, are non-deterministic. 
The working distribution $w_{\scriptscriptstyle \Lambda}=\sum_{i=1}^4 \mu_i\,w_i$ is specified by four barycentric coordinates summing to 1, $\mu_1$, $\mu_2$, $\mu_3$, and $\mu_4$, which define the \enquote{context}. 
Each solution is specified by a simplicial quantum state, $(w_{\scriptscriptstyle \Lambda}, \mathcal{W}_{\scriptscriptstyle \Lambda})$, where $\mathcal{W}_{\scriptscriptstyle \Lambda}=\mathrm{conv}(w_i)$ is the tautological simplex, $\mathrm{conv}(\tilde{\omega}_i)$.

\paragraph{Default context.}
The default context is the completely random state in which the working distribution $w_{\scriptscriptstyle \Lambda}$ is the center of mass $\tilde{c}$ of the simplex with $\mu_1=\mu_2=\mu_3=\mu_4=1/4$. Both its window entropy and its simplicial entropy are equal to 2 bits.

The transcription into a Hilbert space is straightforwards. The density operator is the random matrix $\frac{1}{4}\mathds{1}_4$ and the gauge group is $\mathcal{G}=\mathrm{U}(4)$.

\paragraph{Partial subsystems.}
Partial subsystems depict the restriction of the full constraints to just one bit.
For convenience, rename $\mathsf{X}_1=\mathsf{X}_a$, $\mathsf{X}_2=\mathsf{X}_b$ and $\mathcal{P}=\mathcal{P}_c$. The probability space $\mathcal{P}_c$ is the Kronecker product of the two probability spaces  $\mathcal{P}_a$ and  $\mathcal{P}_b$, i.e., $\mathcal{P}_c=\mathcal{P}_a\otimes\mathcal{P}_b$. 
In addition let $\tilde{\omega}_{ai}$ with $i\in\llbracket 1,2\rrbracket$,  $\tilde{\omega}_{bi}$ with $i\in\llbracket 1,2\rrbracket$ and  $\tilde{\omega}_{ci}$ with $i\in\llbracket 1,4\rrbracket$ denote the bases in $\mathcal{P}_a$,  $\mathcal{P}_b$ an  $\mathcal{P}_c$ respectively, where $\tilde{\omega}_{c1} = \tilde{\omega}_{a1}\otimes\tilde{\omega}_{b1}$, $\tilde{\omega}_{c2} = \tilde{\omega}_{a2}\otimes\tilde{\omega}_{b1}$, $\tilde{\omega}_{c3} = \tilde{\omega}_{a1}\otimes\tilde{\omega}_{b2}$ and $\tilde{\omega}_{c4} = \tilde{\omega}_{a2}\otimes\tilde{\omega}_{b2}$.

Consider the reduction in $\mathcal{P}_a$ and  $\mathcal{P}_b$ of the simplicial quantum state $(w_{\scriptscriptstyle{\Lambda}},\mathcal{W}_{\scriptscriptstyle \Lambda})$, renamed $(w_{c},\mathcal{W}_{c})$, where $\mathcal{W}_{c}=\mathrm{conv}(w_{ci})$ is the tautological simplex in $\mathcal{P}_c$, $w_{ci}=\tilde{\omega}_{ci}$ are its vertices and $w_c=\sum_{i=1}^4 \mu_i\tilde{\omega}_{ci}$ is the working distribution.
  Assume that the four simplicial coefficients, $\mu_i$, are arbitrary. 
 Every vertex, e.g.,  $w_{c2}=\tilde{\omega}_{c2}= \tilde{\omega}_{a2}\otimes\tilde{\omega}_{b1}$, defines a deterministic and thus separable distribution, e.g., $ \Prob_{c2}(\omega_{c})=\Prob_{a2}(\omega_{a})\times\Prob_{b1}(\omega_{b})$, so that the \emph{simplex} $\mathcal{W}_{c}$ is \enquote{separable}, Definition (\ref{separablesimplex}). 
 
 The LP system in  $\mathcal{P}_c$ is just defined by Eq. (\ref{twobitstate}), that is the tautology $I_4$. As a result, the LP system in  $\mathcal{P}_a$ is defined by the marginal of $I_4$, that is the tautology $I_2$.
Let $w_a$ and $\tilde{a}$ denote the marginal of $w_c$ and $\tilde{c}$ in $\mathcal{P}_a$ respectively.
From Eq. (\ref{marginalpure}) and  Proposition (\ref{partialLP}), they read
\begin{align}
\label{marginal2bit}
\begin{aligned}
w_{a}&=\sum_{i=1}^2\sum_{j=1}^2 w_{c,(\omega_{ai};\omega_{bj})}\,\tilde{\omega}_{ai} \\
&= (\mu_1+\mu_3)\tilde{\omega}_{a1}+(\mu_2+\mu_4)\tilde{\omega}_{a2}\\
\tilde{a}&= \frac{1}{2}( \tilde{\omega}_{a1}+\tilde{\omega}_{a2})
\end{aligned}
\end{align}
In $\mathcal{P}_a$, the tautological simplex $\mathcal{W}_{a}=\mathrm{conv}(\tilde{\omega}_{ai})$ is the specific simplex of a simplicial quantum state,  $(w_a, \mathcal{W}_{a})$. It can be regarded as the reduced state in $\mathcal{P}_a$ of the mixed simplicial quantum state $(w_c, \mathcal{W}_{c})$. The marginal $\tilde{a}$ of the center $\tilde{c}$ of $\mathcal{W}_{c}$ is identical to the center $c_a$ of $\mathcal{W}_{a}$.
The same procedure can be used in  $\mathcal{P}_b$ yielding a simplicial quantum state   $(w_b, \mathcal{W}_{b})$ with 
$
w_{b}= (\mu_1+\mu_2)\tilde{\omega}_{b1}+(\mu_3+\mu_4)\tilde{\omega}_{b2}.
$
The global simplicial quantum \emph{state} $(w_c, \mathcal{W}_{c})$ is \enquote{separable}, Definition (\ref{separablesimplicial}).
%

\subsubsection{Singlet state}
\label{purebipartite} 

Consider a 2-bit system subject to the logical constraint,
$$\mathsf{X}_1= \overline{\mathsf{X}}_2,$$
and an additional condition of symmetry, namely in terms of probability, that  $(\mathsf{X}_1=1)$ and $(\mathsf{X}_2=1)$ are equally likely.
\paragraph{LP system in $\mathcal{P}$.}
The hypotheses are translated into the following {specific}  constraints
\begin{equation}
\label{singlet}
\Prob(1;2)=\Prob(-1;-2)=0
\quad;\quad
\Prob(1)=\Prob(2).
\end{equation}
The LP problem comprises the previous universal equations, Eq. (\ref{norm1}$-$\ref{consis21}) together with the specific constraints Eq.  (\ref{singlet}). 
Eliminate $\Prob(\pm 1)$, $\Prob(\pm 2)$ using Eqs.  (\ref{consis12}, \ref{consis21}). 
Define the usual basis $\tilde{\Omega}$ in $\mathcal{P}$.
Now the LP system, Eq.  (\ref{lpeq}) reads
\begin{align}
\label{lpsinglet} 
\begin{aligned}
p_1+p_2+p_3+p_4&=1\\ 
p_1&=0\\
p_4&=0\\
p_1+p_2-p_3-p_4&=0\\
\mathrm{subject~to~} p&\ge 0.
\end{aligned}
\end{align}
The unique solution is
$$p_2=p_3=\frac{1}{2}\quad;\quad p_1=p_4=0$$
Therefore, the solution is a pure simplicial quantum $(w_{\scriptscriptstyle \Lambda},\{w_{\scriptscriptstyle \Lambda}\})$ state with the working distribution $w_{\scriptscriptstyle \Lambda}=(0, 1/2, 1/2, 0)$.
The effective probability space is $\mathbb{W}_1=\mathrm{Span}(w_{\scriptscriptstyle \Lambda})$

\paragraph{Partial subsystems.}
The singlet state is notoriously entangled (or twisted from Definition \ref{separablesimplicial}). Therefore, its marginalization leads to two 1-bit mixed states.  
For convenience, rename again $\mathsf{X}_1=\mathsf{X}_a$, $\mathsf{X}_2=\mathsf{X}_b$, $\mathcal{P}=\mathcal{P}_c$, $\mathcal{P}_c=\mathcal{P}_a\otimes\mathcal{P}_b$ and let $\tilde{\omega}_{a_i}$ with $i\in\llbracket 1,2\rrbracket$,  $\tilde{\omega}_{b_i}$ with $i\in\llbracket 1,2\rrbracket$ and  $\tilde{\omega}_{c_i}$ with $i\in\llbracket 1,4\rrbracket$ denote the bases in $\mathcal{P}_a$,  $\mathcal{P}_b$ an  $\mathcal{P}_c$ respectively, where $\tilde{\omega}_{c_1} = \tilde{\omega}_{a_1}\otimes\tilde{\omega}_{b_1}$, $\tilde{\omega}_{c_2} = \tilde{\omega}_{a_2}\otimes\tilde{\omega}_{b_1}$, $\tilde{\omega}_{c_3} = \tilde{\omega}_{a_1}\otimes\tilde{\omega}_{b_2}$ and $\tilde{\omega}_{c_4} = \tilde{\omega}_{a_2}\otimes\tilde{\omega}_{b_2}$. 
In addition, rename  $p_{a_1b_1}$,  $p_{a_1b_2}$,  $p_{a_2b_1}$ and  $p_{a_2b_2}$ the current coordinates $p_1$, $p_2$, $p_3$ and $p_4$ in $\mathcal{P}_c$. 
The LP system Eq. (\ref{lpsinglet}) is rewritten  as
\begin{align*}
\begin{aligned}
p_{a_1b_1}+p_{a_1b_2}+p_{a_2b_1}+p_{a_2b_2}=1
\ ;\ 
p_{a_1b_1}+p_{a_1b_2}-p_{a_2b_1}-p_{a_2b_2}=0
\ ;\
p_{a_1b_1}=p_{a_2b_2}=0
\end{aligned}
\end{align*}
The working distribution $w_{\scriptscriptstyle \Lambda}$ is renamed $w_{c}=(0, 1/2, 1/2, 0)$, that is
$$w_{c}= \frac{1}{2}\Big(\tilde{\omega}_{a_1}\otimes\tilde{\omega}_{b_2}+\tilde{\omega}_{a_2}\otimes\tilde{\omega}_{b_1}\Big) $$

The marginal of $w_c$  is $w_a =(w_{a,a_1},w_{a,a_2})$, with $ w_{a,a_1} =\Prob_c(\omega_{a_1};\omega_{b_1}) +\Prob_c(\omega_{a_1};\omega_{b_2})= 1/2$ and similarly $w_{a,a_2}=1/2$,  
$$w_{a}= \frac{1}{2}\Big(\tilde{\omega}_{a_1}+\tilde{\omega}_{a_2}\Big) =\Big( \frac{1}{2},  \frac{1}{2}\Big). $$
Now, the simplex $\mathcal{W}_a$ is the convex hull of $\tilde{v}_{\omega_{b_1}}$ and $\tilde{v}_{\omega_{b_2}}$ where
$v_{\omega_{b_1},\omega_{a_1}}=\Prob_c(\omega_{a_1}|\omega_{b_1})={ {\Prob_c(\omega_{a_1};\omega_{b_1})} }/{\Prob_c({\omega_{b_1}})}=0$ and similarly, $v_{\omega_{b_1},\omega_{a_2}}=1$, $v_{\omega_{b2},\omega_{a_1}}=1$ and $v_{\omega_{b_2},\omega_{a_2}}=0$, so that the vertices are the basic vectors $\tilde{\omega}_{a_1}$ and $\tilde{\omega}_{a_2}$. The specific simplex is again the tautological simplex in $\mathcal{P}_a$ and the rank is $r_a=2$. 
 $$\mathcal{W}_a=\mathrm{conv}(\tilde{\omega}_{a_1},\tilde{\omega}_{a_2})$$
By symmetry, the same results are obtained in $\mathcal{P}_b$ by permuting the indexes $a$ and $b$.

\paragraph{Transcription into $\mathcal{H}$.}
The transcription into a Hilbert space is straightforward.
Resume the initial notations.
 Let $\mathcal{H}$ denote the 2-bit Hilbert space spanned by the orthonormal basis $\ket{\overline{1};\overline{2}}$, $\ket{\overline{1};{2}}$, $\ket{{1};\overline{2}}$, $\ket{{1};{2}}$ (where $\mathsf{X}_i$ is replaced by $i$ for simplicity).
As a pure state, the effective unitary gauge subgroup is $\mathcal{G}_\mathrm{eff}=\mathrm{U}(1)$ and   
we have only one significant gauge phase in the current window, say $\phi$. Then, the pure state in $\mathcal{H}$ is
$$\rho=\ket{e}\bra{e}\quad\mathrm{where}\quad \ket{e}=\frac{1}{\sqrt{2}}(\ket{\overline{1}{2}}-e^{i\phi}\ket{{1}\overline{2}})$$
so that
$$
\rho=\ket{e}\bra{e} =\frac{1}{2}
\begin{bmatrix}
   0 & 0 & 0 & 0\\
   0 & 1 & a & 0\\
   0 & a^* & 1 & 0\\
   0 & 0 & 0 & 0
\end{bmatrix}
\quad \mathrm{with~} a = -e^{i\phi}
$$
We recover the singlet state of standard quantum mechanics.
There is also a antiunitary gauge subgroup generated by standard complex conjugation, that is here the swap of $a$ and $a^*$.
Incidentally, a unitary gauge operator acting on $\mathcal{H}$ can be interpreted as a rotation in the Bloch representation of the qubits so that the singlet state is isotropic in this Bloch space, while the antiunitary operator corresponds to a discrete mirror symmetry.

The singlet state has been defined in both $\mathcal{P}$ and $\mathcal{H}$. Therefore, it can perfectly be emulated in the classical realm. A possible implementation is proposed in Ref.~\cite{mf2}. 

\paragraph{}In a principal window, $\ket{e} = [1  , 0 , 0 , 0]^T$ and the density operator is
$\rho=\ket{e}\bra{e} $. 
The Hilbert space is the direct sum of two eigensubspaces of dimension 1 and 3 respectively and thus the full unitary gauge subgroup group is the direct product $\mathcal{G}=\mathrm{U}(1)\times \mathrm{U}(3)$, while the effective unitary gauge subgroup is $\mathcal{G}_\mathrm{eff}=\mathrm{U}(1)$.

\subsubsection{Triplet state}
\label{tripletstate}
Relax the strict constraint on the singlet state $\mathsf{X}_1= \overline{\mathsf{X}}_2,$ as just its average, that is 
$\langle \mathsf{X}_1\rangle=\langle \overline{\mathsf{X}}_2\rangle$. 
This is immediately translated as
\begin{equation}
\label{Bell2} 
\Prob(1)=\Prob(-2). 
\end{equation}
The LP problem comprises the previous universal equations, Eq. (\ref{norm1}-\ref{consis21}) together with this new specific  constraint Eq. (\ref{Bell2}).

\paragraph{LP system in $\mathcal{P}$.}
Eliminate $\Prob(\pm 1)$, $\Prob(\pm 2)$ using Eqs. (\ref{consis12}, \ref{consis21}). 
 We obtain the LP system in $\mathcal{P}$, Eq.  (\ref{lpeq}), as
\begin{align*}
  p_1+p_2+p_3+p_4&=1\\
  p_1- p_4 &=0\\
\mathrm{subject~to~} p&\ge 0.
\end{align*}
The rank of the LP system is $m=2$.
Equivalently, the LP system is specified by the expectation of the observable $A(\omega)$ defined by the covector  $\mathrm{a}=(1, 0, 0, -1)$.  The Bayesian formulation is thus
$$(\Lambda) : \mathrm{~Assign~}\Prob \mathrm{~subject~to~}\langle A \rangle=0.$$

\paragraph{Specific polytope $\mathcal{W}_{\scriptscriptstyle \Lambda}$.}
The specific polytope is actually a simplex with $3$ vertices, say $w_1$,  $w_2$ and $w_3$ in the 4-D probability space $\mathcal{P}$.
To allow easy viewing, it is possible to eliminate  $p_1=\Prob(-1;-2)$. We obtain the equivalent LP system in a new 3-D space $\mathcal{P}'$ as,

\begin{minipage}{0.35\linewidth}
\begin{tikzpicture}[scale=2]
\fill [gray, opacity=0.4](0,0.5,0) -- (1, 0,0) -- (0, 0,1) -- cycle ;
\draw (0,0.5,0) -- (1, 0,0) -- (0, 0,1) -- cycle ;
\draw(0, 0,0) -- (1.2, 0,0);\draw[->, >=latex,very thick] (1.1,0,0)--(1.2, 0,0); 
\draw (0, 0,0) -- (0, 1.2,0);\draw [->, >=latex,very thick](0, 1.1,0) -- (0, 1.2,0); 
\draw (0, 0,0) -- (0, 0,1.3);\draw [->, >=latex,very thick](0, 0,1.2) -- (0, 0,1.3); 
\draw (0, 0,0) node[below]{0};
\draw (0, 0,1) node{$\bullet$};
\draw (1, 0,0) node{$\bullet$};
\draw (1, 0,0) node[below]{1};
\draw (-0.1, 0,1) node [left]{1};
\draw (0, 1, 0) node[left]{1};
\draw (0, 1, 0) node{-};
\draw (0, 0.5, 0) node[left]{0.5};
\draw (0.1, 0.5, 0) node[right]{$w_3'$}; 
\draw (0, 1.2, 0) node[above]{$p_4$};
\draw (1.1, 0.2, 0) node {$w_2'$};
\draw (1.2, -0.1, 0) node {$p_3$};
\draw (0, 0, 1.2) node[below]{$p_2$};
\draw (0.25, 0, 1.1) node{$w_1'$};
\end{tikzpicture}
\end{minipage}
\begin{minipage}{0.6\linewidth}
\begin{align*}
 p_2+p_3+2p_4&=1\\
\mathrm{subject~to~}p&\ge 0.
\end{align*}
The new specific polytope has still three vertices, $w_1'$,  $w_2'$ and $w_3'$, and a continuous set of solutions.  The feasible solutions are located on a triangle  $\mathrm{conv}( w'_1,w'_2,w'_3).$  Two extreme solutions are deterministic, i.e, $w'_1$ and $w'_2$. Alternatives are non deterministic in this window.

Returning to the 4-D vector space $\mathcal{P}$, the extreme points of the simplex, $w_i$ are therefore $w_1=(0,1,0, 0)$, $w_2=(0, 0, 1, 0)$ and $w_3=(1/2, 0, 0, 1/2)$.
While $w_1$ and $w_2$ are deterministic and thus separable, it can be seen that $w_3$ is actually entangled.
\end{minipage}
By default, the working distribution is the point of maximum simplicial entropy, i.e., the center of mass of the polytope $c=(1/3) (w_1+w_2+w_3)=(1/6$, $1/3$, $1/3$, $1/6)$.

Otherwise, we can specify freely a particular simplicial distribution on the vertices, as 
$$\Sigma_\mu=\{\mu_1,\mu_2, \mu_3, \}\quad\mathrm{where~} \mu_1,\mu_2, \mu_3,\ge 0\quad\mathrm{and~} \mu_1+\mu_2+ \mu_3=1,$$
The working distribution is then
\begin{equation}
\label{wl}
w_{\scriptscriptstyle \Lambda}= \sum_{i=1}^3 \mu_i w_i = \Big(\frac{\mu_3}{2}, \mu_1, \mu_2, \frac{\mu_3}{2}\Big)
\end{equation}
\paragraph{Partial subsystems.}  
For the sake of convenience, rename again in this section $\mathsf{X}_1=\mathsf{X}_a$, $\mathsf{X}_2=\mathsf{X}_b$ and $\mathcal{P}=\mathcal{P}_c$ and let $\mathcal{P}_c=\mathcal{P}_a\otimes\mathcal{P}_b$. 
In addition let $\tilde{\omega}_{ai}$ with $i\in\llbracket 1,2\rrbracket$,  $\tilde{\omega}_{bi}$ with $i\in\llbracket 1,2\rrbracket$ and  $\tilde{\omega}_{ci}$ with $i\in\llbracket 1,4\rrbracket$ denote the bases in $\mathcal{P}_a$,  $\mathcal{P}_b$ an  $\mathcal{P}_c$ respectively, where $\tilde{\omega}_{c1} = \tilde{\omega}_{a1}\otimes\tilde{\omega}_{b1}$, $\tilde{\omega}_{c2} = \tilde{\omega}_{a2}\otimes\tilde{\omega}_{b1}$, $\tilde{\omega}_{c3} = \tilde{\omega}_{a1}\otimes\tilde{\omega}_{b2}$ and $\tilde{\omega}_{c4} = \tilde{\omega}_{a2}\otimes\tilde{\omega}_{b2}$. The rank $r$ is renamed $r_c=3$ and the working distribution $w_{\scriptscriptstyle \Lambda}$ is renamed $w_{c}$, that is
\begin{equation}
\label{tripletwc}
w_{c}=
\frac{\mu_3}{2}\,\tilde{\omega}_{a1}\otimes\tilde{\omega}_{b1}+\mu_1\, \tilde{\omega}_{a2}\otimes\tilde{\omega}_{b1} +\mu_2\,\tilde{\omega}_{a1}\otimes\tilde{\omega}_{b2} +\frac{\mu_3}{2}\,\tilde{\omega}_{a2}\otimes\tilde{\omega}_{b2}, 
\end{equation}
while the mass center $c=(1/3) (w_1+w_2+w_3)$  reads
$$
c=\frac{1}{6}\,\tilde{\omega}_{a1}\otimes\tilde{\omega}_{b1}+ \frac{1}{3}\,\tilde{\omega}_{a2}\otimes\tilde{\omega}_{b1} +\frac{1}{3}\,\tilde{\omega}_{a1}\otimes\tilde{\omega}_{b2} +\frac{1}{6}\,\tilde{\omega}_{a2}\otimes\tilde{\omega}_{b2}.
$$
The marginal $w_a =(w_{a,a1},w_{a,a2})$ of $w_c$  in $\mathcal{P}_a$ is easily computed as $ w_{a,a1}=\Prob_c(\omega_{a1}) =\Prob_c(\omega_{a1};\omega_{b1}) +\Prob_c(\omega_{a1};\omega_{b2})$ and similarly for $w_{a,a2} $ and reads,
\begin{equation}
\label{tripletwa}
w_{a}=\Big(\mu_2+\frac{\mu_3}{2}\Big)\,\tilde{\omega}_{a1}+\Big(\mu_1+\frac{\mu_3}{2}\Big)\,\tilde{\omega}_{a2},
\end{equation}
In particular, for $\mu_1=\mu_2=\mu_3=1/3$, the marginal $\tilde{a}\in\mathcal{P}_a$ of the center of mass, $\tilde{c}$, reads
\begin{equation}
\label{tripletmarginalc}
\tilde{a}=\frac{1}{2}\,\tilde{\omega}_{a1}+\frac{1}{2}\,\tilde{\omega}_{a2}
\end{equation}
and therefore, it is the mass center of the tautological simplex $\mathcal{W}_{Ia}=\mathrm{conv}(\tilde{\omega}_{a1},\tilde{\omega}_{a2})$ in $\mathcal{P}_a$.

On the other hand, from Proposition (\ref{partialLPpure}), since  $w_1$ and $w_2$ are separable while $w_3$ is entangled, the partial states of the three extreme points regarded as pure states in isolation $w_1=\tilde{\omega}_{a2}\otimes\tilde{\omega}_{b1}$, $w_2=\tilde{\omega}_{a1}\otimes\tilde{\omega}_{b2}$ and $w_3=(1/2)(\tilde{\omega}_{a1}\otimes\tilde{\omega}_{b1}+\tilde{\omega}_{a2}\otimes\tilde{\omega}_{b2})$ are respectively 
the two pure states $v_{a1}=\tilde{\omega}_{a2}$ and  $v_{a2}=\tilde{\omega}_{a1}$ 
and the simplex $\mathrm{conv}(\tilde{\omega}_{a1},\tilde{\omega}_{a2})$.
%
Finally, Eq. (\ref{tripletwa}) defines directly a simplicial quantum state $(w_a,\mathcal{W}_a)$ in $\mathcal{P}_a$ whose partial simplex is
$$\mathcal{W}_a = \mathrm{conv}(\tilde{\omega}_{a1},\tilde{\omega}_{a2}).$$
From Eq. (\ref{tripletmarginalc}), its center of mass $c_a=(1/2, 1/2)$ is identical to the marginal $\tilde{a}$ of the mass center $\tilde{c}\in\mathcal{P}_c$ although $w_3$ is entangled.

A similar result holds in  $\mathcal{P}_b$ by permuting the indexes $a$ and $b$.

\paragraph{Transcription into $\mathcal{H}$.}
The transcription is straightforward. In a convenient gauge,  the vertex $w_1=(0,1,0, 0)$ is transcribed as $\ket{a_1}\bra{a_1}$ with $\ket{a_1}=(0,1,0, 0)$,  $w_2=(0,0,1, 0)$ is transcribed as $\ket{a_2}\bra{a_2}$ with $\ket{a_2}=(0,0,1, 0)$ and $w_3=(1/2, 0, 0, 1/2)$ is transcribed as $\ket{a_3}\bra{a_3}$ with $\ket{a_3}=(1/\sqrt{2})(e^{i\phi/2},0,0, e^{-i\phi/2})$, so that, irrespective of $\phi$, $\ket{a_1}$, $\ket{a_2}$ and $\ket{a_3}$ are orthonormal. We obtain
$$ 
\rho =
\mu_1
\begin{bmatrix}
  0  & 0 & 0 & 0\\
   0 & 1 & 0 & 0\\
   0 & 0 & 0 & 0\\
   0 & 0 & 0 & 0
\end{bmatrix}
+
\mu_2
\begin{bmatrix}
  0 & 0 & 0 & 0\\
   0 & 0 & 0 & 0\\
   0 & 0 & 1 & 0\\
   0 & 0 & 0 & 0
\end{bmatrix}
+
\frac{\mu_3}{{2}}
\begin{bmatrix}
  1  & 0 & 0 & e^{-i\phi}\\
   0 & 0 & 0 & 0\\
   0 & 0 & 0 & 0\\
   e^{i\phi} & 0 & 0 & 1
\end{bmatrix}
$$
By diagonalization, we obtain $\rho^{(0)}=Diag(\mu_1,\mu_2,\mu_3,0)$.
The Hilbert space is the direct sum of four subspaces of dimension 1 and thus the effective unitary gauge subgroup is the direct product $\mathcal{G}_\mathrm{eff}=\mathrm{U}(1)\times \mathrm{U}(1)\times \mathrm{U}(1)$.

%
\paragraph{EPR pair.} It is possible to single out a particular solution $(\mu_1,\mu_2,\mu_3)$ to obtain a pure state. For instance, consider a setting $\theta$ and set
$$ \mu_1=\mu_2=(1/2) \cos^2\theta/2 \quad;\quad \mu_3= \sin^2\theta/2.$$
\emph{Local settings.}
We can regard $\theta$ as a global setting and put $\theta=\theta_1-\theta_2$, where $\theta_1$ and $\theta_2$  are considered local settings associated with the sub-registers $\mathsf{X}_1$ and $\mathsf{X}_2$ respectively. Then, from Eq. (\ref{wl}) and standard trigonometric identities, the entries $w_{{\scriptscriptstyle \Lambda},i}$ of the working distribution $w_{\scriptscriptstyle \Lambda}$ read,
\begin{center}
\begin{tabular}{CCCCCC}
 w_{{\scriptscriptstyle \Lambda},1} &=& (1/4)(1-\cos(\theta_1-\theta_2); &
 w_{{\scriptscriptstyle \Lambda},2}  &=& (1/4)(1+\cos(\theta_1-\theta_2);\\
 w_{{\scriptscriptstyle \Lambda},3}  &=& (1/4)(1+\cos(\theta_1-\theta_2);&
 w_{{\scriptscriptstyle \Lambda},4}  &=& (1/4)(1-\cos(\theta_1-\theta_2).
\end{tabular}
\end{center}
This is exactly the joint probability distributions corresponding to an EPR pair of spins. For instance if $\theta_1 = \theta_2$, the spins are opposed.

\emph{Contextual measurement.} In standard quantum mechanics, it is accepted that even a pure state depicts a random process.  This assumption is implicit in all experiments checking  the violation of Bell's inequality. In the present model, this means identifying Bayesian estimation with a conventional probability problem. However, since the system is context-dependent, a mere random drawing is inconsistent and the trial requires a stage of classical communication~\cite{mf,mf2}. Precisely, the trial must be unique and then has to be deported in a common site, e.g., at the boundary of the two regions, say Alice and Bob regions. 

A consistent process can be the following:
Alice and Bob have the opportunity to select freely the setting they want, $\theta_1$ and $\theta_2$ respectively. Whenever they want, they independently  send their choice  to the common trial site by  classical communication:

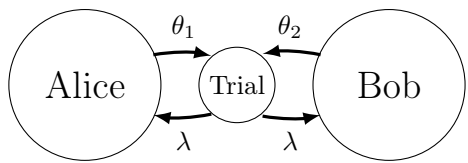
\begin{minipage}{0.45\linewidth}
\begin{tikzpicture}[scale=1]
\draw (-2,0) circle(1);
\draw (-2,0) node {\Large Alice};
\draw (0,0) circle(0.5);
\draw (0,0) node {Trial};
\draw (2,0) circle(1);
\draw (2,0) node {\Large Bob};
\draw [->, >=latex,very thick](-1.1,0.4) arc (100:78:2);
\draw (-.7, 0.5) node[above]{$\theta_1$};
\draw [<-, >=latex,very thick](-1.1,-0.4) arc (-100:-78:2);
\draw (-.7, -0.5) node[below]{$\lambda$};
\draw [->, >=latex,very thick](1.1,0.4) arc (78:100:2);
\draw (.7, 0.5) node[above]{$\theta_2$};
\draw [<-, >=latex,very thick](1.1,-0.4) arc (-78:-100:2);
\draw (.7, -0.5) node[below]{$\lambda$};

\end{tikzpicture}
\end{minipage}
\begin{minipage}{0.5\linewidth}
 Let $\phi$ be the first received setting, either $\theta_1$ or $\theta_2$. Then, a single outcome $\lambda$ is drawn at random on the segment $[0, 2\pi]$ with the so-called \enquote{gauge probability distribution}~\cite{mf2} $p(\lambda) = (1/4)|\cos(\lambda-\phi)|$. This outcome is transmitted to Alice and Bob separately, immediately after receiving their particular choice, $\theta_1$ or $\theta_2$.
\end{minipage}
~\\
The same outcome $\lambda$ is used subsequently by both Alice and Bob to compute their own variable $\mathsf{X}_1 = (1/2)[1+\mathrm{~sgn}\cos(\theta_1-\lambda)]$ for Alice and $\mathsf{X}_2 = (1/2)[1-\mathrm{~sgn}\cos(\theta_2-\lambda)]$ for Bob. It can be shown~\cite{mf,mf2} that the resulting joint probability is precisely the working distribution $w_{\scriptscriptstyle \Lambda}$. Therefore, the Bell-CHSH inequality is instantaneously violated, as soon as the last selection $\theta_1$ or $\theta_2$ is completed.

With regard to the present model, $\theta_i$ can be specified by a number of bits, i.e., by a number of sub-registers belonging to Alice (resp. Bob) region. Then $(\mathsf{X}_1,\theta_1)$ and  $(\mathsf{X}_2,\theta_2)$ form a pair of correlated regions as described in Sec.  (\ref{pairofregisters}). As a result, the correlation between Alice and Bob regions is non-signaling, which is the core of the EPR paradox. The paradox vanishes when one realizes that each party only perceives the marginal probability in her/his own region.

A similar situation is encountered with the PR-Box, just below.

 \subsection{PR-Box}
\label{PRbox}
 Nonlocal boxes were proposed by Khalfi and Tsirelson~\cite{khalfi} and later by Popescu and Rohrlich (PR)~\cite{popescu} to address the question of quantum correlations. 
The PR-box is a particular device  which exceeds the Tsirelson's bound~\cite{tsirelson} of the Bell-CHSH inequality, which is forbidden in quantum bipartite systems. Therefore, the PR-box is usually regarded as \enquote{super-quantum}.
 Tsirelson identified this bound, $2\sqrt{2}$, as a special value derived for two regions from the Grothendieck inequality defined in general topological tensor product spaces~\cite{grothendieck,pisier}, while leaving open the  case of multipartite systems.  Actually, the violation of Tsirelson's bound is only ruled out for bipartite quantum states.
Indeed, it has been shown that arbitrarily large violations of the inequality are already possible for tripartite systems~\cite{perezgarcia}. Now, we shall see that the Tsirelson's inequality is not a quantum limitation of the PR-box either, because the device is basically quadripartite. This is moreover a simple but non trivial illustration of the effectiveness of the present theory. The following results are completely standard but their interpretation is unconventional.

\subsubsection{Description}
Consider a Boolean algebra of four binary variables $\mathsf{X}_1$, $\mathsf{X}_2$, $\mathsf{X}_3$ and $\mathsf{X}_4.$
The definition of  the PR-Box is the following
\begin{align}
\label{eqPRbox}
\begin{aligned}
\Prob(\mathsf{X}_1;\mathsf{X}_2|\mathsf{X}_3;\mathsf{X}_4)=
\begin{cases}
\frac{1}{2}\quad\mathrm{~if~} \mathsf{X}_1\oplus\mathsf{X}_2=\mathsf{X}_3\wedge\mathsf{X}_4\\
0\quad\mathrm{~otherwise}
\end{cases}
\end{aligned}
\end{align}
where $\mathsf{X}_1, \mathsf{X}_2$ are a pair of  output variables and $\mathsf{X}_3, \mathsf{X}_4$ are the input data. The symbol $\oplus$ stands for exclusive-or (XOR). 
Eq.  (\ref{eqPRbox}) can be expanded as
\begin{center}
\begin{tabular}{>{$}c<{$}|>{$}c<{$}>{$}c<{$}>{$}c<{$}>{$}c<{$}   }
\mathsf{X}_1; \mathsf{X}_2\backslash\mathsf{X}_3; \mathsf{X}_4 & 0 0&01&10&11\\
\hline
 0  0 & 1/2 &  1/2 &  1/2 &  0\\
 0  1 & 0 &  0 &  0 &  1/2\\
 1  0 & 0 &  0 &  0 &  1/2\\
 1  1 & 1/2 &  1/2 &  1/2 &  0\\
\end{tabular}
\end{center}
From the chain rule,  we have
$$\Prob(\mathsf{X}_1;\mathsf{X}_2;\mathsf{X}_3;\mathsf{X}_4)= \Prob(\mathsf{X}_1;\mathsf{X}_2|\mathsf{X}_3;\mathsf{X}_4)\times\Prob(\mathsf{X}_3;\mathsf{X}_4) .$$
%
Construct the classical states $\omega_k$ as a conjunction of 4 variables or their negations,  $\omega_k=(\mathsf{Y}_1;\mathsf{Y}_2;\mathsf{Y}_3;\mathsf{Y}_4)$
where $\mathsf{Y}_i\in\{\mathsf{\overline{X}}_i,\mathsf{{X}}_i\}$ and  $k=8x_1+4x_2+2x_3+x_4+1$ with $\mathsf{Y}_i=\mathsf{\overline{X}}_i$ for $x_i=0$ and $\mathsf{Y}_i=\mathsf{{X}}_i$ for $x_i=1$. Then, there are 16 classical states $\omega_k$ for $k\in\llbracket 1,16\rrbracket$.
Finally let $p_k$ denote  $\Prob(\omega_k)$.

Since the conditional probabilities $\Prob(\mathsf{X}_1;\mathsf{X}_2|\mathsf{X}_3;\mathsf{X}_4)$ are definite, we obtain a linear system.
From Eq. (\ref{eqPRbox}), we have $p_4=p_5=p_6=p_7=p_9=p_{10} = p_{11}=p_{16}=0$ and
\begin{align}
\label{equbox}
\begin{aligned}
p_1=p_{13} &= 0.5 \times  \Prob(-3;-4)\\ 
p_2=p_{14} &= 0.5 \times  \Prob(-3;4)\\
p_3=p_{15} &= 0.5 \times \Prob(3;-4)\\
p_8=p_{12} &= 0.5\times \Prob(3;4) \\
\end{aligned} 
\end{align}
Taking the normalization into account, namely,
$$ \Prob(-3;-4)+\Prob(-3;4)+\Prob(3;-4)+\Prob(3;4)=1,$$
we can eliminate all unknowns except $p_1, p_2, p_3, p_8$ to obtain a reduced LP system, 
\begin{align}
\label{reduced}
\begin{aligned}
p_1+ p_2+ p_3+ p_8 =\frac{1}{2}\\
\mathrm{subject~to~} p_i\ge 0.
\end{aligned}
\end{align}
As a LP problem of 4 variables and rank $m=1$, the solutions are located on a simplex with $r=4$ vertices.
Going back to the real-valued probability space, $\mathcal{P}=Span(\omega_k|k\in\llbracket 1,16\rrbracket)$, the dimension of the LP system is $d=16$ and therefore the rank is $m=13$.  The solutions are still located on a simplex $\mathcal{W}_\mathrm{box}$ of $r=d-m+1=4$ vertices, $w_i= (w_{i,j})$. From Eq. (\ref{reduced}), the entries of $w_i$ are
\begin{center}
\begin{tabular}{C|CCCC}
\mathrm{vertex}&w_{i,1}=w_{i,13}&w_{i,2}=w_{i,14}&w_{i,3}=w_{i,15}&w_{i,8}=w_{i,12}\\
\hline
w_1 & 0.5 &  0    &  0   &  0\\
w_2 &  0   &  0.5 &  0   &  0\\
w_3 &  0   &  0    & 0.5 &  0\\
w_4 &  0   &  0    &  0  & 0.5\\
\end{tabular}
\end{center}
Non mentioned entries are zero. A particular working distribution requires the definition of a specific context, e.g., an assignment of the input data $\mathsf{X}_3$ and $\mathsf{X}_4.$ 
As an illustration, we will describe successively the default context and the CHSH systems with deterministic inputs.

\subsubsection{Uniform box}
Define a uniform box as a box with the default distribution, i.e., a quantum state $(g_\mathrm{box},\mathcal{W}_\mathrm{box})$
where $g_\mathrm{box}$ is both the center of mass of the simplex and the working distribution. The $4$ simplicial coordinates $\mu_i$ are all equal to $1/4$ for $i=1$, $2$, $3$, $4$. The $8$ non-zero entries of $g_{\mathrm{box},j}$ are equal to $1/8$ for $j=1$, $2$, $3$, $8$, $12$, $13$, $14$, $15$. It is convenient to reorder the basis vectors in $\mathcal{P}$ as  $(\tilde{\omega}_1$, $\tilde{\omega}_{13}$, $\tilde{\omega}_2$, $\tilde{\omega}_{14}$, $\tilde{\omega}_3$, $\tilde{\omega}_{15}$,  $\tilde{\omega}_8$, $\tilde{\omega}_{12})$, $(\tilde{\omega}_4$, $ \tilde{\omega}_5$, $\tilde{\omega}_6$, $\tilde{\omega}_7$, $\tilde{\omega}_9$, $\tilde{\omega}_{10}$, $\tilde{\omega}_{11}$, $\tilde{\omega}_{16})$. We have then,
\begin{equation}
\label{wbox}
g_{\mathrm{box}}=\frac{1}{8}(1,1,1,1,1,1,1,1,0,0,0,0,0,0,0,0)
\end{equation}
The variables $\mathsf{X}_3$ and $\mathsf{X}_4$ are non-deterministic.
Consider, e.g., the Boolean function $\mathsf{X}_4 =(\omega_2, \omega_4,\omega_6,\omega_8,\omega_{10}, \omega_{12},\omega_{14},\omega_{16})$. The covector $\mathrm{x}_4$ corresponding to the indicator function of $X_4$ in the reordered dual basis is
%
%
$$\mathrm{x}_4=(0, 0,1,1,0,0,1,1,1,0,1,0,0,0,1,0,1)$$  
Therefore, from Eq. (\ref{wbox}), we compute,
$$\langle X_4 \rangle =\langle \mathrm{x}_4\ g_{\mathrm{box}} \rangle =0.5$$
Similarly, $\langle X_3 \rangle =0.5$.

Let us transcribe the quantum state $(g_\mathrm{box},\mathcal{W}_\mathrm{box})$ into a Hilbert space $\mathcal{H}$ with the natural gauge.  Define 
$ \ket{u_i} =  \ket{\sqrt{w}_i}$. By simple inspection, we have $\braket{u_i}{u_j}=\delta_{ij}$.
The quantum state is transcribed in $\mathcal{H}$ as the following density operator of dimension $16$ and of rank $4$,
$$
\rho_\mathrm{box}= \sum_{i=1}^4 \lambda_i\ket{\sqrt{w}_i}\bra{\sqrt{w}_i}
=
\begin{bmatrix}
\mathsf{J}&\mathsf{O}\\
\mathsf{O}&\mathsf{O}\\
\end{bmatrix}
$$
where $\mathsf{O}$ is the zero matrix of dimension $8$ and 
$$
\mathsf{J}
=\frac{1}{8}\ 
\begin{bmatrix}
J&O&O&O\\
O&J&O&O\\
O&O&J&O\\
O&O&O&J
\end{bmatrix}
\qquad\mathrm{~with~}\qquad J =
\begin{bmatrix}
1&1\\
1&1
\end{bmatrix}
\quad\mathrm{~and~}\quad
O =
\begin{bmatrix}
0&0\\
0&0
\end{bmatrix}
$$

\subsubsection{AB-box}
The Tsirelson bound is computed for \emph{deterministic inputs}.  Let $A, B\in\{0,1\}$.
Define the \enquote{AB-box} as the contextual PR-box with $\mathsf{X}_3=A$ and  $\mathsf{X}_4=B$. Now, we can consider four AB-boxes, i.e., 4 distinct working distributions. From Eqs.  (\ref{equbox}, \ref{reduced}), it turns out that these working distributions $w_{AB}$ in the context $(AB)$, are the extreme points $w_i$ of the simplex $\mathcal{W}_\mathrm{box}$, specifically,
\begin{align*} 
w_{AB} =w_i \mathrm{~where~} i = 1, 2, 3, 4 \mathrm{~for~}(AB)=(00) , (01),(10), (11) \mathrm{~respectively}
\end{align*}
corresponding to four pure states.
Therefore, the AB-boxes can be defined in $\mathcal{P}$ and then perfectly emulated in the classical realm. A possible implementation is proposed in Ref.~\cite{mf2}, using a stage of classical communication.

Let us construct explicitly the four pure states in the 16-dimensional Hilbert space $\mathcal{H}$ already defined. 
Let
$$\ket{\psi_{AB}}=\ket{\sqrt{w}_{AB}}$$
denote four wave vectors of $\mathcal{H}$,
where $\sqrt{w}_i$ is the array $(\sqrt{w}_{i,1}, \sqrt{w}_{i,2}, \dots, \sqrt{w}_{i,16})$.
By simple inspection, the four vectors $\ket{\psi_{AB}}$ are orthonormal in $\mathcal{H}$. They can be generated from e.g. $\ket{\psi_{00}}$ by unitary operators $\mathsf{U}_{AB}$, in fact permutation operators, as
$$\ket{\psi_{AB}}= \mathsf{U}_{AB}\ket{\psi_{00}}$$
By construction each vector $\ket{\psi_{AB}}$ is the wave vector of a PR-box in the context $(AB)$.
Let $\rho_{AB}$ denote the density operators acting on $\mathcal{H}$. We have
$$\rho_{AB}=\ket{\psi_{AB}}\bra{\psi_{AB}}= \mathsf{U}_{AB}\ket{\psi_{00}}\bra{\psi_{00}}\mathsf{U}_{AB}^{-1}= \mathsf{U}_{AB}\rho_{00}\mathsf{U}_{AB}^{-1}$$

Irrespective of the context $(AB)$, define a particular observable 
$\mathsf{S}$ as a diagonal Hermitian operator acting on  $\mathcal{H}$, namely,
\begin{equation}
\label{schsh}
\mathsf{S} = Diag (1,  1,    1,    1,  1,   1, -1,  -1, -1,  -1, -1,-1, -1,-1,  1,  1)
\end{equation}
where the entries are given in the reordered basis in $\mathcal{H}$, namely,  $\ket{1}$, $\ket{13}$, $\ket{2}$, $\ket{14}$, $\ket{3}$, $\ket{15}$, $\ket{8}$, $\ket{12}$, $\ket{4}$, $\ket{5}$, $\ket{6}$, $\ket{7}$, $\ket{9}$, $\ket{10}$, $\ket{11}$, $\ket{16}$ so that the 8 last diagonal entries of $\rho_{AB}$ are zero.
The eigenvalues $s_j$ of  $\mathsf{S}$ are $\pm 1$. They are the corresponding components of a covector $\mathrm{s}$ in the probability space $\mathcal{P}$.
It is straightforward to compute the expectation of $S$ with respect to $w_{AB}$ in $\mathcal{P}$ as,
\begin{align}
\label{sab}
 \langle S\rangle_{AB} =\langle \mathrm{s}w_{AB}\rangle=
\begin{cases}
-1 \quad\mathrm{ if }\ A=B=1\\
+1\quad\mathrm{otherwise}
\end{cases}
\end{align}
where $\langle .\rangle_{AB}$ stands for the expectation value with respect to the working distribution in the deterministic context $(AB)$. 

Let $C, D\in\{0,1\}$. Irrespective of the current context $(AB)$, define 4 new observables, i.e., 4 Hermitian operators derived from $\mathsf{S}$ as 
$$\mathsf{S}_{CD} = \mathsf{U}_{CD}^{-1}\ \mathsf{S}\ \mathsf{U}_{CD}.$$
From Eq. (\ref{sab}) we have
\begin{align}
\label{sabcd}
 \langle \mathsf{S}\rangle_{AB} =\tr (\rho_{AB}\mathsf{S})= \tr (\mathsf{U}_{AB}\rho_{00}\mathsf{U}_{AB}^{-1}\mathsf{S}) =\tr (\rho_{00}\mathsf{S}_{AB})=\langle \mathsf{S}_{AB}\rangle_{00}
 =
\begin{cases}
-1 \quad\mathrm{ if }\ A=B=1\\
+1\quad\mathrm{otherwise}
\end{cases}
\end{align}
%

\subsubsection{Bell-CHSH observable}
For $A, B\in\{0,1\}$, define a new observable as $$CHSH =\mathsf{S}_{AB}+ \mathsf{S}_{A'B}+ \mathsf{S}_{AB'}- \mathsf{S}_{A'B'} ,$$ where $A'=1-A$ and $B'=1-B$.
From Eq. (\ref{sabcd}),  in any particular context, e.g. for definiteness in the pure state $\psi_{00}$, compute the expectation,
\begin{align*}
\langle CHSH\rangle\ident\langle CHSH\rangle_{00}
&=
\langle \mathsf{S}_{AB}+ \mathsf{S}_{A'B}+ \mathsf{S}_{AB'}- \mathsf{S}_{A'B'}\rangle_{00} \\
&=
\langle \mathsf{S}_{AB}\rangle_{00}+\langle \mathsf{S}_{A'B}\rangle_{00}+\langle \mathsf{S}_{AB'}\rangle_{00}-\langle \mathsf{S}_{A'B'}\rangle_{00}\\
& =
\langle \mathsf{S}\rangle_{AB}+\langle \mathsf{S}\rangle_{A'B}+\langle \mathsf{S}\rangle_{AB'}-\langle \mathsf{S}\rangle_{A'B'}\\
\end{align*}
Then, still from Eq. (\ref{sabcd}), we obtain,
$$|\langle CHSH\rangle|= 4$$ 

This result might seem surprising because the expectation $\langle CHSH \rangle$ exceeds both the classical and the quantum bounds whereas the device  is achievable in the purely classical realm.  
Indeed,  the assumption of \enquote{local hidden variables} leads to the Bell-CHSH inequality, $|\langle CHSH \rangle|\le 2$. The assumption of a pure bipartite quantum state leads to Tsirelson inequality, $|\langle CHSH \rangle|\le 2\sqrt{2}$. In addition,
 assuming  non-locality, Wim van Dam~\cite{vandam} has proved that the AB-boxes solve the problem of \enquote{communication complexity} \cite{yao, buhrman}, meaning that all distributed computations can be performed with a trivial amount of classical communication, i.e., with one bit.    
 
 Actually,  none of the three assumptions is met.
 The Bell inequality can be violated because the box is context-dependent. The Tsirelson inequality can be violated because the quantum state is quadripartite. The result by van Dam is bypassed because the classical implementation of any context-dependent system requires an implicit stage of communication~\cite{mf2}. Actually, the paradoxical result of  van Dam should be interpreted as another proof of this latter statement.

\section{Discussion}
\label{discussion}
In this section, we briefly discuss some of the issues encountered in the paper and consider the possible implications of the theory. Beyond, we venture some speculations.

\subsection[\enquote{Born's method} versus Bayesian inference]{The \enquote{Born's method} is a technique of Bayesian inference}
\label{discussborn}
The first fresh ingredient implemented in this article is the use of Bayesian inference to compute Boolean expressions.
This method that we called \enquote{Born's method} is a variant of a technique routinely employed in statistical estimation~\cite{boyd}. 
In return, the meaning of probability is vastly different from its usual signification.
Based especially on the works by J. M. Keynes~\cite{keynes} and R.~T.~Cox~\cite{cox},  probability theory is regarded as an extension of the Aristotelian logic to cases where the variables are not wholly definite. Logical rules are thoroughly retained but they are posited with \emph{real-valued numbers} instead of logical symbols.
Technically, the crucial advantage is the unique ability of real numbers to perform optimization, which dramatically boosts the computational power. 

It happens that any Boolean formula can be expressed as a set of \emph{linear} equations in terms of probability, which explains at the outset why quantum mechanics is linear. By contrast,  linearity is regarded as an axiom in standard quantum theory. It leads in particular to the so-called \enquote{no-cloning theorem} which is actually a direct consequence of the  \enquote{Born's method}.

We emphasize that basically this technique has nothing to do with physics and, in fact, we have used physics only as examples of application. In reality,  we have only described a purely mathematical model, namely,  computing Boolean expressions by Bayesian inference.

\subsection{Bayesian versus Frequentist}
\label{frequentist}
As stressed by Jaynes~\cite{jaynes},  Cox's theorem~\cite{cox} is also a pillar of the \enquote{Bayesian} theory of inference as opposed to the \enquote{orthodox} theory where probability is viewed as a \enquote{Frequency}.
 We adopt Jaynes's terminology: \enquote{a Bayesian probability is something that one assigns in order to represent a state of knowledge}, that is to say in the logical domain, whereas a \enquote{Frequency} is a factual property, that is to say in the experimental domain.
The present model is decidedly based on Bayesian probability.


\subsection{Quantum versus classical}
\label{qversusc}

The existence of different observation windows was acknowledged in 1954 by Max Born himself~\cite{born} in the context of the wave-particle duality:
\enquote{Every object that we perceive appears in innumerable aspects. The concept of the object is the invariant of all these aspects.} 
This was called \enquote{the chameleon effect} by L. Accardi and M. Regoli~\cite{accardi1}.
Now, in a Bayesian theater, this is a platitude:
the particular \enquote{aspect} is the current observation window, 
the \enquote{invariant} is the Bayesian prior 
and  the \enquote{perception} is the current working probability distribution. 
In stark contrast, in the \enquote{orthodox} interpretation, the probability is regarded as a factual characteristic of the object and not of the representation. The fact that quantum physics reflects reality much better than classical physics underlines that the human mind captures this reality by Bayesian inference.

The classical description of physical objects assumes that the different \enquote{aspects} of the same object are independent.  
This approximation is nevertheless justified in everyday physics by the fact that the inter-window correlation depends on a dimensional parameter, namely, the Planck constant, which is negligible with the practical units of daily life.
Therefore, the present theory suggests that the \emph{classical limit}  should be defined as the \emph{approximation in which the different observation windows are assumed to provide independent results}. 
In reality,  there is no classical world, but only different levels of approximation~\cite{jaynes10,polkinghorne}.

\subsection{Contextuality and free will}
\label{contextuality}
From Definition (\ref{defcontextuality}), a system is context-dependent when the working probability distribution depends on an exogenous choice. In a way, this exogenous choice can be regarded as the expression of the free will of the observer. 
We encountered two different forms of contextuality, source contextuality and window contextuality. 

\emph{Source contextuality.} First, in the source window, the exogenous choice is to select one particular solution on the specific simplex. This is achieved by introducing a contextual probability distribution, still leaving some uncertainty described by the simplicial entropy in the observation window and the von Neumann entropy in the Bayesian theater. This input is \emph{intrinsic} to the system in that it can be assimilated to the specification of boundary conditions or the setting of Noether constants.

\emph{Window contextuality.} Second, in general systems, the other exogenous choice is to select a particular observation window, i.e., a particular Boolean variable batch, interpreted as a particular point of view on the system. This corresponds to the free choice of a basis in the Hilbert space and is in no way intrinsic to the system.
The Bell-CHSH inequality and the violation of Tsirelson's bound as well as the uncertainty principle and Kochen-Specker's theorem \cite{kochen} bear witness to this window contextuality.

\subsection{Spin-off in physics and beyond}
 
 The present model should have spin-off in different areas, starting with physics, considered the science of observation based on reasoning~\cite{wiener}. The model only deals with logical concepts and can therefore provide only a bare landscape of the world, free from any  specific ontological or  \enquote{ontic} ingredient. Perhaps this is not so essential, especially since genuine ontological elements are undoubtedly unimaginable and therefore unfalsifiable, whereas the candidate \enquote{beables}, whether fire, aether, epicycles, points, vectors, strings or branes are highly problematic or at best purely phenomenological models.
 
This suggests circumventing any specific ontology and following the celebrated Wheeler's doctrine, \enquote{It from bit}. This means that abstract information is the ultimate ingredient while deliberately ignoring any ontological significance.
On this basis, let us submit a few  speculative spin-off in a  purely information-theoretic model.

\emph{Towards new foundations of physics:}
In standard physics the universe is usually represented by a wave vector, that is, a \emph{pure state}. In the present framework, this would describe an information register set to zero, so that the complexity of the world would be just an artifact only due to a sophisticated observation window. 
Paradoxically, for the universe to have a non-trivial content, a pure state is excluded and only a mixed state is acceptable.

Now, the world is reconstructed from a number of observations, expressed for convenience in terms of binary Boolean variables regarded as discrete degrees of freedom. 
Equivalently, the state of the universe can be represented by a gauge group whose invariant observables express symmetry, thus joining Klein's Erlangen program in mathematics~\cite{klein} (see e.g. J.-B.~Zuber~\cite{zuber}).
The deep insight by Steven Weinberg~\cite{weinberg}, namely, \enquote{specifying the symmetry group of Nature may be all we need to say about the physical world}, is fully consistent with this view. 
This should also explain why standard quantum mechanics is so efficient although using problematic prerequisites. In reality, the wave vector would simply be the test witness characterizing the symmetry group.

{Classical physics} describes the universe at a given time as a collection of windows expressed with disparate units and considered approximately independent, so that most of the residual correlation between windows is captured by the so-called \enquote{dimensional analysis}. By contrast, the full correlations are taken into account in quantum physics and the same atlas is conveniently described by a unique Hilbert space, via the iconic Planck constant to restore commensurability between the disparate units. Conversely, the tiny value of this constant in the usual units legitimates the classical approximation in every day physics.
 
In cosmology, the cosmic time should be defined by a monotonic function of the von Neumann entropy of the quantum state. 
Thus,  the arrow of time as well as the \enquote{tendency to disorder} become direct consequences of the maximum entropy principle. As a result, the entropy increases over time and  information is in no case conserved, thereby solving the famous black hole information paradox~\cite{preskill1}. By contrast, in standard physics, reversibility contradicts thermodynamics whereas both are pillows of the theory.
\paragraph{}

\emph{Beyond physics}, this approach is likely to be powerful in all area of reasoning.

The first application concerns Data Science. It provides an explanation of the speedup of both Bayesian computation and quantum computing. 
This explains especially the efficiency of neural networks which are implicit Bayesian calculations.
This efficiency ultimately rests on the unique ability of real numbers to perform optimization unlike discrete implementations.
Therefore,  the results obtained with quantum computers, e.g. for integer factoring~\cite{shor}, can also be achieved by perfectly classical computers~\cite{mf4}, potentially implemented as artificial neuron networks.

Beyond, this suggests that the Church-Turing principle may not be the end of the history and that Bayesian inference could be a more powerful tool than the Turing machine to conceive universal computation as previously suggested, but only for quantum computation,  by D. Deutsch \cite{deutsch}.

Bayesian inferences could even have spin-off in \emph{pure mathematics} because  the means of deducing mathematics from logic could include Bayesian inference and not only deduction. Leopold Kronecker is famous for having declared that \enquote{God made the integers, all else is the work of man.}~\cite{gray}  One step further, one could assert that \enquote{God made logical rules, all else is the work of man.}  At last, more punctually, quantum information could explain the hitherto unknown link~\cite{bouleau} between the theory of potential and probability~\cite{meyer}.

More unexpected for quantum physicists, though suspected by David Bohm and Basil Hiley~\cite{hiley}, other sciences including \emph{soft sciences}  already benefit from this approach. Applications have been described, e.g.,  in cognition and decision making~\cite{aerts, busemeyer, bruza}, psychology~\cite{dzhafarov,dzhafarov1}, social science~\cite{haven} or grammatical language~\cite{clark}. 
Beyond cognition, other emblematic examples could be found in biology, e.g. in both the immune system and 
immunotherapy and even in evolution theory.

\section{Conclusion}
\label{conclusion}
Our goal was to propose an interpretation of quantum formalism. 
Although it is a long-standing issue, whose origins can be traced back to von Neumann~\cite{vonneumann1}, the foundations of quantum mechanics have remained elusive, giving rise to questioning and  discomfort~\cite{ionicioiu}. The probabilistic \enquote{Born interpretation} aroused the Einstein's famous sentence, \enquote{I, at any rate, am convinced that He does not throw dice}~\cite{einstein1}.  Later, in a celebrated lecture~\cite{feynman1}, R.~Feynman gave his equally famous verdict, \enquote{I think I can safely say that nobody understands quantum mechanics}. Let us finally quote the striking Jaynes' opinion: \enquote{A standard of logic that would be considered a psychiatric disorder in other fields, is the accepted norm in quantum theory}~\cite{clearing}.

To address this discomfort, countless approaches have been devised.  Some authors tried to circumvent the conventional logic. Others attempted to reinterpret the experimental results. Finally, some simply denied the existence of a problem. In a tasty paper, updated in 2002~\cite{fuchs}, Christopher  Fuchs enumerated with humor a number of \enquote{religions}: \enquote{The Bohmians~\cite{bohmian}, the Consistent Historians~\cite{griffiths1}, the Transactionalists~\cite{cramer}, the Spontaneous Collapseans~\cite{ghirardi}, the Einselectionists~\cite{zurek}, the Contextual Objectivists~\cite{grangier}, the outright Everettics~\cite{deutsch1,vaidman}, and many more beyond that}.  Recent approaches try to derive quantum logic from \emph{ad hoc} information-theoretic extra principles assumed \enquote{reasonable} or, following R. W. Spekkens~\cite{spekkens1}, propose frameworks claimed \enquote{operational}~\cite{chiribella, hardy, mueller, masanes1, chiribella1, oeckl}, based on the compatibility with specific information processing tasks. 
Epistemic approaches propose generalized probabilistic theories (GPT) comparing quantum and classical probabilities 
\cite{pitowski, janotta}.
Specifically, new frameworks aim to identify the additional axioms needed to derive the quantum formalism from probabilistic constraints, e.g, from \enquote{information causality} or from entropy \cite{pawlowski, barnum}.
Another appealing approach inspired from thermodynamics is to use an entropic method of inference~\cite{caticha1}.
Eventually, a more direct way is to compare quantum states with Bayesian states of knowledge~\citep{fuchs1,leifer,raedt,mermin4}.  

In the present paper, we abstain from introducing any extra axiom but we support the information-theoretic interpretation of quantum formalism based on Bayesian inference theory.

Although using  quantum terminology when appropriate, we have basically dealt with classical information in a classical memory, but at the end, we obtain the exact apparatus of quantum information. This means that quantum information as such is nothing but information itself and therefore independent of any physical content.
 Our major conclusion, as sketched in Sec. (\ref{nutshell}), is somewhat baffling: Quantum information is simply classical information processed by Bayesian inference theory. 

As far as quantum formalism itself is concerned, the current model is the first to logically \emph{deduce} from information theory its fundamental characteristics, almost always posited from the outset as seemingly arbitrary postulates:
 Why is the theory probabilistic? Why is the theory linear? Where does the Hilbert space come from?
Also, most of the emblematic paradoxes, such as entanglement, contextuality, nonsignaling correlation, measurement problem, no-cloning theorem etc., find a perfectly rational explanation. At last the controversial concept of Shannon information conveyed by a wave vector,  or stored in the system is clarified.

Beyond physics, quantum information appears as a multipurpose technique for analyzing a system of logical constraints, in line with classical information.
Whereas classical information is the universal tools of \emph{logic},  quantum information in the universal tool of \emph{inference}. This is perhaps the most important conclusion of this article.

\phantomsection
\addcontentsline{toc}{section}{References}
\bibliography{biblio}

\end{document}